\documentclass[acmsmall]{acmart}
\AtBeginDocument{%
  }

\setcopyright{acmlicensed}
\copyrightyear{2018}
\acmYear{2018}
\acmDOI{XXXXXXX.XXXXXXX}

\acmJournal{JACM}
\acmVolume{37}
\acmNumber{4}
\acmArticle{111}
\acmMonth{8}

\usepackage{acronym}
\usepackage{setspace}
\usepackage{wasysym}
\usepackage{longtable}
\usepackage[figuresright]{rotating}
\usepackage{epsfig,endnotes,color}
\usepackage{graphicx}
\usepackage{pifont}
\usepackage{amsmath}
\usepackage{caption}
\usepackage{subcaption}
\usepackage{url}

\usepackage{CJKutf8}
\usepackage{lscape, latexsym, algorithmic, multirow}
\usepackage[linesnumbered, vlined, ruled]{algorithm2e}
\usepackage{multirow}
\usepackage{mathtools, bbm, color}
\usepackage{booktabs}
\usepackage{amsthm,mathrsfs,amsfonts,dsfont}
\usepackage{hhline}
\usepackage{tabu}
\usepackage{colortbl}
\usepackage{enumerate}
\usepackage{booktabs}
\usepackage{multirow}

\usepackage{array}
\usepackage{adjustbox}
\usepackage{booktabs, 
            multirow, threeparttable}   
\usepackage{siunitx}                    
\usepackage{xparse}                     

\usepackage[graphicx]{realboxes}
\usepackage{float}
\usepackage{tikz}

\newcommand{\circlewithtext}[2][2mm]{%
    \begin{tikzpicture}[baseline=(C.base)]
        \node[draw, circle, minimum size=#1, inner sep=0pt, fill=black, text=white] (C) {\tiny #2};
    \end{tikzpicture}%
}

\definecolor{newblue}{RGB}{144, 191, 213}
\newcommand{\yellowcircle}[2][3mm]{%
    \begin{tikzpicture}[baseline=(C.base)]
        \node[draw, circle, minimum size=#1, inner sep=0pt, fill=newblue, text=black,font=\bfseries] (C) {\small #2};
    \end{tikzpicture}%
}
  
\usepackage{xcolor}
\usepackage{tcolorbox}
\newtcolorbox{mybox}{ colframe=black,colback=gray!15,boxrule=1pt,arc=2pt,left=2pt,right=2pt,top=1pt,bottom=1pt}

\newif\ifcommentcond
\commentcondtrue 

\newif\ifupdatecond
\updatecondfalse 

\newcommand{\surveyPaperNum}{107}

\begin{document}

\title{A Survey of Operating System Kernel Fuzzing}

\author{Jiacheng Xu}
\email{stitch@zju.edu.cn}
\affiliation{%
  \institution{Zhejiang University}
  \city{Hangzhou}
  \country{China}
}

\author{He Sun}
\email{qldxtest@gmail.com}
\affiliation{%
  \institution{Institute for Network Science and Cyberspace, Tsinghua University}
  \city{Beijing}
  \country{China}
}

\author{Shihao Jiang}
\email{sh.jiang@zju.edu.cn}
\affiliation{%
  \institution{Zhejiang University}
  \city{Hangzhou}
  \country{China}
}

\author{Qinying Wang}
\authornote{These authors are co-corresponding authors.}
\email{wangqinying@zju.edu.cn}
\affiliation{%
  \institution{Zhejiang University}
  \city{Hangzhou}
  \country{China}
}

\author{Mingming Zhang}
\email{zhangmm@mail.zgclab.edu.cn}
\affiliation{%
 \institution{Zhongguancun Laboratory}
 \city{Beijing}
 \country{China}
}

\author{Xiang Li}
\email{lixiang@nankai.edu.cn}
\affiliation{%
  \institution{Nankai University}
  \city{Tianjin}
  \country{China}
}

\author{Kaiwen Shen}
\email{kaiwenshen17@gmail.com}
\affiliation{%
  \institution{Institute for Network Science and Cyberspace, Tsinghua University}
  \city{Beijing}
  \country{China}
}

\author{Charles Zhang}
\email{charles@vul337.team}
\affiliation{%
  \institution{Tsinghua University}
  \city{Hangzhou}
  \country{China}
}

\author{Shouling Ji}
\email{sji@zju.edu.cn}
\affiliation{%
  \institution{Zhejiang University}
  \city{Hangzhou}
  \country{China}
}

\author{Peng Cheng}
\authornotemark[1]
\email{lunarheart@zju.edu.cn}
\affiliation{%
  \institution{Zhejiang University}
  \city{Hangzhou}
  \country{China}
}

\author{Jiming Chen}
\email{cjm@zju.edu.cn}
\affiliation{%
  \institution{Zhejiang University}
  \city{Hangzhou}
  \country{China}
}


\begin{abstract}
The Operating System (OS) kernel is foundational in modern computing, especially with the proliferation of diverse computing devices. However, its development also comes with vulnerabilities that can lead to severe security breaches. Kernel fuzzing, a technique used to uncover these vulnerabilities, poses distinct challenges when compared to user-space fuzzing. These include the complexity of configuring the testing environment and addressing the statefulness inherent to both the kernel and the fuzzing process. Despite the significant interest from the community, a comprehensive understanding of kernel fuzzing remains lacking, hindering further progress in the field.
In this paper, we present the first systematic study focused specifically on OS kernel fuzzing. We begin by outlining the unique challenges of kernel fuzzing, which distinguish it from those in user space. Following this, we summarize the progress of 107 academic studies from top-tier venues between 2017 and 2025. To structure this analysis, we introduce a stage-based fuzzing model and a novel fuzzing taxonomy that highlights nine core functionalities unique to kernel fuzzing. Each of these functionalities is examined in conjunction with the methodological approaches employed to address them. Finally, we identify remaining gaps in addressing challenges and outline promising directions to guide forthcoming research in kernel security.
\end{abstract}




\maketitle

\section{Introduction}
OS kernels are central to modern computing systems, enabling communication between software and hardware components. 
Given the OS kernel's central role, its vulnerabilities lead to serious security breaches, including privilege escalation, sensitive data leakage, and remote code execution.
For example,  the \textit{Dirty Cow} vulnerability \cite{dirtycow} in the Linux kernel is infamous for enabling unauthorized privilege escalation, allowing attackers to manipulate and execute code at the root level. The risk posed by such vulnerabilities is amplified by growing and increasingly complex mobile environments, making it critical to secure against these threats.
Meanwhile, fuzzing has proven to be an effective and practical approach for vulnerability discovery.
Therefore, OS kernel fuzzing techniques have attracted significant attention from the research community \cite{chen2022sfuzz, Syzkaller,pandey2019triforce,chen2021syzgen,sun2021healer}. 

Compared to user-space fuzzing, OS kernel fuzzing presents significant and complex challenges for the following reasons.
First, unlike applications operating in controlled and uniform environments, kernel code interacts with a broad array of hardware components, each featuring its own drivers and peculiarities \cite{wu23devfuzz, ma2022print}. This intricate interplay increases the risk of system-wide crashes or instability in the event of kernel faults, making it precarious \cite{marius2018what, maier2019unicorefuzz}.
As a result, creating a consistent and reliable testing environment becomes particularly challenging.
Second, synthesizing test cases for kernels is usually more challenging than for applications. The difficulty stems from the need to handle a wide variety of complex, structured inputs, e.g., system calls (syscalls) \cite{shankara2018moonshine, sun2021healer, hao2023syzdiscribe} and peripherals \cite{feng2020p2im, jiang2021ecmo}, whose specifications are often deeply embedded within the kernel codebase.
Finally, the kernel’s inherent complexity and low-level nature introduce additional obstacles. It is challenging to precisely control kernel actions, monitor its internal state, and accurately interpret its responses to inputs during fuzzing \cite{liu2020fans}. These difficulties are further amplified by challenges related to scalability and lightweight design, which tend to become more pronounced as the fuzzing process grows or evolves.

Owing to these inherent complexities, OS kernel fuzzing techniques have become a major focus of extensive research. The rapidly expanding collection of OS kernel fuzzing techniques shows wide variation in goals and methods across different stages of the fuzzing pipeline.
It is essential to conduct a deeper investigation into their shared characteristics and the specific challenges they aim to address. Additionally, assessing performance trade-offs and uncovering untapped opportunities for future advancements are vital to furthering progress in this field.
However, thus far, no systematic review of the OS kernel has been conducted. Existing surveys mainly focus on general fuzzing techniques and evaluation criteria \cite{klees2018evaluating,wang2020sok,yun2022fuzzing, schl2024prudent}.  Specifically, fuzzing for embedded systems, representing a related yet distinct line of work, is introduced \cite{zhu2022fuzzing}.

To achieve this, we conduct an extensive review of \surveyPaperNum \ OS kernel fuzzing papers published in top-tier conferences between 2017 and August 2025, providing insights into the three research questions:

\begin{itemize}
\item \textbf{RQ1:} What desired functionalities distinguish OS kernel fuzzing from general fuzzing? 

\item \textbf{RQ2:} What are the techniques adopted for kernel fuzzing to satisfy the desired functionalities?

\item \textbf{RQ3:} 
What are the challenges and future opportunities in the field of kernel fuzzing?

\end{itemize}

Although coverage- and crash-based metrics can provide insight into a kernel fuzzer’s effectiveness, they are difficult to compare across different studies because of variations in definitions, experimental setups, and research objectives. Instead of relying solely on these metrics, we focus on the specific functionalities each technique implements, its applicability, and its methodological contributions. This functionality-oriented perspective offers a more practical understanding of a fuzzer’s impact and utility.
With this in mind, we first present the background of OS kernel fuzzing in \textbf{Section~\ref{sec:preliminary}}, followed by our surveying methodology and results in \textbf{Section~\ref{sec:collection}}. In \textbf{Section~\ref{sec:overview}}, we outline the unique challenges of fuzzing in the OS kernel domain. In this regard, we propose a stage-based fuzzing model that decomposes the fuzzing process into discrete steps. For each stage, we describe the essential functionalities and key techniques employed by kernel fuzzers, addressing \textbf{RQ1}. 
Afterward, we provide a discussion of the existing proposals targeting each stage, i.e., \textit{environment preparation} (\textbf{Section~\ref{sec:environ}}), \textit{input model} (\textbf{Section~\ref{sec:input}}) and \textit{fuzzing loop} (\textbf{Section~\ref{sec:loop}}).
Drawing on qualitative assessment criteria, 
we emphasize the implications learned from existing approaches and suggest promising technical solutions, responding to \textbf{RQ2}. 
Additionally, \textbf{Section~\ref{sec:discussion}} addresses the challenges and future directions of kernel fuzzing research, partially informed by a case study. This analysis contributes to answering \textbf{RQ3}.
Finally, we provide a conclusion in \textbf{Section~\ref{sec:conclusion}}.
 
In summary, we make the following key contributions:
\begin{itemize}
\setlength{\itemsep}{0pt}
\setlength{\parsep}{0pt}
\setlength{\parskip}{0pt}
    \item To the best of our knowledge, we present the first systematization of knowledge of OS kernel fuzzing techniques based on the review and analysis of \surveyPaperNum\ leading literature.
    \item  We uncover the unique characteristics of OS kernel fuzzing compared to user-space fuzzing and develop a comprehensive taxonomy encompassing three core stages and nine essential functionalities.
    \item Using this taxonomy, we systematize advancements in the field and analyze gaps between desired functionalities and current practices. 
    \item We identify existing challenges and highlight promising future directions. 
\end{itemize} 
\section{Preliminaries}
\label{sec:preliminary}

\subsection{Role of OS Kernel}

\begin{figure}[h]
    \centering
    \includegraphics[width=0.8\linewidth]{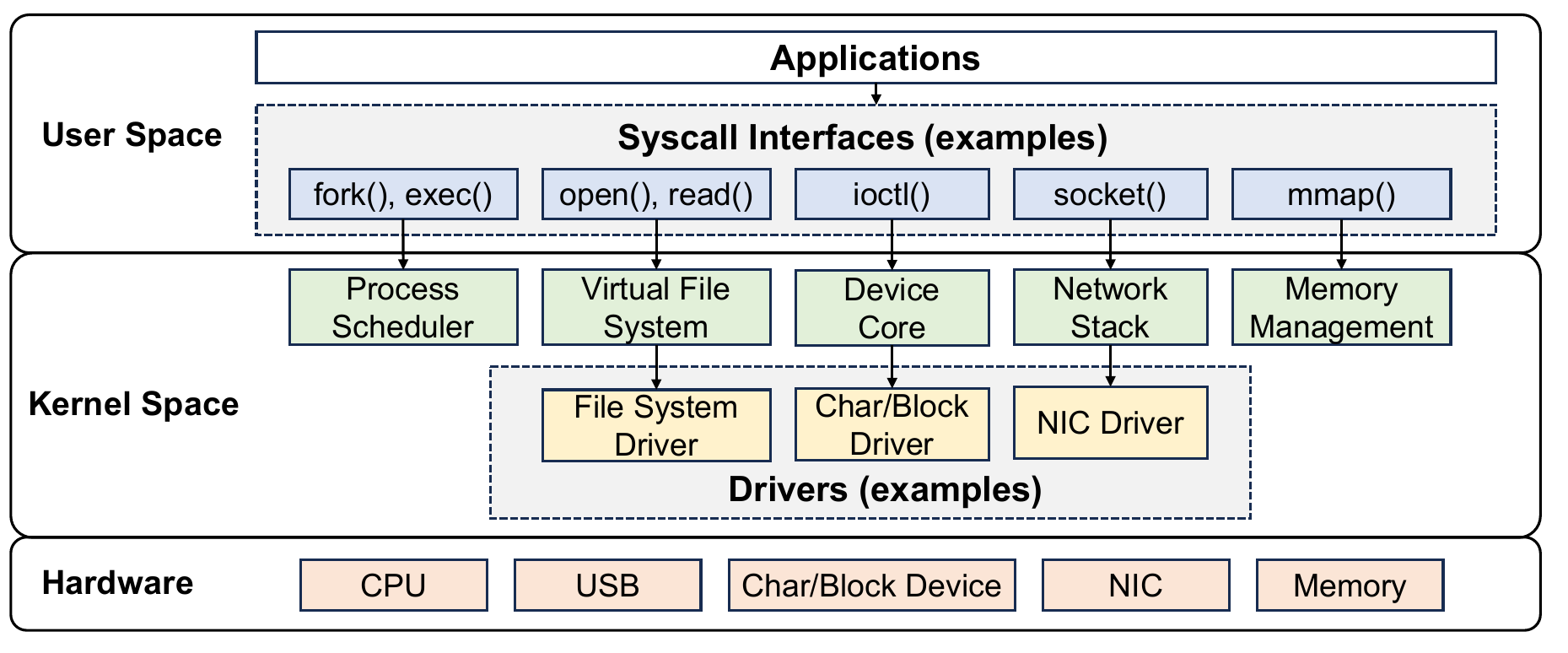}
    \caption{Typical role and architecture of OS kernel in modern computer systems.}
    \label{fig:kernel}
\end{figure}

In modern computer systems, the OS kernel serves as the fundamental component responsible for managing hardware resources and providing essential abstractions to user-space applications. At the highest level, the user-space hosts applications and runtime libraries, which interact with the kernel through system calls (syscalls) to request various services. At the lowest level lies the hardware layer, comprising physical components such as the CPU, memory, and peripheral devices. Situated between these layers, the kernel operates with elevated privileges and functions as a critical mediator, coordinating communication between software and hardware. It performs core tasks such as process scheduling, memory management, file system operations, inter-process communication, and device control \cite{hahm2016os}. These functions ensure that multiple applications can execute concurrently, securely, and efficiently while abstracting away the complexity of direct hardware management. The typical role and position of the OS kernel within this layered architecture are illustrated in Figure~\ref{fig:kernel}. As the backbone of virtually all computing platforms---from traditional servers and desktops to mobile and embedded systems---the kernel constitutes a foundational software infrastructure upon which higher-level services and applications depend. Its reliability is therefore critical not only to system stability but also to performance and security. Therefore, kernels are subject to rigorous testing aimed at improving overall reliability. Among testing methods, fuzzing has emerged as one of the most widely adopted approaches. For example, syzbot \cite{syzbot}, which is integrated into the upstream Linux development process, has reported over 5,000 bugs.

\subsection{Categories of OS Kernels}
With the rapid expansion of computing devices, their scope has extended beyond traditional PCs and servers to encompass a broad spectrum of mobile terminals, embedded systems, and Internet-of-Things (IoT) environments. 
This diversification has led to a corresponding increase in the variety of OS kernels tailored to different application domains.
General-purpose kernels are designed to provide comprehensive functionality and flexibility across multiple platforms, powering millions of devices in diverse domains, particularly within desktop environments. Prominent examples include Linux \cite{linus1991linux}, Windows NT \cite{windows}, and XNU \cite{macos}, all of which emphasize user experience and broad compatibility. In contrast, real-time operating system (RTOS) kernels, such as FreeRTOS \cite{freertos}, prioritize deterministic scheduling and minimal latency to satisfy strict timing requirements in embedded and IoT settings. These kernels are often tightly coupled with heterogeneous hardware, presenting unique challenges for emulation and testing. Complementing these, mobile OS kernels, exemplified by the Android kernel, share certain traits with RTOS. At the same time, they are specifically optimized for energy efficiency, fine-grained access control, and the delivery of rich system services. 
These characteristics, while enhancing functionality, also introduce additional security risks through expanded interfaces. Beyond these categories, there are domain-specific kernels designed for resource-constrained IoT devices \cite{Ioannis2023firmsolo, ioannis2024pandawan, tinyos} as well as for confidential computing \cite{optee, marcel2023teezz}.
The pervasive and foundational nature of OS kernels highlight the critical importance of ensuring their security, reliability, and resilience.

\section{Paper Collection} \label{sec:collection} 

\subsection{Survey Methodology}


We first outline the methodology we follow to collect papers. 
 In this paper, the scope of OS kernels encompasses architectures that provide essential services necessary for system functionality, 
 including hardware abstraction layer, driver model, memory management and scheduling \cite{hahm2016os}.
 We find that these kernels have evolved beyond traditional server or desktop models, taking on more diverse and specialized forms.
 Thus, our analysis includes kernels from general-purpose OS and their customizations \cite{linus1991linux, macos, windows, android, henkel2006embeded},  RTOS \cite{vxworks, barabanov1997linux, mera2021dice}, TEE-OS \cite{taras2019mtower, optee}, ROS \cite{kim2022robofuzz, bai2024multi, shen2024enhancing} and nano ones \cite{freertos, mbedos}, covering a range of environments from desktop to IoT devices.
To ensure a comprehensive survey, we followed these steps: 

\begin{enumerate}
    \item \textbf{Venue selection}. Our study primarily focuses on papers published between 2017 and 2025 at $A\ast$ software engineering, cyber security and computer systems venues ranked by CORE2023 \cite{core2023}. The selected publication venues are listed in Table \ref{tab:venues}.

\setlength{\tabcolsep}{10pt}
\begin{table}[htbp]
\scriptsize
\centering
\caption{The selected top-tier venues from software engineering, cyber security and computer systems.}
\label{tab:venues}
\begin{tabular}{@{}cll@{}}
\toprule
Research Topic                        & \multicolumn{1}{c}{Type} & \multicolumn{1}{c}{Publications} \\ \midrule
\multirow{2}{*}{Software Engineering} & Conferences              & ICSE, ISSTA, FSE, ASE            \\ \cmidrule(l){2-3} 
                                      & Journals                 & TOSEM, TSE                       \\ \midrule
\multirow{2}{*}{Cyber Security}       & Conferences              & USENIX Security, S\&P, CCS, NDSS  \\ \cmidrule(l){2-3} 
                                      & Journals                 & TDSC, TIFS                       \\ \midrule
\multirow{2}{*}{Computer System}      & Conferences              & OSDI, SOSP, USENIX ATC, EuroSys, ASPLOS   \\ \cmidrule(l){2-3} 
                                      & Journals                 & -                                \\ \bottomrule
\end{tabular}
\end{table}
    
    \item \textbf{Keyword match}. We then conducted a preliminary review of the literature and identified a set of keywords relevant to OS kernel fuzzing. These keywords were used to construct the final search query: \textit{("operating system" OR "OS kernel" OR "Android services" OR "rtos" OR "firmware") AND ("fuzzing" OR "fuzzer" OR "testing")}. The search was performed using DBLP, which is a widely recognized bibliographic database in the field of computer science, and the results were restricted to papers published in the previously selected venues. This process resulted in an initial collection of 134 papers.
    \item \textbf{Inclusion / exclusion criteria}. We manually inspected the papers collected in the previous step and retained those that met our inclusion criteria. Specifically, a paper was included in the survey if it proposed a novel fuzzing method explicitly designed for OS kernel or its subsystems. In contrast, studies that focused solely on bare-metal firmware or addressed fuzzing techniques not specific to kernels, such as those targeting general network protocols, were excluded. After applying these criteria, a total of 99 papers were selected for further analysis.
    \item \textbf{Snowballing}. Finally, we performed both forward and backward snowballing to ensure broader coverage and capture relevant studies that may have been missed in the initial search. As a result, a total of \surveyPaperNum\ papers were identified for inclusion in the survey.
\end{enumerate}

\subsection{Survey Result}
\begin{figure}[htbp]
    \centering

  \begin{subfigure}[b]{0.45\textwidth}
    \centering
    \includegraphics[width=0.65\linewidth]{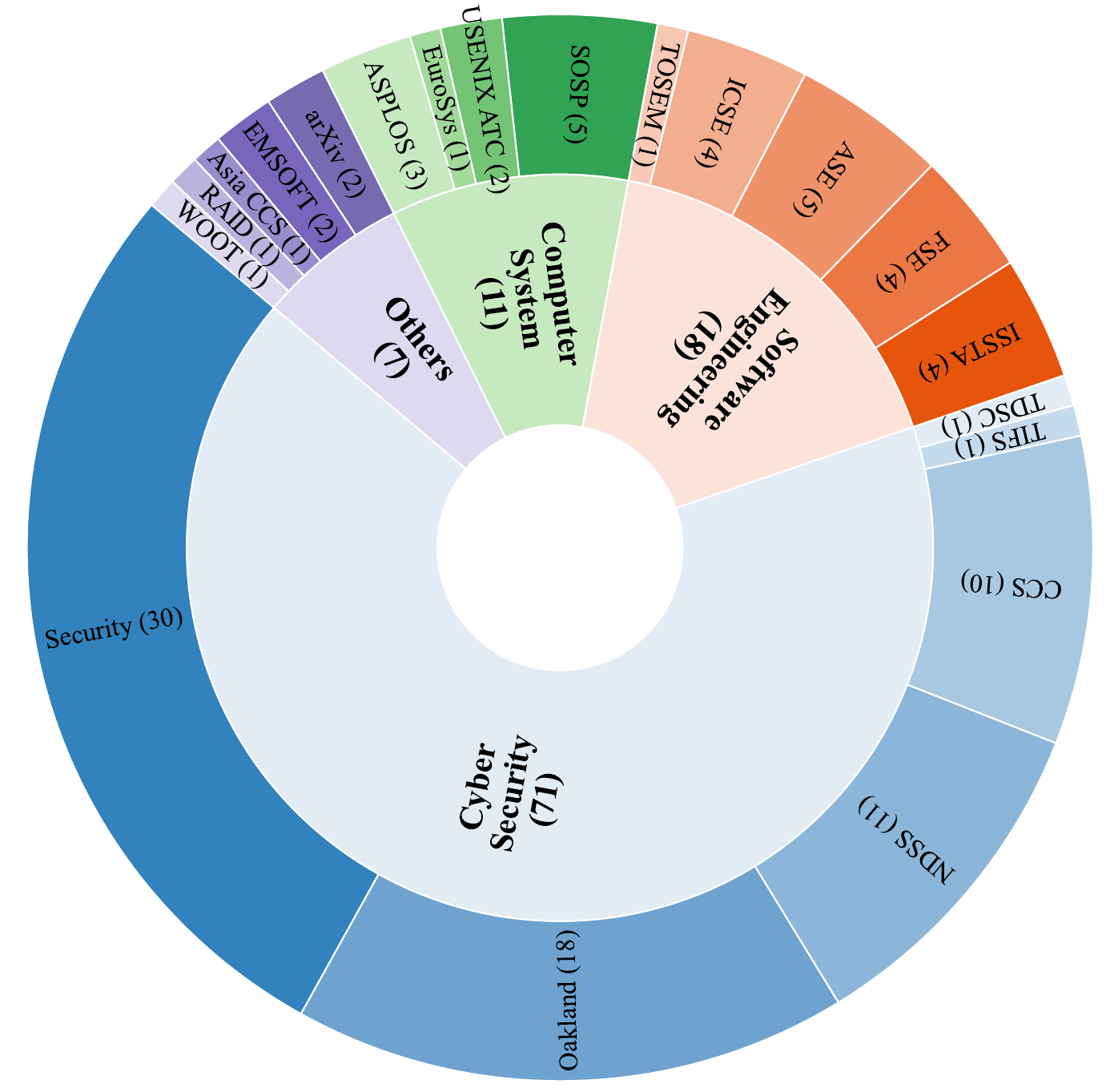}
    \caption{Distribution of papers by venues.}
    \label{fig:stats_venue}
  \end{subfigure}
  \begin{subfigure}[b]{0.45\textwidth}
    \centering
    \includegraphics[width=0.79\linewidth]{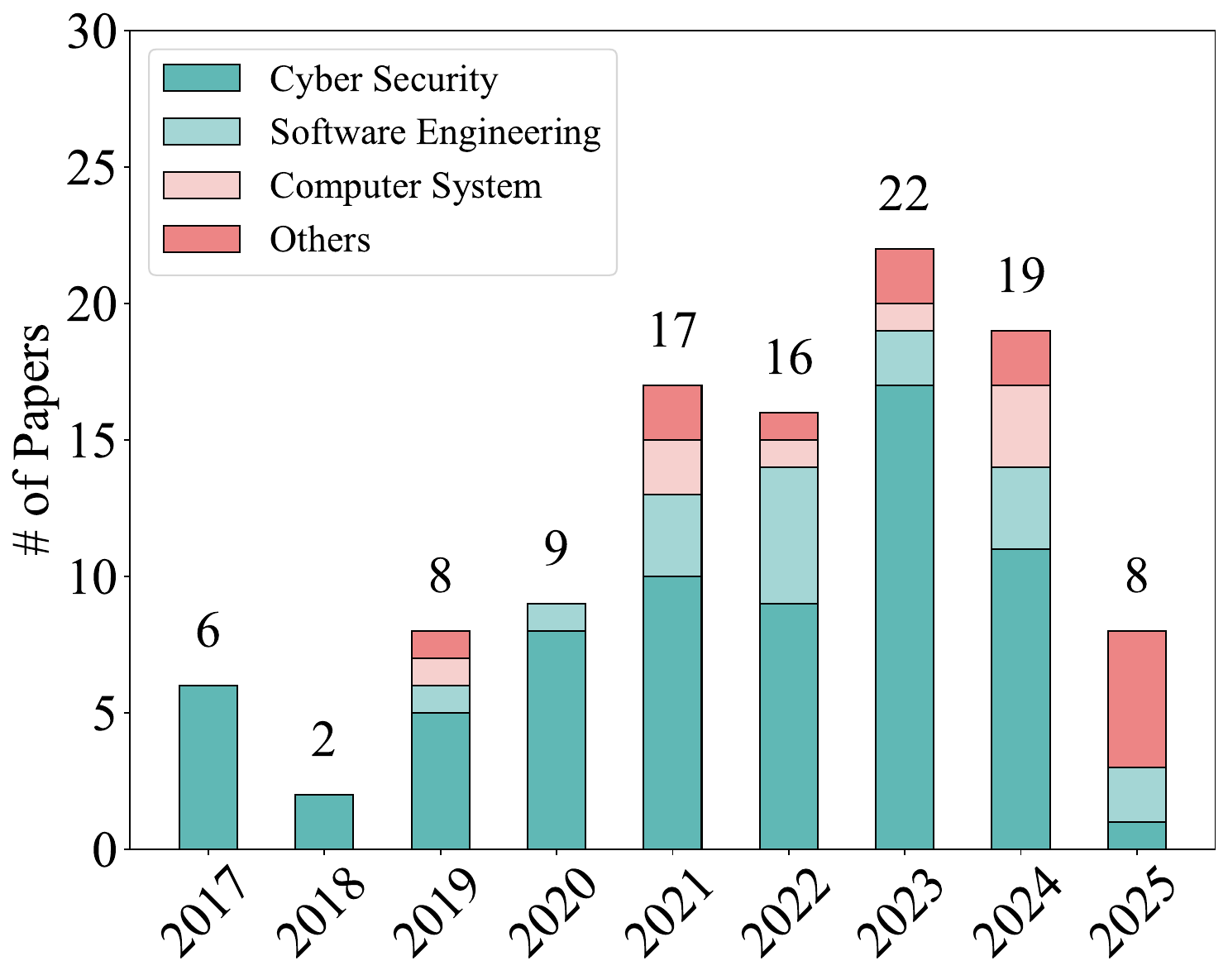}
    \caption{Distribution of papers by years.}
    \label{fig:stats_year}
  \end{subfigure}
    \caption{Statistics of the collected papers.}
    \label{fig:stats}
\end{figure}

At the time of submission, our dataset included papers published between 2017 and August 2025, covering publicly available works (including early-access and preprint versions) from top-tier venues.
Based on these collected papers, we analyze the current research landscape and  trends in OS kernel fuzzing from three perspectives: publication venues and publication years. The statistical results are presented in Figure~\ref{fig:stats}.

\noindent \textbf{Publication venues.} The surveyed literature is distributed across 20 different venues. A majority of OS kernel fuzzing papers are published in cyber security venues reflecting the high security risks associated with kernel vulnerabilities. This accounts for 67\% of the total, as illustrated in  Figure~\ref{fig:stats_venue}. Software engineering venues follow, comprising 17\%, while system conferences account for 9\%. Notably, 28\% of the papers were published at the USENIX Security Symposium, making it the most prominent venue in this research area.

\noindent \textbf{Publication years.} Figure~\ref{fig:stats_year} shows the distribution of papers by publication year. There is a clear upward trend in the number of kernel fuzzing publications, indicating the increasing importance of this research area and the growing attention it has received from the academic community. The apparent decline in 2025 is due to the fact that the year is not yet complete, and many papers have not been publicized at the time of this survey.

\section{Overview of Kernel Fuzzing} \label{sec:overview}

\subsection{Challenges in Kernel Fuzzing} \label{sec:challenges}
As previously discussed, the OS kernel occupies a unique position in the system architecture, bridging between the user space and the hardware layer.
Thus, the complexity and privileged nature of OS kernels introduce distinct challenges in the design and implementation of fuzzing components, distinguishing kernel fuzzing from traditional fuzzing campaigns. To address \textbf{RQ1}, the challenges are categorized as follows, with a side-by-side comparison provided in Table~\ref{tab:challenge}:
\begin{itemize}
    \item \textbf{C1: Constrained running environment}. In user-space, programs under test can typically run and restart in isolation with negligible overhead. By contrast, kernel fuzzing must operate in a constrained environment due to the tight coupling between OS kernels and underlying hardware. Kernels are fundamentally designed to support a broad spectrum of physical devices, and many  of their functionalities depend on the availability of corresponding hardware dependencies, such as peripheral devices and embedded platforms \cite{clements2020halucinator, Ioannis2023firmsolo, wu23devfuzz}. Deploying real devices is costly and does not scale, while emulators offer only partial coverage of device diversity. For example, Linux supports more than 13,000 PCI devices, yet the widely used emulator  \cite{bellard2005qemu} only implements fewer than 130 \cite{zhao2022semantic}.  
    In addition, kernels are often deployed in diverse and challenging contexts, including closed-source systems and resource-constrained scenarios \cite{zhu2024cross, liu2021ifizz, marcel2023teezz}. These conditions significantly complicate the collection of coverage information and hinder the deployment of advanced but heavyweight fuzzing techniques.
    
    \item \textbf{C2: Complex input interface}. 
    User-space fuzzing usually focuses on fixed interfaces with well-documented inputs. 
    However, the OS kernel inherently exposes more rich and highly complex interfaces 
    for interacting with both user space and hardware, including syscalls, filesystems and peripheral I/O. 
    Inputs for these interfaces are deeply structured and constrained by fine-grained semantics \cite{kim2020hfl, yang2025kernelgpt}.
    For example, arguments for \texttt{ioctl} syscall vary widely across devices and often embed nested structures that must satisfy strict rules only encoded in millions of lines of code. Similarly, in subtasks such as directed kernel fuzzing (DKF), it often reveals the characteristics of discretized inputs \cite{tan2023syzdirect}.
    Furthermore, vulnerabilities often manifest through cross-interface interactions, such as the combination of filesystem images with dependent syscalls [165]. Such an interplay between multiple interfaces significantly expands the input space and has no direct parallel in user-space fuzzing, posing substantial challenges in building accurate  and comprehensive input models.

    \item \textbf{C3: High statefulness}. Statefulness exists in both user- and kernel-space programs, but it is far more pervasive and complex in kernels due to their global and long-lived execution context. Kernels often maintain and evolve internal states over time in response to prior interactions \cite{xu2024mock}. For example, as discussed later in Section~\ref{sec:dependency}, triggering a bug may depend on a specific sequence of syscalls \texttt{msync}, \texttt{mlockall} and \texttt{mmap}, reflecting implicit statefulness that is difficult to model. 
    In this setting, branch coverage serves as a useful, though not definitive, fitness metric for greybox fuzzing. It may fail to capture the most informative feedback in kernel-specific scenarios \cite{zhao2023state}.
    Unlike protocol fuzzing, where states are often represented explicitly through finite-state machines (FSM), the definition and recognition of states within kernels remain largely unexplored. Additionally, the stateful nature of kernels introduces practical challenges: each reboot resets the internal state of the target system, forcing fuzzers to reinitiate the exploration process \cite{song2020agamotto}. This leads to significant inefficiencies, as time and computational resources are repeatedly spent rediscovering previously explored states.
\end{itemize}

\begin{table}[htbp]
\scriptsize
\centering
\caption{Comparison of challenges in user-space and kernel fuzzing.}
\label{tab:challenge}
\begin{tabular}{@{}cccc@{}}
\toprule
\textbf{Challenges} & \textbf{\begin{tabular}[c]{@{}c@{}}User-space\\ Fuzzing\end{tabular}}          & \textbf{\begin{tabular}[c]{@{}c@{}}Kernel\\ Fuzzing\end{tabular}}                                             & \textbf{Examples}                                                                                                                       \\ \midrule
\textbf{C1}         & \begin{tabular}[c]{@{}c@{}}minimal setup;\\ ready-to-run programs\end{tabular} & \begin{tabular}[c]{@{}c@{}}coupled with hardware;\\ environment provisioning;\\ resource-limited\end{tabular} & \begin{tabular}[c]{@{}c@{}}hardware-in-the-loop for TEE \cite{wang2024syztrust};\\ hardware-based tracing \cite{schumilo2017kafl}\end{tabular}                          \\ \midrule
\textbf{C2}         & fixed, well-documented inputs                                                  & \begin{tabular}[c]{@{}c@{}}rich, deeply nested inputs;\\ interplay across interfaces\end{tabular}                                                                          & \begin{tabular}[c]{@{}c@{}}spec. of \texttt{ioctl} buried in huge code space \cite{hao2023syzdiscribe}; \\ discretized inputs for DKF \cite{tan2023syzdirect}; \\ interplay of syscalls + filesystem images \cite{kim2019janus}\end{tabular} \\ \midrule
\textbf{C3}         & \begin{tabular}[c]{@{}c@{}}often stateless \\ or FSM-capturable\end{tabular}   & \begin{tabular}[c]{@{}c@{}}long-live, implicit states;\\ hard to model;\\ repeated re-exploration\end{tabular}                 & \begin{tabular}[c]{@{}c@{}}syscall dependency (\texttt{mlockall}$\rightarrow$\texttt{msync}) \cite{hao2022demy};\\ state-oriented fitness \cite{zhao2023state}\end{tabular}                                           \\ \bottomrule
\end{tabular}
\end{table}

\subsection{Key Functionalities}
\begin{figure}[htbp]
    \centering
    \includegraphics[width=0.55\textwidth]{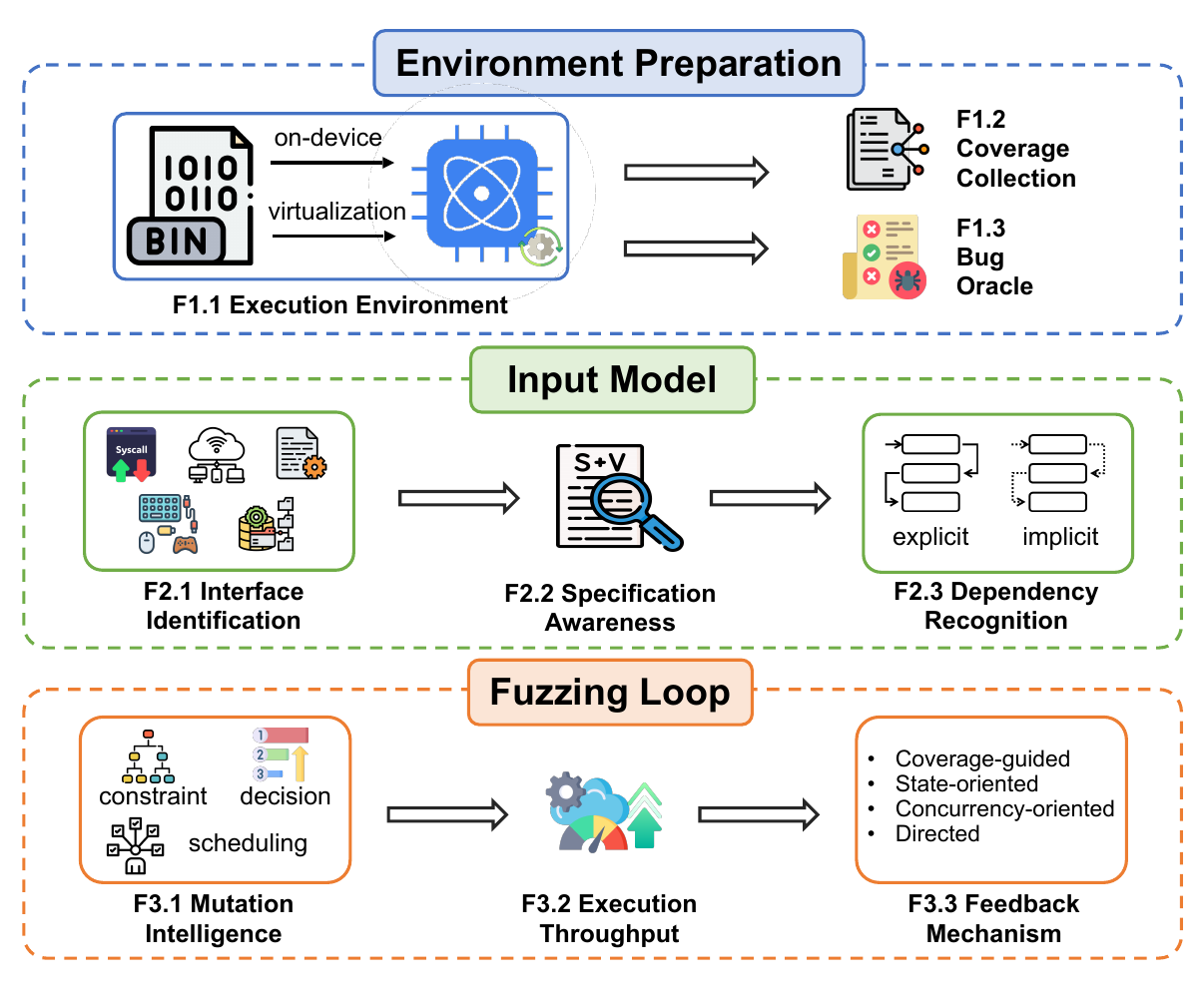}
    \caption{Key stages and functionalities in kernel fuzzing.}
    \label{fig:kernel_fuzzing}
\end{figure}

According to the survey results, we summarize that OS kernel fuzzing largely adheres to the standard methodology of typical fuzzing but is specifically adapted to address the aforementioned challenges. 
It involves generating or mutating test cases that interact with kernel interfaces (e.g., syscalls) and executes them in a controlled environment \cite{Syzkaller}. During execution, the fuzzer utilizes runtime feedback (typically branch coverage) to steer execution toward unexplored space and increase the likelihood of discovering bugs, while continuously monitoring the target kernel for anomalous behavior such as crashes or hangs \cite{shi2019industry, shi2024industry}.
As depicted in Figure~\ref{fig:kernel_fuzzing}, this workflow is typically structured around three core stages: environment preparation, input model and fuzzing loop, each playing a critical role in enabling effective fuzzing of OS kernels. 

To address the constrained environment challenge, the \textbf{environment preparation} establishes a stable and isolated execution context for the kernel, either through virtualization to enable scalable testing across multiple kernel instances \cite{maier2021bsod, schumilo2017kafl, johnson2021jetset}, or on physical devices to ensure faithful representation of real-world behavior \cite{wang2024syztrust, mera2024shift, song2019periscope}.
It also enables coverage tracking, and supports effective detection of potential vulnerabilities \cite{kcov, li2024enhance, li2022μAFL}. In this regard, three functionalities for developed environment preparation include: \textit{execution environment (F1.1)}, \textit{coverage collection (F1.2)}, and \textit{bug oracle (F1.3)}.

To handle the complexity of kernel input interfaces, an \textbf{input model} must first identify the relevant entry points through which the kernel is exercised (\textit{interface identification, F2.1}), and then capture the structural and semantic rules governing these interfaces (\textit{specification awareness, F2.2}), including expected data formats and constraints \cite{hu2021achyb, chen2020test, jake2017difuze, chen2024syzgenplus}. Such a model further incorporates the key functionality of \textit{dependency recognition (F2.3)}, enabling fuzzers to generate test cases that are not only syntactically valid but also semantically meaningful.

Building on the prepared environment and the defined input model, the \textbf{fuzzing loop} continuously feeds generated or mutated test cases into the target kernel and leverages diverse feedback metrics specifically tailored to kernel fuzzing. In particular, to address the challenges posed by the kernel’s high degree of statefulness, the fuzzing loop aims to  reduce the overhead of state re-initialization and improve execution throughput \cite{song2020agamotto, jung2025moneta, zheng2019firm, yang2024thunder}. Furthermore, it incorporates state-oriented feedback to strengthen the ability of state exploration \cite{zhao2023state, wang2024syztrust, liu2024leverage}. Overall, a proficient fuzzing loop centers on three functionalities: \textit{mutation intelligence (F3.1)}, \textit{execution throughput (F3.2)}, and \textit{feedback mechanism (F3.3)}.

\begin{figure}[htbp]
    \centering
    \includegraphics[width=0.5\linewidth]{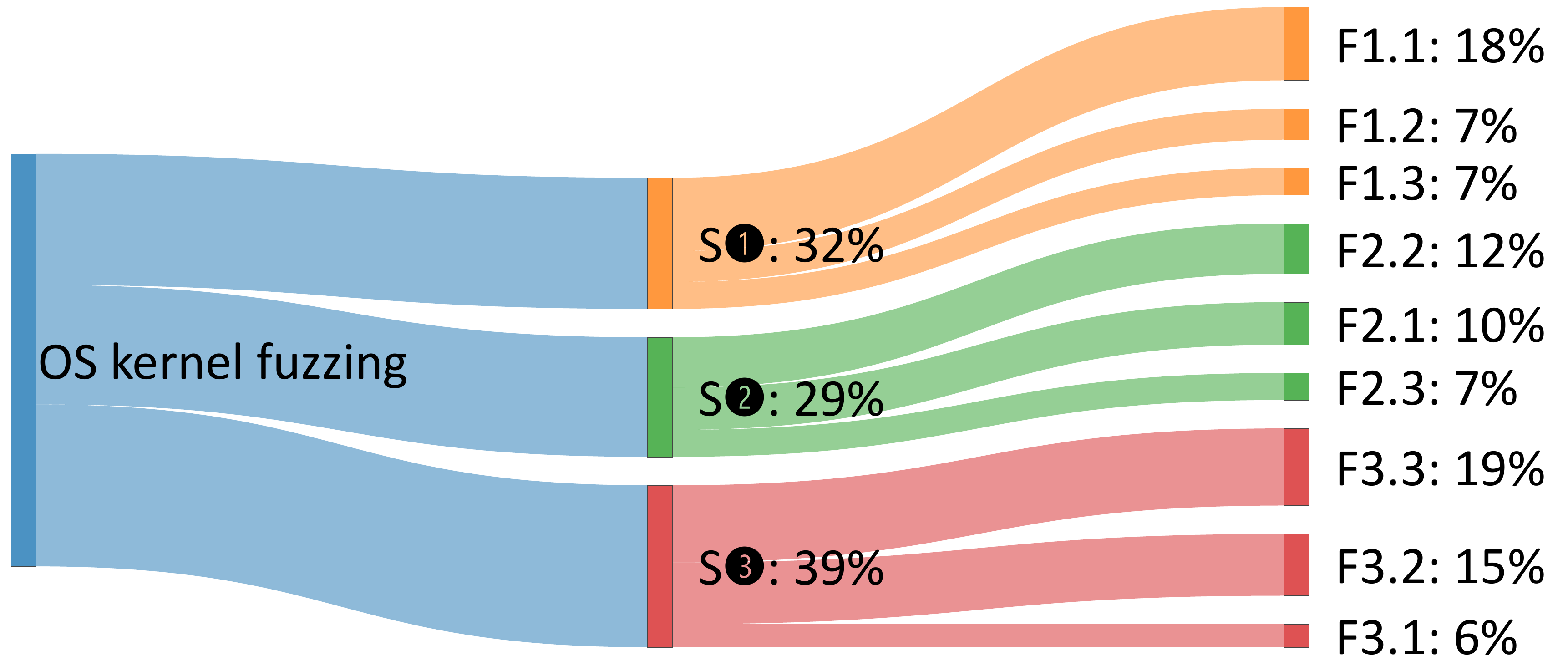}
    \caption{Distribution of papers by target stages and functionalities.}
    \label{fig:stats_func}
\end{figure}

We also study surveyed papers based on their primary targeted functionalities, with the results presented in Figure~\ref{fig:stats_func}. In terms of core components, approximately 32\% of the research concentrates on environment preparation for customized kernels. Additionally, 40\% of the studies focus on optimizing the fuzzing loop, while 28\% aim to improve the correctness and completeness of the input space. As for functionalities, environment preparation (18\%) and feedback mechanisms (19\%) emerge as prominent areas of interest, both representing aspects that require further research and practical advancement. The distribution of targeted functionalities also closely aligns with the key challenges previously identified (in Section~\ref{sec:challenges}) in OS kernel fuzzing. Furthermore, it is noteworthy that 59\% of the papers specifically target Linux kernel fuzzing. This trend is largely driven by the availability of Syzkaller \cite{Syzkaller}, which has established a mature and widely adopted infrastructure, thereby facilitating subsequent fuzzing research and optimization efforts.

Based on the above study, we answer \textbf{RQ1} as follows. We emphasize that \textbf{C1}, \textbf{C2} and \textbf{C3} distinguish OS kernel fuzzing from conventional fuzzing approaches. To address these challenges, we categorize nine key functionalities across three stages, among which \textit{execution environment (F1.1)}, \textit{specification awareness (F2.2)} and \textit{feedback mechanism (F3.3)} emerge as the most extensively studied aspects.

\subsection{Key Techniques}
In response to \textbf{RQ2}, we examine how existing techniques implement and advance the key functionalities identified above. 
Due to the aforementioned challenges,  these studies have inspired a diverse range of technical innovations. 
These techniques can be organized into three categories, aligned with the stages of our model:

\begin{itemize}
    \item \textbf{Environment-centric techniques.} These approaches focus on strengthening the fuzzing environment in terms of usability, scalability, and observability. To improve deployment, they use \textit{on-device fuzzing}, which executes tests directly on the target hardware, and \textit{rehosting-based fuzzing}, which migrates the execution to an emulated environment (\textit{F1.1}). For monitoring, they rely on \textit{invasive instrumentation}, which modifies the target to collect fine-grained runtime data, and \textit{non-invasive tracing}, which observes execution with minimal interference (\textit{F1.2}). To expand bug detection, they employ \textit{oracles} that identify not only \textit{memory corruption} but also \textit{non-crash bugs} (\textit{F1.3}).

    \item \textbf{Input-centric techniques.} These methods focus on constructing and refining input models to improve fuzzing effectiveness. For test input, they employ \textit{primary interface fuzzing}, which targets a single input channel (e.g., syscalls, peripherals, or filesystems), as well as \textit{multiple-interface fuzzing}, which combines interfaces to uncover interaction behaviors (\textit{F2.1}). To increase input validity, they utilize \textit{static inference}, which extracts constraints from source code, and \textit{dynamic analysis}, which derives constraints from runtime behaviors (\textit{F2.2}). To capture stateful behaviors, they address \textit{explicit dependencies}, where input relationships are explicitly defined, and \textit{implicit dependencies}, where hidden semantic relations manifest through shared internal states (\textit{F2.3}).
 
    \item \textbf{Loop-centric techniques.} These approaches aim to improve the fuzzing loop by enhancing execution efficiency and exploration focus. To accelerate execution, they adopt \textit{virtualization enhancements}, which optimize VM performance, and \textit{system snapshots}, which enable efficient restoration of kernel states (\textit{F3.1}). To strengthen feedback, they apply \textit{constraint solving} for precise path exploration, \textit{thread scheduling} for concurrent execution control, and \textit{decision intelligence} to prioritize promising test cases (\textit{F3.2}). To guide exploration, they employ \textit{directed fuzzing}, which steers inputs toward given targets, \textit{state-oriented fitness}, which focuses on diverse program states, and \textit{concurrency-oriented fitness}, which exposes thread interleavings systematically (\textit{F3.3}).
\end{itemize}

Together, these techniques provide complementary solutions to improve efficiency, coverage, and bug-finding capability. In the following sections, we analyze each functionality in depth and derive a series of \textbf{implications} that summarize the strengths, limitations, and directions of these techniques. Each implication is grounded in the corresponding functionality discussed above (addressing \textbf{RQ2}) and extends the analysis toward unresolved challenges and future research opportunities (addressing \textbf{RQ3}).
\section{Environment Preparation} \label{sec:environ} 


\subsection{Execution Environment} \label{sec:exec}
Two primary approaches provide the execution environment for OS kernel fuzzing: on-device fuzzing \cite{song2019periscope,li2022μAFL,wang2024syztrust,eisele2023fuzzing} and emulation-based fuzzing \cite{song2020agamotto,keil2007stateful,talebi2018charm,pan2017digtool,schumilo2014don,renzelmann2012symdrive,maier2019unicorefuzz}. We present a summary of the literature and the corresponding solutions employed within the environment preparation in Table \ref{tab:env}.
\subsubsection{On-device Fuzzing}

An on-device fuzzer ensures the target kernel's continuous and stable operation due to \textbf{C1} by executing it on actual devices. It employs a user-space application or debugging system for coverage collection and bug detection, connecting directly to the kernel. This fuzzer effectively identifies defects related to unique hardware properties or configurations.

\begin{figure}[htbp]
\centering
\begin{subfigure}[b]{0.4\linewidth} 
    \centering
    \includegraphics[width=0.65\textwidth]{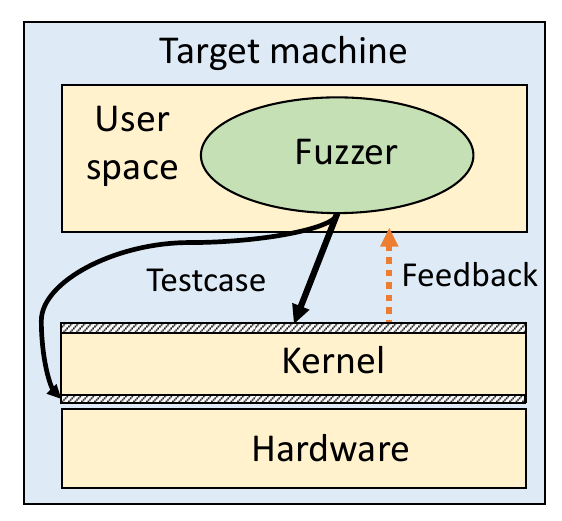}
    \caption{Local fuzzer}
    \label{fig:userspacefuzzer}
\end{subfigure}
\begin{subfigure}[b]{0.45\linewidth} 
    \centering
    \includegraphics[width=0.65\textwidth]{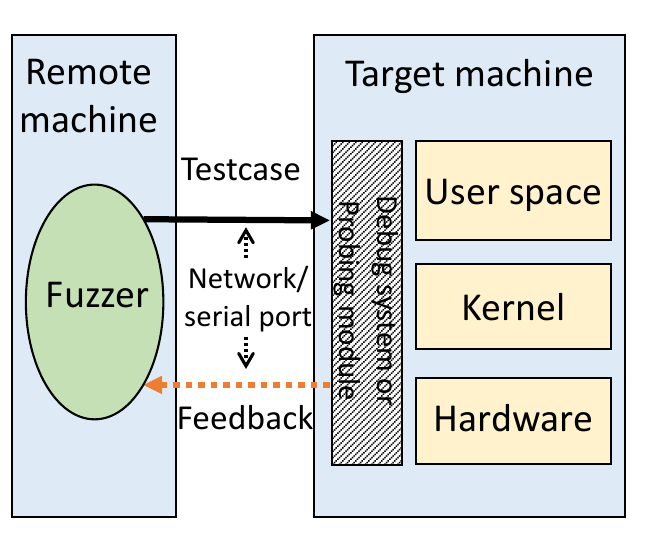}
    \caption{Remote fuzzer}
    \label{fig:remotefuzzer}
\end{subfigure}
\caption{Two typical on-device fuzzers.}
\label{fig:ondevicefuzzer}
\end{figure}

\textbf{Local fuzzer.} This type of fuzzer runs in the user space on a local machine and utilizes the exposed kernel interface to fuzz the kernel \cite{schumilo2021nyx}, as illustrated in Figure \ref{fig:ondevicefuzzer}. It has limited ability to control and monitor the target kernel because of its low privilege. Even worse, it loses all execution information when the kernel crashes.

\textbf{Remote fuzzer.} The remote fuzzer connects to the target machine that is loaded with a kernel via serial ports \cite{song2019periscope, wang2024syztrust,eisele2023fuzzing,li2022μAFL} or network \cite{nico2023drone}. This fuzzer requires a debugging system or probing module to be deployed in the target kernel, and the debug feature is utilized to control the target firmware. For instance, SyzTrust~\cite{wang2024syztrust} and $\mu$AFL~\cite{li2022μAFL} utilize ARM Coresight architecture to control the execution of embedded OSes, and PeriFuzz~\cite{song2019periscope} designed their probing framework to manage the hardware boundary of a kernel. 

Most of these approaches are open source and capable of supporting closed-source OSes. However, they often require intrusive control over OS execution. While the native execution environment of on-device fuzzers provides high stability and fidelity, it is limited in capacity and input execution speed when applied to RTOS on resource-constrained devices, such as ARM Cortex-M chips with clock speeds between 10 MHz and 600 MHz. Furthermore, on-device fuzzing is typically OS-specific and necessitates real, debug-enabled devices or elevated privileges for the fuzzer, which results in increased costs.

\subsubsection{Emulation-based Fuzzing}

Loading the kernel into a virtual environment offers a more scalable approach with complete control for fuzzing OS kernels, providing a costless and effective solution for kernel introspection. However, the challenge lies in maintaining stability and fidelity to ensure that the emulated kernel operates consistently and without interruption. According to the previous work \cite{fasano2021sok}, there are two principal ways to construct the virtual environment: hardware emulation system and rehosted embedded system. Regarding the rehosted embedded system, we further categorize these rehosting techniques into hardware-in-the-loop, high-level emulation, MMIO modeling.

\textbf{Full emulation-based fuzzing.} Full hardware emulation replicates the functionalities of specific hardware accurately, allowing unmodified kernel execution and fuzzing when peripherals are adequately emulated \cite{fasano2021sok}. 
Full emulation aims to implement as many peripherals as possible, providing relatively high stability and fidelity compared to rehosting. When the target emulator is open-source, developers have full control over both the emulator and the running kernel inside it, thereby maximizing the capability for introspection and analysis. Although there are three major emulators for OS kernels, including VMWare \cite{walters1999vmware}, VirtualBox \cite{khan2022virtualization}, and QEMU~\cite{bellard2005qemu}, full emulation-based fuzzing predominantly utilizes QEMU. This preference is due to its effectiveness in handling general-purpose OS and embedded Linux environments. QEMU is particularly advantageous because it is open source and compatible with a range of fuzzing tools \cite{Syzkaller, pandey2019triforce, schumilo2017kafl}, making it the preferred choice. Regarding RTOS and TEE, QEMU offers only limited support. Consequently, existing fuzzers designed for these specialized kernels often rely on rehosting techniques. Since these kernels interact closely with the hardware or specialized architectures, achieving full emulation requires significant effort, making rehosting a more feasible approach for fuzzing in such environments. The speed of kernel operation and fuzzing depends on the machine running the emulator and whether any acceleration techniques are deployed. For instance, emulating a kernel on a machine with similar performance may be slower due to instruction translation overhead, while emulating low-performance kernels on high-performance machines can improve speed.

\textbf{Rehosting-based fuzzing.} While full emulation-based fuzzing is primarily designed for general-purpose OSes and their variants, rehosting-based fuzzing offers a complementary approach for RTOS such as Amazon FreeRTOS \cite{freertos}, ARM Mbed \cite{mbedos}, Zephyr \cite{zephyr}, and LiteOS \cite{cao2008liteos}. Unlike full emulation, a rehosted embedded system focuses on modeling only the essential features of target kernels required for fuzzing or dynamic analysis. This approach also provides full control over the target kernel and a comprehensive introspection.
Based on our survey of state-of-the-art rehosting techniques, we identified three primary strategies: hardware-in-the-loop \cite{koscher2015surrogates, talebi2018charm}, high-level hooking \cite{clements2020halucinator, feng2020p2im, jiang2021ecmo, li2021library}, and MMIO modeling \cite{gustafson2019toward, harrison2020partemu, zhou2021automatic, cao2020device, johnson2021jetset}. Hardware-in-the-loop, while useful, suffers from lower stability due to potential delays in forwarding hardware data, which can cause crashes during the operation and fuzzing of RTOS. Additionally, the speed bottleneck in this approach is tied to the execution speed of the hardware itself.
The latter, MMIO modeling and high-level hooking, face significant challenges related to fidelity and stability, particularly in accurately simulating hardware behavior, such as DMA and interrupt emulation, and handling complex peripherals \cite{mera2021dice}. When fuzzing a kernel within a rehosted system, crashes may occur due to the absence of certain feature models. Similarly to full emulation, the speed of kernel operation and fuzzing depends on the machine running the emulator and used acceleration techniques.

\begin{mybox}
    \textbf{Implication \yellowcircle{1}: OS-Agnostic Rehosting.}
    Current kernel fuzzing environments struggle to balance stability, overhead, and introspection but face significant challenges in achieving OS-agnostic execution, particularly for RTOS and TEE. Fidelity remains a critical limitation, as fuzzers often fail or get stuck due to incomplete or inaccurate emulation. This highlights the need for more faithful rehosting techniques that can generalize across diverse OS platforms. Promising directions include lightweight full-system emulation tailored for specialized environments and improved rehosting frameworks that enhance stability and fidelity while minimizing engineering effort.
\end{mybox}

\subsection{Coverage Collection} \label{sec:covcollect}
Coverage is a key indicator for evaluating the fuzzing effectiveness.
To collect coverage, there are two principal ways: invasive instrumentation and non-invasive tracing.

\begin{table}[htbp]
\centering
\scriptsize
\caption{Coverage collection methodologies and their applications}
\label{tab:cover}
\begin{tabular}{@{}m{1.5cm} m{3.7cm} m{1.1cm} m{5.5cm}@{}}
\toprule
\textbf{Methodology} & \textbf{Technique} & \textbf{Literature} & \textbf{Use Case} \\ \midrule
\multirow{3}{*}{\begin{tabular}[c]{@{}c@{}}Invasive\\ Instrumentation\end{tabular}} 
& Compilers add instrumentation code to record coverage during compilation 
&  \cite{Syzkaller} \cite{song2019periscope} \cite{peng2020usb} \cite{hao2023syzdiscribe} \cite{tan2023syzdirect} \cite{jeong2019razzer}
& Applicable to kernels whose source code is available. It is mostly adopted for easy usage and high flexibility. The overhead is usually low. \\ \cmidrule(l){2-4} 

& Emulators inject instrumentation code to trace coverage during dynamic binary translation 
& \cite{maier2019unicorefuzz} \cite{lee2020partemu} \cite{clements2020halucinator} \cite{Ioannis2023firmsolo} \cite{hui2023greehouse} \cite{ioannis2024pandawan}
& Applicable to kernels that are binary-only but emulatable. The overhead is significantly higher than source-based instrumentation. \\ \cmidrule(l){2-4} 

& Instrument a compiled kernel without altering its original functionality. 
&  \cite{yin2023kext}
& Applicable to binary-only kernels. The technique may compromise the kernel structure and thus is less adopted. \\ \midrule

\multirow{2}{*}{\begin{tabular}[c]{@{}c@{}}Non-invasive\\ Tracing\end{tabular}}
& Leverage built-in features of the kernel to trace coverage 
&  \cite{chen2021syzgen} \cite{chen2024syzgenplus}
& Depends on specific types of kernels. \\ \cmidrule(l){2-4} 

& Leverage low-level hardware to collect coverage 
& \cite{schumilo2017kafl} \cite{wang2024syztrust} \cite{li2022μAFL}
& Applicable to both source-based and binary kernels, especially on devices. It is architecture-specific and provides limited types of coverage. \\ \bottomrule
\end{tabular}
\end{table}



\subsubsection{Invasive Instrumentation}
In essence, invasive instrumentation modifies target code during compilation or runtime, and exposes interfaces for tracking executed portions.
Table \ref{tab:cover} presents a detailed comparison of coverage collection techniques and their use cases.

\textbf{Source-based instrumentation.} 
Instrumenting the kernel during compilation is the most effective and intuitive method for coverage collection. For example, KCOV enhances fuzzing by injecting signals into basic blocks, significantly improving bug discovery through code coverage \cite{kcov, Syzkaller, shankara2018moonshine, wang2021syzvegas, sun2021healer, jeong2019razzer, xu2024mock}. However, this approach primarily targets bugs reachable via syscall inputs and has limitations in non-deterministic areas and non-syscall handlers. Solutions like PeriScope \cite{song2019periscope} and USBfuzz \cite{peng2020usb} have advanced coverage by focusing on fine-grained and remote collection methods, enabling more effective fuzzing in areas like driver operations and interrupts.



\textbf{Dynamic instrumentation.} 
When source code is unavailable, dynamic instrumentation provides an alternative for modifying a kernel without requiring access to its source. This approach typically involves an emulator that translates raw machine instructions into an intermediate representation. During this translation, instrumentation code is inserted to enable coverage collection \cite{bellard2005qemu}. A key prerequisite is that the kernel must be booted within the emulator. Hence, dynamic instrumentation is commonly integrated with emulation tools such as AFL-QEMU \cite{afl}, enabling analysis of binary-only kernels for Linux \cite{clements2020halucinator, maier2019unicorefuzz, hui2023greehouse, ioannis2024pandawan}, RTOS \cite{mera2021dice}, and TEE \cite{lee2020partemu}.

\textbf{Binary rewriting.} 
For binary only kernels, binary rewriting \cite{nagy2021zafl, zhang2021stochfuzz, dinesh2020retro} provides a viable alternative. 
Adapting binary rewriting for kernel fuzzing presents additional hurdles, such as significant overhead that reduces fuzzing efficiency \cite{maier2021bsod} and the complexity and ambiguities associated with static methods \cite{dinesh2020retro, zhang2021stochfuzz}. Nonetheless, recent innovations \cite{yin2023kext} for macOS kernel extensions show promise, achieving cost-effective static binary rewriting by leveraging macOS's features for efficient coverage instruction injection.

\subsubsection{Non-invasive Tracing} 
As a result of \textbf{C1}, instrumentation is often restricted because the kernel runs in privileged mode. The strong coupling with hardware leads to a lack of  middleware support, thereby limiting observability.
In such cases, fuzzers rely on limited interfaces, typically using debug checkpoints or hardware assistance. 


\textbf{Debug checkpoint.}  
Similar to binary rewriting, tracing coverage through debug checkpoints involves setting checkpoints within the target and invoking them to gather feedback during execution. The key differences lie in their scalability and whether they leverage built-in kernel features. Tracing coverage via debug checkpoints is a target-specific approach that depends heavily on the target kernel's support. For example, 
SyzGen \cite{chen2021syzgen} uses macOS debugging tools to address challenges posed by closed-source kernels. Another example is on-device fuzzers that utilize hardware-based debugging, where feedback collection is closely tied to hardware features, such as embedded system debug units \cite{eisele2023fuzzing}.

\textbf{Hardware assistance.} 
Hardware components like CPU have direct access to kernel space and every instruction, facilitating the acquisition of detailed feedback. Techniques like Intel PT and ARM ETM are widely used for capturing execution information, offering two key advantages.
First, they ensure greater completeness and robustness in coverage collection. In contrast, user-space solutions are inherently limited, as they can only begin after the kernel or drivers have initialized and therefore miss the coverage generated during the bootstrap phase \cite{zhao2022semantic,li2022μAFL, hao2022demy}.
Second, these tools provide execution information from arbitrary OS code including closed-source ones, serving as OS-agnostic feedback mechanisms that support kernel fuzzing across multiple platforms \cite{schumilo2017kafl, aschermann2019redqueen, bugs2020, wang2024syztrust}. 
However, despite these hardware-based advantages, hardware-assisted approaches are limited to architecture-specific targets, similar to debug checkpoints.

\begin{mybox}
    \noindent \textbf{Implication \yellowcircle{2}: Coverage Disparity.}
    Source-based instrumentation has become the dominant approach in kernel fuzzing, adopted by approximately 77\% of greybox fuzzers, owing to its flexibility and ability to collect fine-grained, customizable feedback (as detailed in Section \ref{sec:feedback}).
    In contrast, binary-only kernels predominantly rely on OS- or architecture-specific techniques, 95\% of which yield only coarse coverage metrics. This underscores the need for techniques that can deliver richer coverage feedback without requiring kernel source access.
\end{mybox}

\renewcommand{\arraystretch}{1.4}
\begin{table}[htbp]
\scriptsize
\centering
\caption{Summary of kernel fuzzers and their solutions used in environment preparation.}
\label{tab:env}
\begin{threeparttable}
\begin{tabular}{|c|l|l|@{}c@{}|l|lll|}
\hline
\multirow{2}{*}{\textbf{Year}} & \multirow{2}{*}{\textbf{Fuzzer}} & \multirow{2}{*}{\textbf{Func.}} & \multirow{2}{*}{\textbf{Type}} & \multirow{2}{*}{\textbf{Target}} & \multicolumn{3}{c|}{\textbf{Env. Preparation}} \\ \cline{6-8} 
 &  &  &  &  & \multicolumn{1}{c|}{\begin{tabular}[c]{@{}c@{}}Exec.\\ Env.\end{tabular}} & \multicolumn{1}{c|}{\begin{tabular}[c]{@{}c@{}}Coverage\\ Collection\end{tabular}} & \multicolumn{1}{c|}{\begin{tabular}[c]{@{}c@{}}Bug\\ Oracle\end{tabular}} \\ \hline
2017 & Digtool \cite{pan2017digtool} & \circlewithtext{1.1} \circlewithtext{1.3} & \CIRCLE & Win. & \multicolumn{1}{l|}{CE} & \multicolumn{1}{l|}{-} & LA \\ \hline
2017 & kAFL \cite{schumilo2017kafl} & \circlewithtext{1.1} \circlewithtext{1.2} & \LEFTcircle & Win. \& Linux & \multicolumn{1}{l|}{GPVM} & \multicolumn{1}{l|}{HA} & FS \\ \hline
2017 & Syzkaller \cite{Syzkaller} & \circlewithtext{1.1} \circlewithtext{1.2} \circlewithtext{1.3} & \LEFTcircle & General & \multicolumn{1}{l|}{GPVM} & \multicolumn{1}{l|}{SI} & San. \& FS \& DD \\ \hline
2018 & Charm \cite{talebi2018charm} & \circlewithtext{1.1} & \LEFTcircle & Android & \multicolumn{1}{l|}{HIL} & \multicolumn{1}{l|}{SI} & San. \& FS \\ \hline
2019 & PeriScope \cite{song2019periscope} & \circlewithtext{1.1} & \LEFTcircle & Android & \multicolumn{1}{l|}{OD} & \multicolumn{1}{l|}{SI} & San.  \\ \hline
2019 & Hydra \cite{kim2019finding} & \circlewithtext{1.3} & \LEFTcircle & Linux & \multicolumn{1}{l|}{-} & \multicolumn{1}{l|}{SI} & San. \& SC \\ \hline
2019 & Unicorefuzz \cite{maier2019unicorefuzz}& \circlewithtext{1.1} \circlewithtext{1.2} & \LEFTcircle & Linux & \multicolumn{1}{l|}{GPVM} & \multicolumn{1}{l|}{DI} & San. \\ \hline
2020 & EASIER \cite{pu2020exvivo} & \circlewithtext{1.1} \circlewithtext{1.2}  & \LEFTcircle & Android & \multicolumn{1}{l|}{Rehost} & \multicolumn{1}{l|}{DI} & FS \\ \hline
2020 & HALucinator \cite{clements2020halucinator}&  \circlewithtext{1.1} & \LEFTcircle & RTOS & \multicolumn{1}{l|}{CE} & \multicolumn{1}{l|}{DI} & DS \\ \hline
2020 & PARTEMU \cite{lee2020partemu}& \circlewithtext{1.1}  & \LEFTcircle & TEE & \multicolumn{1}{l|}{CE} & \multicolumn{1}{l|}{DI} & FS  \\ \hline
2020 & USBFuzz \cite{peng2020usb}& \circlewithtext{1.1}  & \LEFTcircle & Win. \& Linux & \multicolumn{1}{l|}{Rehost} & \multicolumn{1}{l|}{SI} & San.  \\ \hline
2021 & Aafer et al. \cite{aafer21smarttvs}& \circlewithtext{1.2} & \CIRCLE & Android & \multicolumn{1}{l|}{OD} & \multicolumn{1}{l|}{-} & DT \\ \hline
2021 & FirmGuide \cite{liu2021firmguide} & \circlewithtext{1.1} & \CIRCLE & Emb. Linux  & \multicolumn{1}{l|}{Rehost} & \multicolumn{1}{l|}{-} & FS \\ \hline
2021 & ECMO \cite{jiang2021ecmo}& \circlewithtext{1.1} & \CIRCLE & Emb. Linux & \multicolumn{1}{l|}{Rehost} & \multicolumn{1}{l|}{-} & - \\ \hline
2021 & DICE \cite{mera2021dice} & \circlewithtext{1.1} & \LEFTcircle & RTOS & \multicolumn{1}{l|}{CE} & \multicolumn{1}{l|}{DI} & San. \\ \hline
2022 & $\mu$AFL \cite{li2022μAFL} & \circlewithtext{1.2} & \LEFTcircle & RTOS & \multicolumn{1}{l|}{HID} & \multicolumn{1}{l|}{HA} & FS \& TO \\ \hline
2022 & EQUAFL \cite{zhang2022efficent} & \circlewithtext{1.1} & \CIRCLE & Emb. Linux & \multicolumn{1}{l|}{CE} & \multicolumn{1}{l|}{DI} & FS \& TO  \\ \hline
2022 & RoboFuzz \cite{kim2022robofuzz} & \circlewithtext{1.3} & \LEFTcircle & ROS & \multicolumn{1}{l|}{HIL} & \multicolumn{1}{l|}{SI} & San. \& DT \& SC  \\ \hline
2022 & Tardis \cite{shen2022tardis}& \circlewithtext{1.2} & \LEFTcircle & RTOS & \multicolumn{1}{l|}{GPVM} & \multicolumn{1}{l|}{SI} & - \\ \hline
2023 & TEEzz \cite{marcel2023teezz}& \circlewithtext{1.1} & \LEFTcircle & TEE & \multicolumn{1}{l|}{OD} & \multicolumn{1}{l|}{SI} & FS  \\ \hline
2023 & FirmSolo \cite{Ioannis2023firmsolo}& \circlewithtext{1.1} & \CIRCLE & Emb. Linux & \multicolumn{1}{l|}{CE} & \multicolumn{1}{l|}{DI} & FS  \\ \hline
2023 & KextFuzz \cite{yin2023kext}& \circlewithtext{1.2} & \LEFTcircle & MacOS & \multicolumn{1}{l|}{OD} & \multicolumn{1}{l|}{BR} & FS \\ \hline
2023 & Greenhouse \cite{hui2023greehouse} & \circlewithtext{1.1} & \CIRCLE & Emb. Linux & \multicolumn{1}{l|}{Rehost} & \multicolumn{1}{l|}{DI} & FS  \\ \hline
2023 & BoKASAN \cite{cho2023bokasan}& \circlewithtext{1.3} & \CIRCLE & Win. \& Linux & \multicolumn{1}{l|}{GPVM} & \multicolumn{1}{l|}{-} & DS \\ \hline
2024 & BVF \cite{sun2024finding}& \circlewithtext{1.3} & \LEFTcircle  & Linux & \multicolumn{1}{l|}{GPVM} & \multicolumn{1}{l|}{SI} & San. \\ \hline
2024 & R2D2 \cite{shen2024enhancing}& \circlewithtext{1.2} & \LEFTcircle & ROS & \multicolumn{1}{l|}{GPVM} & \multicolumn{1}{l|}{SI} & San. \& FS \\ \hline
2024 & IPEA \cite{shi2024facilitating}& \circlewithtext{1.2} \circlewithtext{1.3} & \LEFTcircle & RTOS & \multicolumn{1}{l|}{HIL} & \multicolumn{1}{l|}{SI} & San.  \\ \hline
2024 & Monarch \cite{lyu2024monarch} & \circlewithtext{1.3} & \LEFTcircle & Linux & \multicolumn{1}{l|}{GPVM} & \multicolumn{1}{l|}{SI} & San. \& SC \\ \hline
2024 & SyzTrust \cite{wang2024syztrust} & \circlewithtext{1.2} & \LEFTcircle & TEE & \multicolumn{1}{l|}{HIL} & \multicolumn{1}{l|}{HA} &  FS \& TO \\ \hline
2024 & Pandawan \cite{ioannis2024pandawan}& \circlewithtext{1.1} & \LEFTcircle & Emb. Linux & \multicolumn{1}{l|}{Rehost} & \multicolumn{1}{l|}{DI} & FS \\ \hline
\end{tabular}
\end{threeparttable}
\begin{tablenotes}[flushleft]
\scriptsize
\item[] \circlewithtext{1.1}: The satisfied functionality of a method; \Circle\ : Whitebox fuzzing. \LEFTcircle\ : Greybox fuzzing. \CIRCLE\ : Blackbox fuzzing; 
\item[] Win: Windows; Emb. Linux: Embedded Linux; CE: Custom Emulation; GPVM: General-Purpose Virtual Machine; HIL: Hardware In the Loop; OD: On Device; HA: Hardware Assistance; SI: Source-based Instrumentation; DI: Dynamic Instrumentation; DT: Differential Testing; BR: Binary Rewriting; LA: Log Analysis; San: Sanitizer; FS: Fatal Signal; DD: Deadlock; SC: Semantic Checker; TO: Timeout; -: Not Specified.
\end{tablenotes}
\end{table}

\subsection{Bug Oracle} 
\label{sec:bugaccess}
Before starting fuzzing, it is necessary to design a bug oracle to detect bugs.
Generally, these bug oracles target two primary types of issues: memory corruption and non-crash bugs. In addition, we discuss solutions for bug triage, focusing on strategies to effectively manage crashes.

\subsubsection{Oracle for Memory Corruption}
We categorize bug oracles for memory corruption into fatal signals and sanitizers.


\textbf{Fatal Signal.} 
Fatal signals are among the most widely used and intuitive bug oracles for detecting critical errors, such as illegal memory access. These signals manifest in various forms, including segmentation faults \cite{schumilo2017kafl, pu2020exvivo, Ioannis2023firmsolo, marcel2023teezz}, general protection faults \cite{Syzkaller, gong2025snowflow, lee2020partemu}, kernel panics \cite{liu2021firmguide, yin2023kext, hui2023greehouse, aafer21smarttvs}, and task hangs \cite{zhang2022efficent}. Additionally, kernel-level exception handling mechanisms \cite{hao2022demy, wang2024syztrust} are also used to indicate abnormal behavior.
In on-device or hardware-in-the-loop testing scenarios, where direct inspection of execution output or heavyweight analysis is challenging, timeouts are often employed as a practical proxy \cite{li2022μAFL, zhang2022efficent, wang2024syztrust}.
However, relying solely on fatal signals is insufficient, as they may fail to detect logical or silent bugs that do not immediately disrupt system execution.

\textbf{Sanitizer.} 
Sanitizers have been the de facto oracles widely used by almost all kernel fuzzers. They detect bugs by instrumenting code and monitoring runtime behavior, with each type of bug requiring specific sanitizers. 
The research community has developed several kernel sanitizers, addressing vulnerabilities like use-after-free / out-of-bounds \cite{kasan}, data races \cite{kcsan} and undefined behaviors \cite{ubsan}.
While effective, these sanitizers have limitations. First, they usually rely on source code instrumentation, which introduces significant overhead and is unsuitable for binary-only systems \cite{shi2024facilitating, dinesh2020retro}. Therefore, ongoing efforts aim to develop more efficient structures \cite{jeon2020fuzzan} and extend support to closed-source cases \cite{cho2023bokasan, pan2017digtool}. For example, BoKASAN \cite{cho2023bokasan} leverages the Linux kernel’s \textit{ftrace} feature to insert hooks into critical functions, enabling dynamic instrumentation of binary-only kernels. Additionally, recent research \cite{sun2024finding} points out a fact that sanitizers do not cover the entire kernel, leaving many vulnerabilities undetected. It would be interesting to figure out how far these sanitizers are.



\subsubsection{Oracle for Non-crash Bugs} 
Detecting silent bugs can be challenging since they do not always lead to a crash. It adds barriers to bug discovery. We summarize existing oracle for non-crash bugs into two types.


\textbf{Differential testing.} Differential testing addresses the challenge of detecting silent bugs by executing the same test case across multiple kernel versions or configurations. It is primarily designed to identify correctness issues, based on the assumption that the kernel should exhibit consistent behavior across different environments when given identical inputs.
To support this approach, RoboFuzz \cite{kim2022robofuzz} detects correctness bugs in the ROS by comparing execution results between simulators and real-world deployments. Similarly, physical discrepancies have been used as indicators of potential bugs in another case \cite{aafer21smarttvs}. The primary challenge hindering the application of differential testing in kernels is the lack of suitable reference counterparts for comparison.

\textbf{Semantic checkers.} 
Traditional sanitizer usually target low-level memory errors like use-after-free or out-of-bounds accesses. By contrast, semantic bug detection focuses on detecting logic errors or high-level correctness, such as violations of properties and specifications \cite{lyu2024monarch, kim2022robofuzz, kim2019finding}. It is used for ensuring that kernel behavior aligns with expected logical rules and operational semantics. For instance, in filesystem modules, semantic oracles verify if desired states or properties have been violated. Monarch's \cite{lyu2024monarch} detects semantic violations in memory, persistence, fault, and concurrency scenarios by comparing runtime states against symbolic execution based on POSIX specifications and distributed fault models.  However, these checkers are often scenario-specific, limiting supported types of vulnerabilities and their general applicability.

\subsubsection{Bug Triage.}

Beyond detecting crashes via bug oracles, kernel fuzzing critically relies on bug triage to manage the large volume of reports it produces. Triage typically consists of prioritization, minimization, and deduplication \cite{manes2021art}. Compared with user space, prioritization in the kernel is substantially more complex. First, pervasive concurrency and statefulness make many bugs difficult to reproduce, rendering reproduction-based ranking unreliable \cite{alek2023syzbot}. Second, a report that appears “low-risk” may still expose severe consequences due to the diverse interactions within the kernel’s large codebase \cite{zou2022syzscope}. To address these challenges, recent efforts focus on assessing potential impact through proximity-based search \cite{lin2022grebe, zou2022syzscope, yuan2023ddrace} or exploitability analysis \cite{zou2024syzbridge}. Minimization in kernel fuzzing resembles that in user space. For example, Syzkaller reduces syscall sequences while preserving crash triggers \cite{Syzkaller, xu2024mock}. Deduplication inherits similar difficulties as prioritization. Besides, stacktrace–based heuristics popular in user space \cite{lyu2019mopt} are often insufficient for deduplication of kernel crashes, since memory layout and thread interleavings cause the same underlying bug to manifest with subtly different behaviors \cite{mu2022indepth}. Despite these advances, existing approaches primarily focus on memory corruption reports and thus heavily rely on information provided by sanitizers (e.g., stack traces and memory addresses). However, their effectiveness in detecting and analyzing non-crash bugs remains largely unexplored.

\begin{mybox}

    \textbf{Implication \yellowcircle{3}: Beyond Memory Corruption.} Enhancing kernel bug oracles requires a balance between detecting a variety of bug types, from memory corruption to more subtle non-crash issues. The detection of memory corruption vulnerabilities  is relatively well-developed, with ongoing efforts primarily aimed at improving usability and minimizing overhead. In contrast, identifying semantic-aware bugs presents greater challenges, especially with the emergence of new classes of kernel vulnerabilities \cite{jakob2025uncovering, sun2024validating}. Tackling non-crash bugs necessitates sophisticated approaches like differential testing and indicators for logical errors.
\end{mybox}

\section{Input Model}
\label{sec:input} 



Establishing the input model is a subsequent step after setting up a reliable testing environment. Blind fuzzing is inefficient for navigating the kernel's complex structure due to the vast input space. Therefore, a systematic approach to defining input synthesis is essential. In this section, we review studies on techniques for specifying desired input, emphasizing the importance of interface identification, specification awareness, and dependency recognition. Table \ref{tab:input} provides a summary of the literature and the corresponding solutions for the input model stage.

\subsection{Interface Identification}
The OS kernel, serving as the intermediary between user space and hardware, provides a myriad of interfaces. 
For a fuzzer to automatically and effectively detect vulnerabilities, it must first identify the interfaces that align with its objectives.
\subsubsection{Primary Interfaces}

There are primarily five types of fuzzing interfaces exposed by OS kernels, consisting of:

\textbf{Syscall.} As the primary interface between user space and kernel space, syscalls serve as the fundamental mechanism for invoking kernel-level functionality. In the latest Linux 6.14 release, the kernel exposes nearly 500 syscalls, not including the numerous variants that are controlled by command parameters (e.g., through the \textit{cmd} argument). To capture this complexity, Syzkaller defines more than 8,000 syscall descriptions manually using its domain-specific language. These syscalls are vital for OS functionality and provide standardized interfaces for user space to perform a wide range of tasks. As such, they form the principal input surface for kernel fuzzing and are typically exercised in the form of syscall sequences. Given the extensive and evolving syscall interface, along with the limitations of manual specification, considerable research has been devoted to automating the extraction and analysis of syscall semantics and interfaces~\cite{jake2017difuze, liu2020fans, hao2023syzdiscribe}.

\textbf{Peripheral devices.} 
Peripheral devices communicate with the kernel through mechanisms such as MMIO \cite{feng2020p2im, mera2021dice, jake2017difuze, shen2022drifuzz}, Port I/O \cite{pu2020exvivo, talebi2018charm, peng2020usb}, DMA \cite{mera2021dice, song2019periscope, ma2022print, wu23devfuzz}, and interrupts \cite{zhao2022semantic, shen2022drifuzz, jake2017difuze}. These channels form a foundational interface for injecting test cases, facilitating the exploration of vulnerabilities within the OS–hardware communication layer. Thorough testing of these interfaces is critical, as they can serve as potential vectors for security breaches. Given the high cost and complexity of interacting with real hardware, practical approaches include device-based techniques \cite{huster2024boldly, talebi2018charm, jake2017difuze, song2019periscope, mera2024shift} and device-free modeling \cite{shen2022drifuzz, wu23devfuzz, ma2022print, zhao2022semantic}.



\textbf{Filesystem.} 
Filesystems are fundamental components of an OS kernel, crucial for managing user files and maintaining data consistency during system crashes. They are typically structured, complex binary blobs mounted as disk images. Users interact with a mounted filesystem image through a set of file operations (e.g., syscalls). Some studies focus on mutating images as binary inputs \cite{schumilo2017kafl, aschermann2019redqueen}, while others limit themselves to generating operations \cite{Syzkaller, chen2020test}.

\textbf{Network.} 
Network access is a fundamental feature in modern OS kernels, including streamlined or minimalistic ones. It provides a practical interface for testing and analysis, especially in resource-constrained systems such as embedded devices and IoT, where other interfaces may be constrained or unavailable \cite{nico2023drone,chen2018iotfuzzer}.

\textbf{Configuration.} In highly configurable OS kernels, configuration plays a critical role in shaping the input space. For instance, the default configuration of Linux kernels excludes approximately 80\% of code changes from the compiled binary \cite{necip2024maxmizing}. Because configuration changes necessitate recompiling the kernel, configuration testing must strike a balance between bootability, code coverage, and build time \cite{hasanov2025alittle}.


\subsubsection{Multi-dimensional Input}

Traditional methods typically concentrate on a single primary interface.  This approach, while effective, tends to overlook vulnerabilities within the interplay between various interfaces (\textbf{C2}). 
The need for a more holistic approach has become increasingly apparent \cite{wu23devfuzz}.
As revealed by Janus \cite{kim2019janus}, relying solely on one aspect of the filesystem—either the disk image or file operations—inevitably neglects the other. This results in incomplete and less effective testing. Therefore, Janus proposes a two-dimensional strategy that explores input space from both sides. Such a inherent synergy is not only applicable to syscall-filesystem but is also highly relevant in other scenarios, such as syscall-driver \cite{Scharnowski2023hoedur, jang2023reusb, xu2024saturn, ma2022print, pu2020exvivo}. For example, driver fuzzing is another area where a multi-dimensional approach is necessary. Device drivers often interact with various kernel interfaces, including syscalls and peripheral interfaces. 
Recent work explores USB drivers through record-and-replay techniques \cite{jang2023reusb} and host–gadget synergy \cite{xu2024saturn}, further highlighting the importance of considering cross-interface behavior. Given the existence of thousands of complex devices and drivers beyond USB, a general methodology to systematically capture and test the wide range of interface interactions in OS kernels is highly desired.

\begin{mybox}
    \textbf{Implication \yellowcircle{4}: Interface Interactions.} The diversity of kernel input interfaces underscores the complexity and breadth of input model in kernel fuzzing. 
    Beyond primary interfaces, studies have shown using one-sided input alone can miss up to 77\% of kernel code that would otherwise be explored through cross-interface interactions \cite{xu2024saturn, kim2019janus}. Although several efforts have investigated specific cross-interface scenarios, these attempts remain fragmented and domain-specific. The gap underscores the need for fuzzing approaches that explicitly model and exercise interface interactions, rather than treating interfaces in isolation. 
\end{mybox}

\subsection{Specification Awareness}\label{sec:grammar}

Recognizing specification requirements is critical for fuzzing as kernel interfaces expect inputs adhering to defined structures and formats. 
However, being grammar-aware while generating input is a complex task.
First, kernels interact with various devices and software, 
leading to diverse input formats. 
Second, beyond syntax, understanding input semantics is also important, as syntactically correct inputs can vary greatly in their effects.
Despite the emergence of specification free techniques \cite{bulekov2020nogrammar}, specification-based approaches remain the mainstream \cite{liu2020fans, Dawoud2021bring, marcel2023teezz, chen2021syzgen, hao2023syzdiscribe}. To address these challenges, researchers have focused on extracting grammar specifications as an integral component of input generation. 
These methods can generally be categorized into static and dynamic.

\subsubsection{Static Inference}
Before the advent of automation, developers typically relied on their domain expertise to manually craft syscall specifications \cite{Syzkaller}. This manual process was both labor-intensive and error-prone, significantly limiting the scalability of testing across a broader range of syscall targets. These challenges have driven the development of static specification inference techniques, which analyze kernel source code to infer expected input structures and identify constraints without requiring code execution \cite{zhao2022semantic, wu23devfuzz, sun2025syzparam, liu2020fans}.
For instance, DIFUZE \cite{jake2017difuze} pioneered the use of heuristic static analysis to extract syscall specifications from kernel driver source code. Building upon this foundation, SyzDescribe \cite{hao2023syzdiscribe} enhances both the depth and accuracy of inference by constructing a principled response model of kernel drivers. In certain cases, documentation can also support the analysis by providing supplementary information \cite{choi2021ntfuzz}. These rule-based approaches, while effective, often require continuous maintenance to accommodate evolving kernel targets and struggle with handling edge cases \cite{yang2025kernelgpt}.
To address these limitations, LLM-based techniques have presented high potential. KernelGPT \cite{yang2025kernelgpt} leverages the code-understanding capabilities of LLMs to adaptively extract specifications from source code with minimal reliance on external knowledge. It achieves a 6$\times$ increase in specification generation while maintaining 93.3\% accuracy, substantially outperforming traditional methods.
However, LLMs face their own challenges. Their difficulty in capturing implicit semantics arises because such semantics are not explicitly encoded in source code, but instead distributed across control flows, hidden in indirect function calls, or dependent on runtime conditions (e.g., device-specific logic in DRM subsystems). Current LLMs rely heavily on textual and structural cues from the training data, which makes it hard for them to infer these latent dependencies without additional program analysis support. 
Beyond static structure, recent study \cite{sun2025syzparam} highlights the influence of runtime parameters exposed via \textit{sysfs}, underscoring the complexity of the specification inference.
Despite the progress of automated approaches, specification generation for subsystems such as eBPF \cite{hung2024brf} and complex protocols \cite{nico2023drone} remains an open problem. These domains still depend heavily on manual analysis due to the scarcity of explicit specifications, the prevalence of implicit semantics, and the need for runtime context that neither static rules nor LLMs alone can fully resolve.

\subsubsection{Dynamic Analysis}
In contrast to static inference, dynamic approaches typically adopt an iterative strategy that leverages runtime information collected from executing targets to complement static analysis. Based on the method of data collection, existing dynamic techniques can be broadly categorized into two types: log analysis \cite{chen2021syzgen, han2017imf, jang2023reusb, aafer21smarttvs} and dynamic probing \cite{sun2022ksg, yin2023kext, zhu2024cross}. Log analysis serves as an indirect strategy, particularly useful for binary-only kernels where direct program inspection is inconvenient.
For example, targeting macOS, SyzGen \cite{chen2021syzgen} starts by generating initial syscall templates through trace analysis and subsequently refines argument types and constraints using symbolic execution. Similar log-based techniques have also been applied to analyze interactions in USB protocols \cite{peng2020usb} and the Android ecosystem \cite{aafer21smarttvs}. Dynamic probing has become feasible due to the enhanced observability features integrated into modern operating system kernels. In Linux, eBPF can be repurposed to hook into syscall probes and access the corresponding file operations \cite{sun2022ksg}. Similarly, macOS provides a kernel extension wrapper that offers comparable functionality and serves as an entry point for inference when combined with taint analysis \cite{yin2023kext}. Compared to static inference, dynamic approaches yield more accurate and realistic outputs. However, they may also suffer from false negatives, as certain drivers or components may not be active or accessible during runtime analysis.

\begin{mybox}
    \textbf{Implication \yellowcircle{5}: Hybrid Generation.} 
    Specifications are crucial for grammar-aware fuzzing but remain difficult to generate due to diverse formats and implicit semantics. Static inference offers broad coverage but lacks adaptability. LLM-based methods improve  generality by up to 6$\times$ \cite{yang2025kernelgpt} but cannot handle indirect call cases, and dynamic analysis provides realism but overlooks inactive paths. These complementary strengths and weaknesses indicate that no single approach is sufficient. Hybrid solutions combining static, dynamic, and LLM-based techniques are needed to achieve comprehensive and accurate specification generation.
\end{mybox}

\renewcommand{\arraystretch}{1.4}
\begin{table}[htbp]
\scriptsize
\begin{center}
\caption{Summary of kernel fuzzers and their solutions used in input model.}
\label{tab:input}
\begin{threeparttable}
\begin{tabular}{|c|l|l|@{}c@{}|l|lll|}
\hline
\multirow{2}{*}{\textbf{Year}} & \multirow{2}{*}{\textbf{Fuzzer}} & \multirow{2}{*}{\textbf{Func.}} & \multirow{2}{*}{\textbf{Type}} & \multirow{2}{*}{\textbf{Target}} & \multicolumn{3}{c|}{\textbf{Input Model}} \\ \cline{6-8} 
 &  &  &  &  & \multicolumn{1}{c|}{\begin{tabular}[c]{@{}c@{}}Interface\\ Ident.\end{tabular}} & \multicolumn{1}{c|}{\begin{tabular}[c]{@{}c@{}}Spec.\\ Awareness\end{tabular}} & \multicolumn{1}{c|}{\begin{tabular}[c]{@{}c@{}}Dependency\\ Recognition\end{tabular}} \\ \hline
2017 & IMF \cite{han2017imf} & \circlewithtext{2.2} \circlewithtext{2.3} & \CIRCLE & MacOS & \multicolumn{1}{l|}{Sys.} & \multicolumn{1}{l|}{LA} & Exp. \\ \hline
2017 & Syzkaller \cite{Syzkaller} & \circlewithtext{2.2} \circlewithtext{2.3} & \CIRCLE & General & \multicolumn{1}{l|}{Sys.} & \multicolumn{1}{l|}{Manual} & Exp. \\ \hline
2017 & DIFUZE \cite{jake2017difuze} & \circlewithtext{F.2} & \CIRCLE & Linux \& Android & \multicolumn{1}{l|}{Sys.} & \multicolumn{1}{l|}{RA} & - \\ \hline
2018 & MoonShine \cite{shankara2018moonshine} & \circlewithtext{2.3} & \LEFTcircle & Linux & \multicolumn{1}{l|}{Sys.} & \multicolumn{1}{l|}{-} & Exp. \& Imp. \\ \hline
2019 & Janus \cite{kim2019janus} & \circlewithtext{2.1} & \LEFTcircle & Linux & \multicolumn{1}{l|}{Sys. \& FS} & \multicolumn{1}{l|}{-} &  \\ \hline
2020 & Dogfood \cite{chen2020test} &\circlewithtext{2.3} &\LEFTcircle  & Linux & \multicolumn{1}{l|}{Sys.} & \multicolumn{1}{l|}{-} & Exp.  \\ \hline
2020 & EASIER \cite{pu2020exvivo} & \circlewithtext{2.1} \circlewithtext{2.2} & \LEFTcircle & Linux & \multicolumn{1}{l|}{Sys. \& Periph.} & \multicolumn{1}{l|}{DP} & - \\ \hline
2020 & FANS \cite{liu2020fans} & \circlewithtext{2.2} \circlewithtext{2.3} & \CIRCLE  & Android & \multicolumn{1}{l|}{Sys.} & \multicolumn{1}{l|}{RA} & Exp. \\ \hline
2020 & SyzGen \cite{chen2021syzgen} & \circlewithtext{2.2} \circlewithtext{2.3} & \LEFTcircle &  MacOS & \multicolumn{1}{l|}{Sys.} & \multicolumn{1}{l|}{LA \& SE} & Exp. \\ \hline
2020 & ACHyb \cite{hu2021achyb} & \circlewithtext{2.3} & \LEFTcircle & Linux & \multicolumn{1}{l|}{Sys.} & \multicolumn{1}{l|}{-} & Exp.  \\ \hline
2021 & DYNAMO \cite{Dawoud2021bring}& \circlewithtext{2.2} & \LEFTcircle & Android & \multicolumn{1}{l|}{Sys.} & \multicolumn{1}{l|}{RA} & - \\ \hline
2021 & NTFUZZ \cite{choi2021ntfuzz} & \circlewithtext{2.2} \circlewithtext{2.3} & \CIRCLE & Win. & \multicolumn{1}{l|}{Sys.} & \multicolumn{1}{l|}{RA} & Exp.  \\ \hline
2021 & HEALER \cite{sun2021healer}& \circlewithtext{2.3} & \LEFTcircle & Linux & \multicolumn{1}{l|}{Sys.} & \multicolumn{1}{l|}{-} & Exp. \& Imp. \\ \hline
2021 & PrIntFuzz \cite{ma2022print}& \circlewithtext{2.1} \circlewithtext{2.2} & \LEFTcircle & Linux & \multicolumn{1}{l|}{Sys. \& Periph.} & \multicolumn{1}{l|}{RA} & - \\ \hline
2022 & Dr. Fuzz \cite{zhao2022semantic} & \circlewithtext{2.2} & \LEFTcircle & Linux & \multicolumn{1}{l|}{Periph.} & \multicolumn{1}{l|}{RA \& SE} & - \\ \hline
2022 & KSG \cite{sun2022ksg}& \circlewithtext{2.2} & \LEFTcircle & Linux & \multicolumn{1}{l|}{Sys.} & \multicolumn{1}{l|}{DP} & - \\ \hline
2023 & Schiller et al. \cite{nico2023drone}& \circlewithtext{2.1} \circlewithtext{2.2} & \CIRCLE & Android \& RTOS & \multicolumn{1}{l|}{Network} & \multicolumn{1}{l|}{Manual} & - \\ \hline
2023 & FuzzNG \cite{bulekov2020nogrammar}& \circlewithtext{2.2} & \LEFTcircle & Linux & \multicolumn{1}{l|}{Sys.} & \multicolumn{1}{l|}{Spec. Free} & - \\ \hline
2023 & SyzDescribe \cite{hao2023syzdiscribe}& \circlewithtext{2.2} & \LEFTcircle & Linux & \multicolumn{1}{l|}{Sys.} & \multicolumn{1}{l|}{RA} & Exp. \\ \hline
2023 & DEVFUZZ \cite{wu23devfuzz}& \circlewithtext{2.1} & \LEFTcircle & Win. \& Linux & \multicolumn{1}{l|}{Sys. \& Periph.} & \multicolumn{1}{l|}{RA \& SE} & - \\ \hline
2023 & TEEzz \cite{marcel2023teezz}& \circlewithtext{2.2} \circlewithtext{2.3} & \LEFTcircle & TEE & \multicolumn{1}{l|}{Sys.} & \multicolumn{1}{l|}{RA} & Exp. \\ \hline
2023 & ReUSB \cite{jang2023reusb}& \circlewithtext{2.1} \circlewithtext{2.2} & \LEFTcircle & Linux & \multicolumn{1}{l|}{Sys. \& Periph.} & \multicolumn{1}{l|}{LA} & - \\ \hline
2023 & KextFuzz \cite{yin2023kext}& \circlewithtext{2.2} & \LEFTcircle & MacOS & \multicolumn{1}{l|}{Sys.} & \multicolumn{1}{l|}{DP} & - \\ \hline
2023 & BRF \cite{hung2024brf}& \circlewithtext{2.2} & \LEFTcircle & Linux & \multicolumn{1}{l|}{Sys.} & \multicolumn{1}{l|}{Manual} & - \\ \hline
2024 & MOCK \cite{xu2024mock}& \circlewithtext{2.3} & \LEFTcircle & Linux & \multicolumn{1}{l|}{Sys.} & \multicolumn{1}{l|}{-} & Exp. \& Imp. \\ \hline
2024 & Countdown \cite{bai2024countdown}& \circlewithtext{2.3} & \LEFTcircle & Linux & \multicolumn{1}{l|}{Sys.} & \multicolumn{1}{l|}{-} & Exp. \& Imp. \\ \hline
2024 & SyzGen++ \cite{chen2024syzgenplus}& \circlewithtext{2.2} 
\circlewithtext{2.3} & \LEFTcircle & Linux \& MacOS & \multicolumn{1}{l|}{Sys.} & \multicolumn{1}{l|}{LA \& SE} & Exp. \\ \hline
2024 & SATURN \cite{xu2024saturn}& \circlewithtext{2.1} & \LEFTcircle & Linux & \multicolumn{1}{l|}{Sys. \& Periph.} & \multicolumn{1}{l|}{-} & - \\ \hline
2025 & KernelGPT \cite{yang2025kernelgpt}& \circlewithtext{2.2} & \LEFTcircle & Linux & \multicolumn{1}{l|}{Sys.} & \multicolumn{1}{l|}{LLMA} & - \\ \hline
2025 & Hasanov et al. \cite{hasanov2025alittle}& \circlewithtext{2.1} & \LEFTcircle & Linux & \multicolumn{1}{l|}{Config.} & \multicolumn{1}{l|}{-} & - \\ \hline
2025 & SyzParam \cite{sun2025syzparam}& \circlewithtext{2.2} & \LEFTcircle & Linux & \multicolumn{1}{l|}{Sys.} & \multicolumn{1}{l|}{RA} & - \\ \hline
\end{tabular}
\end{threeparttable}
\end{center}
\begin{tablenotes}[flushleft]
\scriptsize
\item[] \circlewithtext{1.1}: The satisfied functionality of a method; \Circle\ : Whitebox fuzzing. \LEFTcircle\ : Greybox fuzzing. \CIRCLE\ : Blackbox fuzzing; 
\item[] Win: Windows; Sys: Syscall; FS: File System Image; Periph: Peripheral; Config: Configuration; LA: Log Analysis; RA: Rule-based Analysis; DP: Dynamic Probing; SE: Symbolic Execution; LLMA: LLM-based Analysis; Exp: Explicit; Imp: Implicit; -: Not Specified or Not Targeted.
\end{tablenotes}
\end{table}

\subsection{Dependency Recognition}\label{sec:dependency}
One of the most critical characteristics of OS kernels is their statefulness, as discussed in \textbf{C3}.
This nature necessitates a coordinated organization of test cases, referred to as explicit / implicit dependency.

\begin{figure}[htbp]
    \centering
    \includegraphics[width=0.8\textwidth]{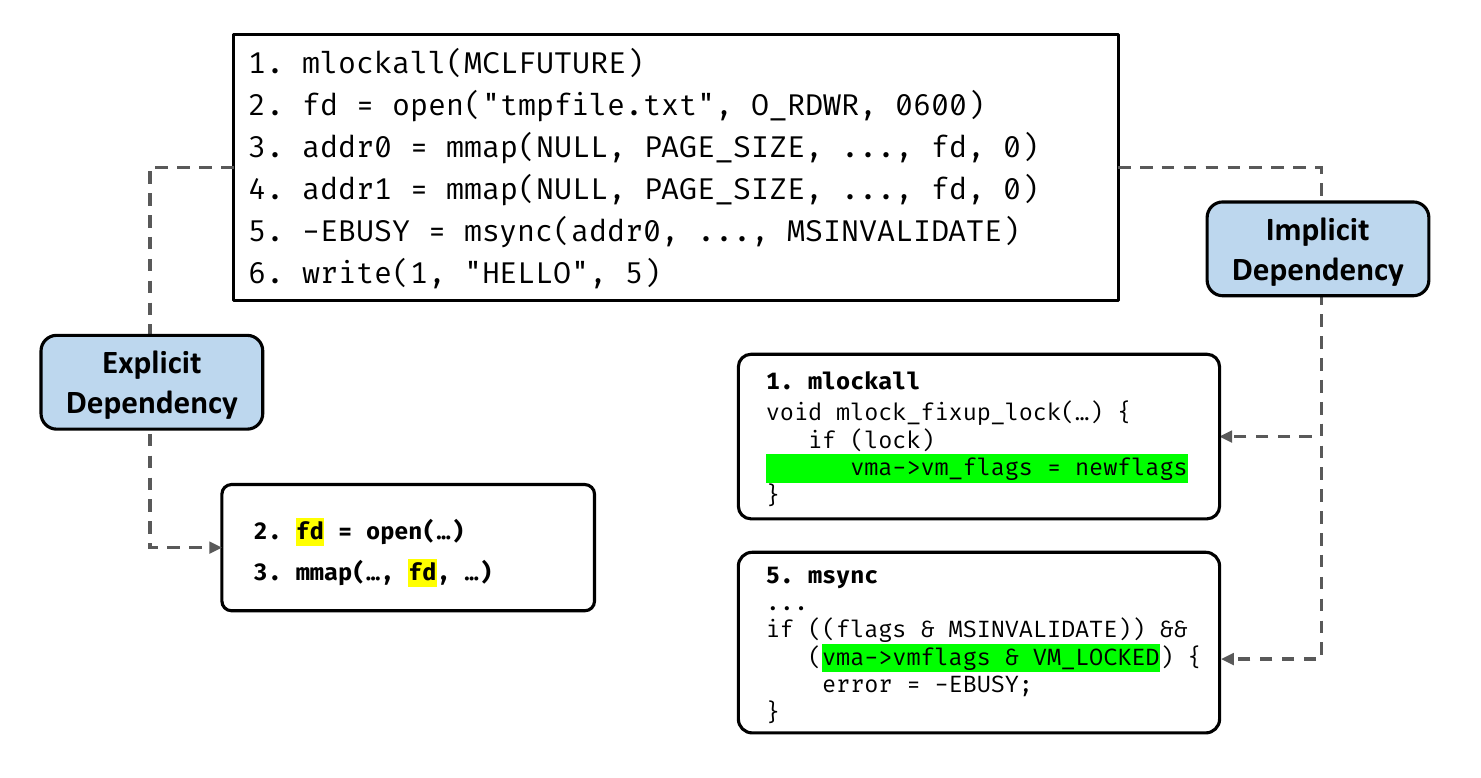}
    \caption{Explicit and implicit dependencies between syscalls.}
    \label{fig:dep}
\end{figure}

\subsubsection{Explicit Dependency}
Explicit dependencies refer to the direct relationships where the output of one syscall directly influences the input of another, such as in resource assignment. In this context, syscalls that generate outputs are identified as producers, while those that consume these outputs are considered consumers. We define a syscall \textit{$c_i$} as explicitly dependent on another syscall \textit{$c_j$} when \textit{$c_i$} is a consumer and \textit{$c_j$} is a producer. 
If \texttt{open} is not executed or fails, subsequent syscalls like \texttt{mmap} cannot execute successfully. Beyond return values, syscalls can also accept parameters derived from other syscalls. Some studies have sought to identify explicit dependencies among syscalls through methods such as trace inference \cite{han2017imf, chen2021syzgen, chen2024syzgenplus}, layered model building \cite{chen2020test} or producer-consumer analysis \cite{Syzkaller, shankara2018moonshine, sun2021healer, xu2024mock}. For example, IMF \cite{han2017imf} records the input and output values of hooked syscalls and then applies heuristic inference to logs, focusing on the order and value of entries.
In the realm of producer-consumer analysis, fuzzers like Syzkaller perform type-based analysis on specifications, assigning higher priority to producer-consumer pairs. 

\subsubsection{Implicit Dependency}
Implicit dependency, in contrast, is more subtle and mandates a sequence of syscalls without involving explicit producer-consumer relationships.
It stems from the kernel's inherently stateful nature, characterized by extensive shared data structures and resources that can be accessed through multiple system calls. For instance, memory operations like \texttt{mlockall} and \texttt{msync} have no relevance in parameters or return value, but they operate on shared variables implicitly (shown in Figure~\ref{fig:dep}), thereby creating implicit dependencies \cite{shankara2018moonshine}. These dependencies are challenging to identify because they are often obscured within the vast and complex kernel codebase. Researchers currently use static analysis, dynamic analysis, and a combination of both to uncover these dependencies. Several studies have proposed static analysis techniques for identifying potential dependency pairs in kernel code, particularly when syscalls operate on shared global variables \cite{shankara2018moonshine, kim2020hfl, marius2023actor}. Although these methods can be informative, they are prone to generating false positives \cite{jeong2019razzer}. 
More recent research has advanced the field through dynamic dependency recognition, which offers more reliable results. For instance, HEALER \cite{sun2021healer} and MOCK \cite{xu2024mock} infer runtime dependencies by minimizing coverage and employ context-free and context-aware models, respectively
Furthermore, refcount has been explored as an additional means of representing implicit dependencies for mutation guidance \cite{bai2024countdown}.
Despite these advancements, the aforementioned methods are seldom applicable to closed-source targets. In fact, heuristic-based approaches remain dominant due to their practicality and usability.

\begin{mybox}
    \textbf{Implication \yellowcircle{6}: Dependency Integration.} Prior works have explored capturing or modeling explicit and implicit dependencies, yet no consensus has been reached among researchers. Ideally, these dependencies should be incorporated into specifications like \textit{Syzlang}, which currently lacks support for them. Extending \textit{Syzlang} to integrate these dependencies presents a promising direction.
\end{mybox}

\section{Fuzzing Loop}
\label{sec:loop} 

Once the environment is configured and inputs defined, fuzzers initiate the fuzzing process. Traditional methods face kernel-specific challenges such as statefulness and concurrency. Functionalities required at this stage include execution throughput, mutation intelligence, and feedback mechanisms. Table \ref{tab:fuzz} provides a summary of the literature and the corresponding solutions employed within the fuzzing loop.

\subsection{Execution Throughput} 

Execution speed has a significant impact on performance of both user- and kernel-space fuzzing. However, the emulation-based architectures commonly employed in kernel fuzzing, together with its statefulness, necessitate solutions that differ substantially from those used in user space.
Note that the techniques discussed here are developed based on their native environments and do not alter execution functionality, distinguishing them from those in \textit{F1.1}.

\subsubsection{Virtualization Enhancement} 
The introduction of a VM layer in kernel fuzzing enables improvements in execution throughput through enhanced virtualization efficiency.

\textbf{Accelerated virtualization.} Virtualization acceleration techniques \cite{yu2020ava} have been widely studied in the community. Existing fuzzing methods enable high performance virtualization through hardware assistance \cite{schumilo2017kafl} and user-mode emulation \cite{zheng2019firm}. However, these approaches are generally architecture- and fuzzer-specific, and thus limit their application.


\textbf{Efficient synchronization.} As the memory space of host and guest VMs is mutually isolated, their communication incurs significant overheads. For example, Syzkaller runs the fuzzer and executor inside the VM and synchronizes the state via RPC. Subsequent works mitigate the problem by proposing more efficient synchronization mechanisms, such as shared memory \cite{sun2021healer, yang2024thunder, liu2023horus} and data transfer \cite{liu2023horus}.

\subsubsection{System Snapshot} 
The accumulated internal states may corrupt the kernel or interfere with subsequent executions. Hence, it is time-consuming but necessary to reboot the system regularly.  The snapshot techniques save time and increase throughput by taking proper system snapshots and restoring them when necessary. The typical practice is to fork an initialized VM as a new instance \cite{pandey2019triforce, Syzkaller}.

\textbf{Lightweight snapshot.} The native QEMU snapshot dumps all the CPU registers and the memory space and thus poses into files. Nevertheless, such a faithful snapshot may pose non-negligible overheads. A lightweight snapshot tailored to fuzzing is heavily desired. To achieve this, existing methods selectively restore memory pages on a Copy-on-Write principle \cite{zheng2019firm}, or customize the snapshot function upon QEMU/KVM for fuzzing adaptation \cite{schumilo2021nyx, bulekov2020nogrammar, gong2021snowboard}. Since this process operates at the emulation level, it can benefit multiple OS kernels that are virtualizable.

\textbf{Checkpoint policy.} It is typical of fuzzers to take a startup snapshot and restore it when necessary. However, the input executions undergo several similar phases besides startup. Hence, by properly creating continuous checkpoints \cite{song2020agamotto, jung2025moneta, yuan2023ddrace}, fuzzers can skip repeated steps and have direct access to the state that is established by time-consuming operations.

\begin{mybox}
    \noindent \textbf{Implication \yellowcircle{7}: Throughput Gain.}
    Studies have shown that secondary operations, such as data transfer, can account for up to 54\% of fuzzing time~\cite{yang2024thunder, liu2023horus}, significantly degrading overall performance.
    Consequently, optimizing virtualization for enhanced kernel fuzzer interaction and improving the bootstrap process for rapid recovery is key to increasing throughput. Future directions should aim at developing more universally applicable virtualization enhancements, creating lightweight snapshot techniques for fuzzing, and devising effective checkpoint policies to minimize redundant operations.
\end{mybox}

\subsection{Mutation Intelligence} \label{sec:mutation}

Although the input model reduces the search space, blind fuzzing still struggles to find bugs due to the complexity and micro-level variations in test cases.
To address this, existing strategies are structured around three key phases:  constraint solving, thread scheduling and decision intelligence. 


\subsubsection{Constraint Solving} 
While random fuzzing excels under lenient conditions, it struggles with stringent branch constraints, such as magic bytes and checksums, requiring extensive efforts to meet specific conditions. Integrating symbolic execution \cite{yun2018qsym}  with fuzzing has significantly boosted the ability to tackle complex constraints in user-space fuzzing, a strategy equally beneficial for kernel fuzzing. Hybrid approaches combining symbolic execution and fuzzing have been applied in kernel environments for interface recovery and value inference \cite{chen2021syzgen, sun2022ksg, zhao2023state, hao2022demy}, although scaling these methods for real-time use in complex kernels presents challenges, including indirect control transfers and path explosion \cite{kim2020hfl}. Solutions specifically designed for kernel fuzzing aim to overcome these obstacles through indirect control flow transformation \cite{kim2020hfl} and selective strategies \cite{chen2022sfuzz, aschermann2019redqueen}. Besides, we also call for 
we emphasize the need to enhance the accessibility of the field, particularly given the current lack of available dynamic constraint-solving solutions.

\subsubsection{Thread Scheduling} 
Concurrency-related kernel vulnerabilities emerge from the inherent complexity and unpredictability of non-deterministic kernel scheduling \cite{pabla2009completely, pedro2014ski}. Detecting these vulnerabilities requires careful consideration of both test inputs and specific thread interleavings, making precise control over threads essential for identifying concurrency issues.
Rather than modifying the kernel scheduler—a process that is both labor-intensive and risky—the prevalent alternative is to control threads using delay injection \cite{yuan2023ddrace, jiang2022context, meng2020krace} or using hypervisor-level control \cite{dae2024ozz, jeong2019razzer}. 
Introducing delays between threads provides a straightforward means to influence thread scheduling \cite{yuan2023ddrace, jiang2022context, meng2020krace}. However, this approach offers limited control over complex thread interactions and often results in significant performance degradation due to the added latency.
To perform preemption, the hypervisor-based approaches either utilize hardware breakpoints as scheduling points and trap the thread in infinite loops \cite{jeong2019razzer, jeong2023segfuzz}, or suspend thread execution on vCPU \cite{gong2021snowboard, gong2023snowcat}.
Despite their effectiveness in enforcing thread control, such methods typically demand extensive modifications to both the hypervisor and the kernel. Consequently, they also suffer from high overhead and limit their scalability and usability severely \cite{xu2025concur}. In addition to the aforementioned approaches, research regarding RTOS has explored exposing concurrency bugs through task priority manipulation \cite{shen2021rtkaller}. More recently, SECT \cite{xu2025concur} introduced a novel direction for thread scheduling. Rather than relying on external mechanisms, SECT targets the scheduling source itself by implementing a custom  scheduler. This scheduler leverages the Linux eBPF feature and thus enables lightweight and flexible thread control.

\subsubsection{Decision Intelligence} 
To maximize bug discovery under limited computational resources, fuzzers must make intelligent decisions when selecting seeds and mutation operators for each iteration. This phase represents the area of greatest similarity between user-space and kernel-space fuzzing. It often involves the use of runtime feedback (e.g., branch coverage) and optimization techniques such as reinforcement learning \cite{wang2021rein, zhang2022mobfuzz, yue2020ecofuzz}, information entropy \cite{marcel2020boost}, simulated annealing \cite{zhang2022exploit, li2025yesterday}, and particle swarm optimization \cite{lyu2019mopt}. These strategies are employed by fuzzers to support a range of testing objectives, including coverage maximization \cite{wang2021syzvegas, xu2024mock, gong2025snowflow}, concurrency analysis \cite{gong2023snowcat}, vulnerability pattern detection \cite{lee2024syzrisk}, and reachability exploration \cite{zhang2022exploit, li2023gfuzz}.
For instance, SyzVegas \cite{wang2021syzvegas} utilizes a reinforcement learning algorithm to prioritize seeds and mutation strategies that yield higher coverage rewards. Similarly, SyzRisk \cite{lee2024syzrisk} intends to allocate more fuzzing energy to inputs with higher probabilities of matching known vulnerability patterns. 
Although the challenges of decision intelligence are similar in both user space and kernel space, algorithms for the kernel domain have not been studied as extensively as their user-space counterparts \cite{}.
It would be an interesting topic to investigate the effectiveness of user-space scheduling algorithms in the context of kernel fuzzing and explore novel strategies specifically tailored to unique characteristics of kernel environments.



\begin{mybox}
    \noindent \textbf{Implication \yellowcircle{8}: Native-feature Intelligence.} 
    While existing strategies have made notable progress, their intrusive nature and high overhead have constrained their practical applicability. These limitations make them unsuitable as a robust foundation for addressing challenges such as constraint solving and thread scheduling.
    Recent studies suggest that leveraging native kernel features offers a promising direction (e.g., achieving 11.4$\times$ speed-up in thread scheduling \cite{xu2025concur}), toward developing lightweight, scalable solutions that enhance mutation intelligence.
    Realizing this potential will require closer collaboration between the community and researchers.
\end{mybox}

\renewcommand{\arraystretch}{1.4}
\begin{table}[htbp]
\scriptsize
\begin{center}
\caption{Summary of kernel fuzzers and their solutions used in fuzzing loop.}
\label{tab:fuzz}
\begin{threeparttable}
\begin{tabular}{|c|l|l|@{}c@{}|l|lll|}
\hline
\multirow{2}{*}{\textbf{Year}} & \multirow{2}{*}{\textbf{Fuzzer}} & \multirow{2}{*}{\textbf{Func.}} & \multirow{2}{*}{\textbf{Type}} & \multirow{2}{*}{\textbf{Target}} & \multicolumn{3}{c|}{\textbf{Fuzzing Loop}} \\ \cline{6-8} 
 &  &  &  &  & \multicolumn{1}{c|}{Thpt.} & \multicolumn{1}{c|}{\begin{tabular}[c]{@{}c@{}}Mutation\\ Intel.\end{tabular}} & \multicolumn{1}{c|}{\begin{tabular}[c]{@{}c@{}}Feedback\\ Mechanism\end{tabular}} \\ \hline
2017 & CAB-Fuzz \cite{atc17cabfuzz} & \circlewithtext{3.2} & \CIRCLE & Win.  & \multicolumn{1}{l|}{-}  & \multicolumn{1}{l|}{CE} & -   \\ \hline
2019 & RedQueen \cite{aschermann2019redqueen} & \circlewithtext{3.2} & \LEFTcircle & Linux & \multicolumn{1}{l|}{-} & \multicolumn{1}{l|}{ISC} & Branch \\ \hline
2019 & Razzer \cite{jeong2019razzer} & \circlewithtext{3.2} & \LEFTcircle & Linux & \multicolumn{1}{l|}{-}  & \multicolumn{1}{l|}{HC} & Branch   \\ \hline
2019 & FIRM-AFL \cite{zheng2019firm} & \circlewithtext{3.1} & \LEFTcircle &  Emb. Linux & \multicolumn{1}{l|}{LS}  & \multicolumn{1}{l|}{-} & Branch   \\ \hline
2020 & HFL  \cite{kim2020hfl} & \circlewithtext{3.2} & \LEFTcircle & Linux & \multicolumn{1}{l|}{-} & \multicolumn{1}{l|}{CE} & Branch   \\ \hline
2020 & Krace \cite{meng2020krace}& \circlewithtext{3.3} & \LEFTcircle & Linux & \multicolumn{1}{l|}{-}    & \multicolumn{1}{l|}{DI} & Branch \& Inst. Pair \\ \hline
2020 & Agamotto \cite{song2020agamotto}& \circlewithtext{3.1} & \LEFTcircle  & Linux & \multicolumn{1}{l|}{DC} & \multicolumn{1}{l|}{-} & Branch \\ \hline
2021 & SyzVegas \cite{wang2021syzvegas} & \circlewithtext{3.2} & \LEFTcircle &  Linux & \multicolumn{1}{l|}{-} & \multicolumn{1}{l|}{SS \& MS} & Branch   \\ \hline
2021 & Snowboard \cite{gong2021snowboard}  & \circlewithtext{3.1} & \LEFTcircle & Linux & \multicolumn{1}{l|}{Snap.}    & \multicolumn{1}{l|}{HC} & Branch   \\ \hline
2021 & Rtkaller \cite{shen2021rtkaller} & \circlewithtext{3.2} & \LEFTcircle & RTOS & \multicolumn{1}{l|}{-}    & \multicolumn{1}{l|}{PA} & Branch   \\ \hline
2022 & SFuzz \cite{chen2022sfuzz}& \circlewithtext{3.2} & \LEFTcircle & RTOS & \multicolumn{1}{l|}{-}    & \multicolumn{1}{l|}{CE \& TA} & Branch  \\ \hline
2022 & RoboFuzz \cite{kim2022robofuzz} & \circlewithtext{3.3} & \LEFTcircle & ROS & \multicolumn{1}{l|}{-}  & \multicolumn{1}{l|}{-} & Branch \& Semantic   \\ \hline
2022 & Conzzer \cite{jiang2022context} & \circlewithtext{3.1} \circlewithtext{3.3} & \LEFTcircle & Linux  & \multicolumn{1}{l|}{-}    & \multicolumn{1}{l|}{DI} & \begin{tabular}[c]{@{}l@{}}Branch \&\\ Concur. Call Pair\end{tabular}   \\ \hline
2022 & GREBE  \cite{lin2022grebe} & \circlewithtext{3.3} & \LEFTcircle & Linux & \multicolumn{1}{l|}{-}    & \multicolumn{1}{l|}{-} & Distance   \\ \hline
2022 & StrawFuzz \cite{zhang2022exploit} & \circlewithtext{3.2} \circlewithtext{3.3} & \LEFTcircle & Android & \multicolumn{1}{l|}{-}  & \multicolumn{1}{l|}{SS \& MS} & Distance \& Key Var. \\ \hline
2022 & FuzzUSB \cite{kim2022fuzzusb}  & \circlewithtext{3.3} & \LEFTcircle & Linux & \multicolumn{1}{l|}{-}    & \multicolumn{1}{l|}{-} & Branch \& FSM \\ \hline
2022 & StateFuzz \cite{zhao2023state} & \circlewithtext{3.3} & \LEFTcircle & Linux & \multicolumn{1}{l|}{-}    & \multicolumn{1}{l|}{-} & Branch \& State Var.   \\ \hline
2023 & Snowcat \cite{gong2023snowcat}  & \circlewithtext{3.2} & \LEFTcircle & Linux & \multicolumn{1}{l|}{-}    & \multicolumn{1}{l|}{SS} & Branch   \\ \hline
2023 & Thunder. \cite{yang2024thunder} & \circlewithtext{3.1} & \LEFTcircle & Linux & \multicolumn{1}{l|}{VE}    & \multicolumn{1}{l|}{-} & Branch   \\ \hline
2023 & SyzDirect \cite{tan2023syzdirect} & \circlewithtext{3.3} & \LEFTcircle & Linux & \multicolumn{1}{l|}{-}    & \multicolumn{1}{l|}{-} & Distance   \\ \hline
2023 & PLA  \cite{ryan2023precise} & \circlewithtext{3.2} & \LEFTcircle & Linux & \multicolumn{1}{l|}{-}    & \multicolumn{1}{l|}{SS} & Branch   \\ \hline
2023 & SegFuzz \cite{jeong2023segfuzz}  & \circlewithtext{3.3} & \LEFTcircle & Linux & \multicolumn{1}{l|}{-}    & \multicolumn{1}{l|}{HC} & Branch \& Segment   \\ \hline
2023 & KLAUS \cite{wu2023mitigating}  & \circlewithtext{3.2} \circlewithtext{3.3} & \LEFTcircle &  Linux  & \multicolumn{1}{l|}{-}    & \multicolumn{1}{l|}{-} & Distance    \\ \hline
2023 & DDRace \cite{yuan2023ddrace} & \circlewithtext{3.3} & \LEFTcircle & Linux & \multicolumn{1}{l|}{DC}    & \multicolumn{1}{l|}{DI} & Distance \& Inst. Pair   \\ \hline
2023 & ACTOR \cite{marius2023actor}& \circlewithtext{3.3} & \LEFTcircle &  Linux & \multicolumn{1}{l|}{-}    & \multicolumn{1}{l|}{-} & Branch \& Action   \\ \hline
2023 & G-Fuzz  \cite{li2023gfuzz}  & \circlewithtext{3.3}  & \LEFTcircle & gVisor \& Linux  & \multicolumn{1}{l|}{-}    & \multicolumn{1}{l|}{SS} & Branch \& Distance   \\ \hline
2023 & Horus \cite{liu2023horus} & \circlewithtext{3.1} & \LEFTcircle &  Linux  & \multicolumn{1}{l|}{VE}   & \multicolumn{1}{l|}{-} & Branch   \\ \hline
2024 & SyzTrust \cite{wang2024syztrust} & \circlewithtext{3.3} & \LEFTcircle & TEE & \multicolumn{1}{l|}{-}    & \multicolumn{1}{l|}{-} & Branch \& State Var.  \\ \hline
2024 & KBinCov \cite{liu2024leverage}  & \circlewithtext{3.3} & \LEFTcircle & Linux & \multicolumn{1}{l|}{-}    & \multicolumn{1}{l|}{-} & Branch \& State Var.   \\ \hline
2024 & SyzRisk \cite{lee2024syzrisk} & \circlewithtext{3.1} & \LEFTcircle &  Linux  & \multicolumn{1}{l|}{-}    & \multicolumn{1}{l|}{SS} & Branch   \\ \hline
2025 & SECT \cite{xu2025concur}& \circlewithtext{3.2} & \LEFTcircle & Linux  & \multicolumn{1}{l|}{-}    & \multicolumn{1}{l|}{eBPFS} & Branch   \\ \hline
\end{tabular}
\end{threeparttable}
\end{center}
\begin{tablenotes}[flushleft]
\scriptsize
\item[] \circlewithtext{1.1}: The satisfied functionality of a method; \Circle\ : Whitebox fuzzing. \LEFTcircle\ : Greybox fuzzing. \CIRCLE\ : Blackbox fuzzing; 
\item[] Win: Windows; Emb. Linux: Embedded Linux; LS: Lightweight Snapshot; DC: Dynamic Checkpoint; VE: Virtualization Enhancement; CE: Concolic Execution; ISC: Input-to-State Correspondence; HC: Hypervisor-level Control; SS: Seed Scheduling; MS: Mutation Scheduling; PA: Priority Assignment; TA: Taint Analysis; DI: Delay Injection; eBPFS: eBPF-based Scheduling; FSM: Finite State Machine; -: Not Specified or Not Targeted.
\end{tablenotes}
\end{table}

\subsection{Feedback Mechanism} \label{sec:feedback}

As discussed in Section \ref{sec:covcollect}, existing kernel fuzzing proposals have improved feedback acquisition using both invasive instrumentation and non-invasive tracing. Establishing an efficient feedback mechanism for OS kernel fuzzing requires defining clear testing goals and proper fitness metrics. 

\subsubsection{Testing Goals} 
Fuzzers discover vulnerabilities with primary testing objectives, expanding code coverage (coverage-guided fuzzing) and prioritizing specific code locations (directed fuzzing).

\textbf{Coverage-guided fuzzing.} Kernel fuzzers typically employ a coverage-centric strategy, aiming to maximize execution path diversity—a signal that has proven effective for uncovering bugs \cite{wang2019sensitive}. These fuzzers assess input quality using fitness metrics, with basic block coverage \cite{jake2017difuze} and branch coverage \cite{shi2019industry, dae2024ozz} being the most commonly used. Additionally, state- and concurrency-oriented metrics have been developed to address kernel-specific characteristics, which will be discussed in more detail later. Due to the inherent complexity of operating system kernels, effective fuzzing requires a multi-dimensional feedback mechanism that integrates control flow analysis, data flow tracking, state exploration, and concurrency probing.

\textbf{Directed fuzzing.} Directed greybox fuzzing (DGF) have exhibited potential in tasks like patch testing \cite{lee2021constraint, marcel2017aflgo} and crash reproduction \cite{lin2022grebe, zou2022syzscope}. Unlike coverage-centric fuzzing, DGF prioritizes seeds closer to specific target points, either manually set or indicated by sanitizers. However, its adoption in the kernel has been limited due to unique challenges of kernels. As mentioned in \textbf{C2}, one of the challenges arises from the testing interface. The input of DGF in user space are continuous and can be arbitrarily mutated. In contrast, kernel-space DGF inputs consist of a discrete set of syscalls, from which a relevant subset that actually reaches the target location must be identified. To address the challenge, existing work conducts static analysis from target code locations to assess reachability. Some of these efforts focus on the specific parameter values required by syscalls \cite{tan2023syzdirect, shi2024industry, you2017semfuzz}, while others emphasize optimizations tailored for particular subsystems \cite{li2023gfuzz}. 
Due to the large-scale codebases, a further challenge specific to kernel fuzzing is the high computational overhead and accuracy degradation associated with distance calculations. For instance, the state-of-theart DGF method in user space, AFLGo \cite{marcel2017aflgo}, takes more than 16 hours to calculate the static distance for
gVisor \cite{gvisor} on average \cite{li2023gfuzz}.
In response, recent fuzzers have introduced various mitigation strategies, including enhanced distance metrics \cite{li2023gfuzz, yuan2023ddrace}, key object filtering \cite{lin2022grebe, zou2022syzscope}, and improved exploration–exploitation balancing techniques \cite{li2023gfuzz, zhang2022exploit}.
For example, G-Fuzz \cite{li2023gfuzz} proposes a lightweight, directed fuzzing framework for gVisor. Its primary contribution lies in a fine-grained and efficient mechanism that substantially reduces the cost of distance computation while improving the handling of indirect calls through more precise analysis.
These efforts highlight a blueprint for integrating DGF with various security tasks, with further applications (e.g., impact and exploitation accessment \cite{zou2022syzscope, lin2022grebe}) in kernel security remaining a less unexplored topic.

\subsubsection{Diverse Fitness} \label{sec:fitness} Classic code coverage lacks sensitivity to complex kernel conditions such as statefulness and thread interleaving. Most fuzzers focus on diverse fitness metrics beyond classic code coverage to approximate the kernel under test comprehensively. These metrics, like block or edge coverage, guide fuzzers towards desired aspects of the target kernel, including state, concurrency. In response to \textbf{RQ2}, Table \ref{tab:fitness} presents a detailed comparison of fitness metrics and their use cases.

\textbf{State-oriented fitness.} As previously noted in \textbf{C3}, the high degree of statefulness constitutes a significant challenge for OS kernel fuzzing. Kernel state encompasses the execution context, including occupied resources like registers and variables, distinct from user-space programs. OS kernels retain their values over time, accumulating internal states. This stateful nature sets kernel fuzzing apart from application fuzzing, requiring specific states to trigger vulnerabilities \cite{zhao2023state, liu2024leverage}. Effective fuzzers navigate this complexity, targeting diverse and deep states. While some works \cite{shankara2018moonshine, chen2021syzgen, sun2021healer, marcel2023teezz, xu2024mock} have examined states indirectly, a systematic approach with state-oriented fitness is still required to  explore the space.
Based on the type of state modeling, existing work can be divided into state-machine-based and state-variable-based.
The state-machine-based approaches aim to identify concrete state machines embedded within the kernel under test. For instance, USB gadget stacks commonly implement various finite state machines (FSMs). To leverage this structure, FuzzUSB \cite{kim2022fuzzusb} combines static analysis with symbolic execution to detect state transitions in USB drivers. Based on these transitions, it infers potential state machines and retains inputs that trigger previously unexplored states. 
However, the state-machine-based approaches typically impose strict structural assumptions on the target system, and in many cases, explicit FSMs may not be present. In contrast, state-variable-based approaches offer a more practical alternative, as they do not rely on the existence of well-defined FSMs and can adapt to a broader range of system behaviors. The central idea behind this approach is to approximate kernel states using critical variables and monitor their value changes as an indicator of state coverage \cite{zhao2023state, liu2024leverage, wang2024syztrust}. For example, StateFuzz \cite{zhao2023state} employs static analysis to identify candidate state variables, focusing specifically on those that can be traced back to global variables and are accessed by multiple operations. Unlike traditional coverage-guided fuzzers, StateFuzz retains an input if it triggers a new value range or an extreme value for a recognized state variable. Similarly, SyzTrust \cite{wang2024syztrust} identifies relevant state variables by summarizing critical structures associated with TEEs.
Although these methods have shown success, their reliance on static analysis or heuristic modeling can lead to false positives and compromise state integrity.
Open questions remain regarding the efficiency and soundness of state approximation. 

\begin{table}[htbp]
\scriptsize
\centering
\caption{Diverse feedback fitness and its applications}
\label{tab:fitness}
\begin{tabular}{@{}m{1.5cm} m{3cm} m{1.7cm} m{5.5cm}@{}}
\toprule
\textbf{Methodology} & \textbf{Technique} & \textbf{Literature} & \textbf{Use Case} \\ \midrule
\multirow{2}{*}{\begin{tabular}[c]{@{}c@{}}State-oriented\\ Fitness\end{tabular}} & Use FSM to model kernel states & \cite{kim2022fuzzusb} & Applicable to modules with clear state machines \\ \cmidrule(l){2-4} 
 & Use state variables as an abstraction of kernel states & \cite{zhao2023state} \cite{wang2024syztrust} \cite{liu2024leverage}  & Applicable to targets where states should be approximated \\ \midrule
\multirow{3}{*}{\begin{tabular}[c]{@{}c@{}}Concurrency-\\ oriented\\ Fitness\end{tabular}} & Function-level & \cite{jiang2022context} & Coarse-grained metric \\ \cmidrule(l){2-4} 
 & Instruction-level & \cite{meng2020krace} & High granularity but limited semantic metric \\ \cmidrule(l){2-4} 
 & Segment-level & \cite{jeong2023segfuzz} & Between function-level and instruction-level \\ \bottomrule
\end{tabular}
\end{table}

\textbf{Concurrency-oriented fitness.} The widespread use of parallelization in OS kernels leads to a rise in concurrency bugs, like data races and deadlocks. Detecting these issues remains particularly challenging due to their inherently non-deterministic nature. While some techniques have been proposed to make thread scheduling more controllable \cite{gong2021snowboard, xu2025concur, jeong2019razzer}, they typically operate without feedback from execution and thus manage thread scheduling in a largely uninformed manner. Traditional code coverage metrics fail to capture unique behaviors resulting from thread interleavings. It is evident that concurrency-oriented fitness is highly desired to facilitate a systematic exploration.

To effectively characterize the concurrency space, existing studies have defined metrics across multiple levels of granularity, including function-level, instruction-level, and segment-level \cite{meng2020krace, yuan2023ddrace, jiang2022context, jeong2023segfuzz}.
The function-level metric represents the coarsest granularity for analyzing concurrency, with Conzzer \cite{jiang2022context} serving as a representative example. The core intuition behind Conzzer is that if a function $func_a$ executes concurrently with another function $func_b$, then both the caller and callee of $func_a$ are also likely to execute concurrently with $func_b$. Based on this insight, Conzzer introduces a \textit{concurrent call pair} metric to capture and describe combinations of potential concurrent functions.
Differently, Krace \cite{meng2020krace} operates at the instruction level and proposes a fine-grained metric known as \textit{alias coverage}. This metric captures the memory locations accessed by concurrently executed instructions. While it offers detailed insight into low-level interactions, it does not fully capture the semantic characteristics of concurrency bugs. Notably, both Conzzer and Krace primarily focus on kernel filesystems, and their effectiveness in scaling to the broader kernel space remains an open question.
To bridge the gap between function-level and instruction-level analysis, SegFuzz \cite{jeong2023segfuzz} presents an intermediate metric called \textit{segment}. SegFuzz defines \textit{segment} as a group of instructions that access shared memory objects. It strikes a balance between granularity and semantic relevance, offering a more practical representation of concurrent behavior.
Despite these advancements, fuzzers targeting concurrency-oriented feedback continue to face significant challenges, particularly in terms of limited flexibility and substantial performance overhead \cite{xu2025concur}.

\begin{mybox}
    \noindent \textbf{Implication \yellowcircle{9}: Multi-feedback Prioritization.}
    Existing OS kernel fuzzers mainly optimize the feedback mechanism based on the characteristics of target components. 
    Exploring more targeted fitness metrics to uncover specific types of vulnerabilities is a valuable direction. However, as the use of multiple fitness metrics increases, the prioritization of feedback in the context of multi-feedback fuzzing has not yet been thoroughly studied, despite being a critical factor influencing the testing efficiency \cite{wang2021syzvegas, xu2024mock}.
\end{mybox}

\section{Challenges and Opportunities} 
\label{sec:discussion} 

We have outlined potential directions of existing techniques in the preceding sections. To answer \textbf{RQ3}, in this section, we present a more detailed exploration of future directions that could enhance specific aspects of the fuzzing process and further improve kernel security.

\textbf{Interactive driver fuzzing.} As emphasized in  implication \yellowcircle{4}, the attack surface of kernel drivers arises from both user space and peripherals. User-space programs interact with drivers via the syscall interface, such as \texttt{ioctl}, while devices connect with drivers through the peripheral interface. Both interfaces significantly impact the functionality of the drivers. Prior work \cite{hao2022demy} has demonstrated that some dependencies cannot be resolved without efforts from both sides. Nevertheless, existing works primarily concentrate on either user space or peripheral interactions when testing drivers, often feeding inputs from a single source. 
While earlier bugs have been systematically mitigated \cite{wu23devfuzz}, the intricate internal states arising from the interactions between these two interfaces have received comparatively little attention, resulting in numerous vulnerabilities remaining unresolved. Recent studies \cite{jang2023reusb, xu2024saturn} have taken one step forward in the direction, although their approaches are limited to specific  and can not scale. A potential solution is to develop a chronological driver model that focuses on code affected by both interfaces and to create a dual-interface fuzzing framework that simultaneously analyzes interactions from user space and peripherals while monitoring state changes.

 \textbf{Harnessing scheduler for concurrency.} 
Kernel concurrency vulnerabilities are inherently more challenging to uncover compared to sequential ones due to the unpredictable nature of kernel scheduling \cite{pedro2014ski, gong2021snowboard}. Despite advancements brought by various fuzzing techniques in \textit{F3.2} and \textit{F3.3}, these methods often necessitate significant modifications on kernels or emulators. These invasive customizations significantly hamper scalability and impose substantial performance overhead \cite{dae2024ozz, jeong2023segfuzz, gong2021snowboard}. Recently, the introduction of the \texttt{sched\_ext} \cite{sched_ext} feature has opened new avenues for addressing this issue. Originally designed to enable flexible and extensible scheduler logic, \texttt{sched\_ext} allows developers to modify scheduling behavior using pluggable eBPF programs \cite{jia2023programmable}. As discussed in the implication \yellowcircle{8}, by designing schedulers specifically tailored for concurrency exploration, it becomes possible to precisely control thread interleaving in a customizable manner. At the same time, this approach retains the performance benefits of native execution and ensures forward compatibility \cite{xu2025concur}. Further investigation is needed to explore the combination of \texttt{sched\_ext} with concurrency-oriented fitness to improve lightweightness and generality.

\textbf{In-domain benchmark construction.}
An in-domain benchmark is essential for fair and accurate evaluation, particularly given the rapid growth of kernel fuzzing techniques. Inspired by benchmarks developed for application fuzzing \cite{hazimeh2021magma, metzman2021fuzzbench, natella2021profuzzbench}, an effective benchmark for kernel fuzzing should possess the following attributes:
(a) Diversity: The benchmark should encompass a wide variety of bugs distributed across different modules.
(b) Verifiability: It should employ reliable and measurable metrics.
(c) Evolvability: As the kernel continuously evolves with the introduction of new features, the benchmark must also adapt to reflect the kernel development.
Yet, the evaluation of kernel fuzzers is complicated by additional factors, as noted in \textit{F1.1} and \textit{F2.1}. 
A practical starting point would involve creating a benchmark specifically for Linux kernel, which typically offers superior infrastructure and has a broader impact.
One potential approach is to combine with syzbot \cite{syzbot}. It includes a wide range of real-world bug reports from various types and different modules. These bugs are also accompanied detailed patch history and status, facilitating efficient triage. Additionally, syzbot's continuous nature inherently supports the benchmark's evolvability, allowing it to stay aligned with the ongoing development of Linux kernel.

 \textbf{LLM integration.} LLMs have shown significant potential across a wide range of tasks \cite{zhao2025fix, li2024enhance, xu2025opt, jueon2024fuzzing, zhou2025benchmark}, owing to their ability to understand and generate both natural language and code. 
Recent studies have also shed light on LLM integration with fuzzing workflows \cite{xu2025ckgfuzzer, xia2024fuzz4all}, including syscall specifications generation \cite{yang2025kernelgpt} and dependency modeling \cite{zhang2025unlocking}. While solutions represent an important step toward LLM-assisted fuzzing, significant limitations remain. As revealed in implication \yellowcircle{5}, a key challenge arises from the prevalence of indirect calls in the kernel \cite{liu2024improve}. Current LLM-based approaches derive specifications by analyzing relevant source code snippets. However, indirect calls are difficult to resolve from source code alone, and such methods fail in cases where critical semantics are embedded in indirect invocations (e.g., in DRM subsystems). At the same time, purely static analysis techniques, although more effective at handling indirect calls, lack the ability to capture textual information. Addressing this issue requires a balanced integration of LLMs with program analysis techniques, rather than relying solely on LLMs.

\textbf{Impact of Rust.} After years of discussion and development, the integration of Rust into the Linux kernel has become a reality. Although Rust code currently comprises only a small fraction of the codebase, plans are already in place for incorporating more extensive Rusty components, especially in drivers \cite{li2024empirical}. In light of this growing adoption, it is important to assess the implications for existing kernel fuzzing infrastructure and strategies.
One key research direction involves examining the extent to which current frameworks effectively support the discovery of vulnerabilities in Rusty kernel code.
For instance, the kernel coverage tool KCOV was originally designed with GCC in mind, and its compatibility and accuracy when applied to Rust code warrant further investigation. Moreover, as the focus of vulnerabilities in Rusty kernel components shifts from traditional memory corruption issues to logic errors \cite{li2024rust}, new categories of bugs may emerge. This evolution underscores the need for enhanced oracles capable of detecting logic vulnerabilities, which remains an open area for future research.

\section{Conclusion} %
\label{sec:conclusion} 
In this work, we conduct a systematic study of \surveyPaperNum \ OS kernel fuzzing papers published between 2017 and August 2025 in top-tier venues. We propose a comprehensive taxonomy of OS kernel fuzzing by introducing a stage-based fuzzing model and defining the desired functionalities at each stage. Leveraging this taxonomy, we analyze how contemporary techniques implement these functionalities, examine the gaps in current approaches, and explore potential solutions. Furthermore, we identify critical challenges faced by existing OS kernel fuzzing methodologies and highlight promising future research directions.
\section{Acknowledge} %
\label{sec:ack}

We are grateful to the editors and the anonymous reviewers for their thoughtful feedback and constructive guidance, which significantly improved the quality of this work. This research was partially supported by the National Science Foundation of China (NSFC) under Grant No. 62293511, No. U244120033, U24A20336, 62172243, 62402425 and 62402418, the China Postdoctoral Science Foundation under No. 2024M762829, the Zhejiang Provincial Natural Science Foundation under No. LD24F020002, the "Pioneer and Leading Goose" R\&D Program of Zhejiang under No. 2025C02033 and 2025C01082, and the Zhejiang Provincial Priority-Funded Postdoctoral Research Project under No. ZJ2024001.

\bibliographystyle{ACM-Reference-Format}
\bibliography{references/ref}


\begin{thebibliography}{192}


\ifx \showCODEN    \undefined \def \showCODEN     #1{\unskip}     \fi
\ifx \showISBNx    \undefined \def \showISBNx     #1{\unskip}     \fi
\ifx \showISBNxiii \undefined \def \showISBNxiii  #1{\unskip}     \fi
\ifx \showISSN     \undefined \def \showISSN      #1{\unskip}     \fi
\ifx \showLCCN     \undefined \def \showLCCN      #1{\unskip}     \fi
\ifx \shownote     \undefined \def \shownote      #1{#1}          \fi
\ifx \showarticletitle \undefined \def \showarticletitle #1{#1}   \fi
\ifx \showURL      \undefined \def \showURL       {\relax}        \fi
\providecommand\bibfield[2]{#2}
\providecommand\bibinfo[2]{#2}
\providecommand\natexlab[1]{#1}
\providecommand\showeprint[2][]{arXiv:#2}

\bibitem[vxw(1987)]%
        {vxworks}
 \bibinfo{year}{1987}\natexlab{}.
\newblock \bibinfo{title}{VxWorks}.
\newblock \bibinfo{howpublished}{\url{https://www.windriver.com/products/vxworks}}.
\newblock


\bibitem[fre(2003)]%
        {freertos}
 \bibinfo{year}{2003}\natexlab{}.
\newblock \bibinfo{title}{FreeRTOS}.
\newblock \bibinfo{howpublished}{\url{https://www.freertos.org/}}.
\newblock


\bibitem[opt(2014)]%
        {optee}
 \bibinfo{year}{2014}\natexlab{}.
\newblock \bibinfo{title}{OP-TEE: a Trusted Execution Environment}.
\newblock \bibinfo{howpublished}{\url{https://github.com/OP-TEE/optee_os}}.
\newblock


\bibitem[kas(2015)]%
        {kasan}
 \bibinfo{year}{2015}\natexlab{}.
\newblock \bibinfo{title}{The Kernel Address Sanitizer (KASAN)}.
\newblock \bibinfo{howpublished}{\url{https://docs.kernel.org/dev-tools/kasan.html}}.
\newblock


\bibitem[kcs(2015)]%
        {kcsan}
 \bibinfo{year}{2015}\natexlab{}.
\newblock \bibinfo{title}{The Kernel Concurrency Sanitizer (KCSAN)}.
\newblock \bibinfo{howpublished}{\url{https://docs.kernel.org/dev-tools/kcsan.html}}.
\newblock


\bibitem[ubs(2015)]%
        {ubsan}
 \bibinfo{year}{2015}\natexlab{}.
\newblock \bibinfo{title}{The Undefined Behavior Sanitizer (UBSAN)}.
\newblock \bibinfo{howpublished}{\url{https://docs.kernel.org/dev-tools/ubsan.html}}.
\newblock


\bibitem[dir(2016)]%
        {dirtycow}
 \bibinfo{year}{2016}\natexlab{}.
\newblock \bibinfo{title}{CVE-2016-5195}.
\newblock \bibinfo{howpublished}{\url{https://nvd.nist.gov/vuln/detail/cve-2016-5195}}.
\newblock


\bibitem[zep(2016)]%
        {zephyr}
 \bibinfo{year}{2016}\natexlab{}.
\newblock \bibinfo{title}{Zephyr: a new generation, scalable, optimized, secure RTOS for multiple hardware architectures}.
\newblock \bibinfo{howpublished}{\url{https://zephyrproject.org/}}.
\newblock


\bibitem[bug(2020)]%
        {bugs2020}
 \bibinfo{year}{2020}\natexlab{}.
\newblock \bibinfo{title}{Bugs on the Windshield: Fuzzing the Windows Kernel}.
\newblock
\urldef\tempurl%
\url{https://research.checkpoint.com/2020/bugs-on-the-windshield-fuzzing-the-windows-kernel/}
\showURL{%
\tempurl}


\bibitem[Aafer et~al\mbox{.}(2021)]%
        {aafer21smarttvs}
\bibfield{author}{\bibinfo{person}{Yousra Aafer}, \bibinfo{person}{Wei You}, \bibinfo{person}{Yi Sun}, \bibinfo{person}{Yu Shi}, \bibinfo{person}{Xiangyu Zhang}, {and} \bibinfo{person}{Heng Yin}.} \bibinfo{year}{2021}\natexlab{}.
\newblock \showarticletitle{Android {SmartTVs} Vulnerability Discovery via {Log-Guided} Fuzzing}. In \bibinfo{booktitle}{\emph{30th USENIX Security Symposium (USENIX Security 21)}}. \bibinfo{publisher}{USENIX Association}, \bibinfo{pages}{2759--2776}.
\newblock
\showISBNx{978-1-939133-24-3}
\urldef\tempurl%
\url{https://www.usenix.org/conference/usenixsecurity21/presentation/aafer}
\showURL{%
\tempurl}


\bibitem[Angelakopoulos et~al\mbox{.}(2023)]%
        {Ioannis2023firmsolo}
\bibfield{author}{\bibinfo{person}{Ioannis Angelakopoulos}, \bibinfo{person}{Gianluca Stringhini}, {and} \bibinfo{person}{Manuel Egele}.} \bibinfo{year}{2023}\natexlab{}.
\newblock \showarticletitle{{FirmSolo}: Enabling dynamic analysis of binary Linux-based {IoT} kernel modules}. In \bibinfo{booktitle}{\emph{32nd USENIX Security Symposium (USENIX Security 23)}}. \bibinfo{publisher}{USENIX Association}, \bibinfo{address}{Anaheim, CA}, \bibinfo{pages}{5021--5038}.
\newblock
\showISBNx{978-1-939133-37-3}
\urldef\tempurl%
\url{https://www.usenix.org/conference/usenixsecurity23/presentation/angelakopoulos}
\showURL{%
\tempurl}


\bibitem[Angelakopoulos et~al\mbox{.}(2024)]%
        {ioannis2024pandawan}
\bibfield{author}{\bibinfo{person}{Ioannis Angelakopoulos}, \bibinfo{person}{Gianluca Stringhini}, {and} \bibinfo{person}{Manuel Egele}.} \bibinfo{year}{2024}\natexlab{}.
\newblock \showarticletitle{Pandawan: Quantifying Progress in Linux-based Firmware Rehosting}. In \bibinfo{booktitle}{\emph{33rd USENIX Security Symposium (USENIX Security 24)}}. \bibinfo{publisher}{USENIX Association}, \bibinfo{address}{Philadelphia, PA}, \bibinfo{pages}{5859--5876}.
\newblock
\showISBNx{978-1-939133-44-1}
\urldef\tempurl%
\url{https://www.usenix.org/conference/usenixsecurity24/presentation/angelakopoulos}
\showURL{%
\tempurl}


\bibitem[ARM(2013)]%
        {mbedos}
\bibfield{author}{\bibinfo{person}{ARM}.} \bibinfo{year}{2013}\natexlab{}.
\newblock \bibinfo{title}{Mbed OS: a platform operating system designed for the internet of things}.
\newblock \bibinfo{howpublished}{\url{https://github.com/ARMmbed/mbed-os}}.
\newblock


\bibitem[Aschermann et~al\mbox{.}(2019)]%
        {aschermann2019redqueen}
\bibfield{author}{\bibinfo{person}{Cornelius Aschermann}, \bibinfo{person}{Sergej Schumilo}, \bibinfo{person}{Tim Blazytko}, \bibinfo{person}{Robert Gawlik}, {and} \bibinfo{person}{Thorsten Holz}.} \bibinfo{year}{2019}\natexlab{}.
\newblock \showarticletitle{{REDQUEEN:} Fuzzing with Input-to-State Correspondence}. In \bibinfo{booktitle}{\emph{26th Annual Network and Distributed System Security Symposium, {NDSS} 2019, San Diego, California, USA, February 24-27, 2019}}. \bibinfo{publisher}{The Internet Society}.
\newblock
\urldef\tempurl%
\url{https://www.ndss-symposium.org/ndss-paper/redqueen-fuzzing-with-input-to-state-correspondence/}
\showURL{%
\tempurl}


\bibitem[Bai et~al\mbox{.}(2024a)]%
        {bai2024multi}
\bibfield{author}{\bibinfo{person}{Jia-Ju Bai}, \bibinfo{person}{Hao-Xuan Song}, {and} \bibinfo{person}{Shi-Min Hu}.} \bibinfo{year}{2024}\natexlab{a}.
\newblock \showarticletitle{Multi-Dimensional and Message-Guided Fuzzing for Robotic Programs in Robot Operating System}. In \bibinfo{booktitle}{\emph{Proceedings of the 29th ACM International Conference on Architectural Support for Programming Languages and Operating Systems, Volume 2}} (La Jolla, CA, USA) \emph{(\bibinfo{series}{ASPLOS '24})}. \bibinfo{publisher}{Association for Computing Machinery}, \bibinfo{address}{New York, NY, USA}, \bibinfo{pages}{763–778}.
\newblock
\showISBNx{9798400703850}
\href{https://doi.org/10.1145/3620665.3640425}{doi:\nolinkurl{10.1145/3620665.3640425}}


\bibitem[Bai et~al\mbox{.}(2024b)]%
        {bai2024countdown}
\bibfield{author}{\bibinfo{person}{Shuangpeng Bai}, \bibinfo{person}{Zhechang Zhang}, {and} \bibinfo{person}{Hong Hu}.} \bibinfo{year}{2024}\natexlab{b}.
\newblock \showarticletitle{CountDown: Refcount-guided Fuzzing for Exposing Temporal Memory Errors in Linux Kernel}. In \bibinfo{booktitle}{\emph{Proceedings of the 2024 on ACM SIGSAC Conference on Computer and Communications Security}} (Salt Lake City, UT, USA) \emph{(\bibinfo{series}{CCS '24})}. \bibinfo{publisher}{Association for Computing Machinery}, \bibinfo{address}{New York, NY, USA}, \bibinfo{pages}{1315–1329}.
\newblock
\showISBNx{9798400706363}
\href{https://doi.org/10.1145/3658644.3690320}{doi:\nolinkurl{10.1145/3658644.3690320}}


\bibitem[Barabanov(1997)]%
        {barabanov1997linux}
\bibfield{author}{\bibinfo{person}{Michael Barabanov}.} \bibinfo{year}{1997}\natexlab{}.
\newblock \showarticletitle{A linux-based real-time operating system}.
\newblock  (\bibinfo{year}{1997}).
\newblock


\bibitem[Bellard(2005)]%
        {bellard2005qemu}
\bibfield{author}{\bibinfo{person}{Fabrice Bellard}.} \bibinfo{year}{2005}\natexlab{}.
\newblock \showarticletitle{QEMU, a fast and portable dynamic translator}. In \bibinfo{booktitle}{\emph{Proceedings of the Annual Conference on USENIX Annual Technical Conference}} (Anaheim, CA) \emph{(\bibinfo{series}{ATEC '05})}. \bibinfo{publisher}{USENIX Association}, \bibinfo{address}{USA}, \bibinfo{pages}{41}.
\newblock


\bibitem[B\"{o}hme et~al\mbox{.}(2020)]%
        {marcel2020boost}
\bibfield{author}{\bibinfo{person}{Marcel B\"{o}hme}, \bibinfo{person}{Valentin J.~M. Man\`{e}s}, {and} \bibinfo{person}{Sang~Kil Cha}.} \bibinfo{year}{2020}\natexlab{}.
\newblock \showarticletitle{Boosting Fuzzer Efficiency: An Information Theoretic Perspective}. In \bibinfo{booktitle}{\emph{Proceedings of the 28th ACM Joint Meeting on European Software Engineering Conference and Symposium on the Foundations of Software Engineering}} (Virtual Event, USA) \emph{(\bibinfo{series}{ESEC/FSE 2020})}. \bibinfo{publisher}{Association for Computing Machinery}, \bibinfo{address}{New York, NY, USA}, \bibinfo{pages}{678–689}.
\newblock
\showISBNx{9781450370431}
\href{https://doi.org/10.1145/3368089.3409748}{doi:\nolinkurl{10.1145/3368089.3409748}}


\bibitem[B\"{o}hme et~al\mbox{.}(2017)]%
        {marcel2017aflgo}
\bibfield{author}{\bibinfo{person}{Marcel B\"{o}hme}, \bibinfo{person}{Van-Thuan Pham}, \bibinfo{person}{Manh-Dung Nguyen}, {and} \bibinfo{person}{Abhik Roychoudhury}.} \bibinfo{year}{2017}\natexlab{}.
\newblock \showarticletitle{Directed Greybox Fuzzing}. In \bibinfo{booktitle}{\emph{Proceedings of the 2017 ACM SIGSAC Conference on Computer and Communications Security}} (Dallas, Texas, USA) \emph{(\bibinfo{series}{CCS '17})}. \bibinfo{publisher}{Association for Computing Machinery}, \bibinfo{address}{New York, NY, USA}, \bibinfo{pages}{2329–2344}.
\newblock
\showISBNx{9781450349468}
\href{https://doi.org/10.1145/3133956.3134020}{doi:\nolinkurl{10.1145/3133956.3134020}}


\bibitem[Bulekov et~al\mbox{.}(2023)]%
        {bulekov2020nogrammar}
\bibfield{author}{\bibinfo{person}{Alexander Bulekov}, \bibinfo{person}{Bandan Das}, \bibinfo{person}{Stefan Hajnoczi}, {and} \bibinfo{person}{Manuel Egele}.} \bibinfo{year}{2023}\natexlab{}.
\newblock \showarticletitle{No Grammar, No Problem: Towards Fuzzing the Linux Kernel without System-Call Descriptions}. In \bibinfo{booktitle}{\emph{30th Annual Network and Distributed System Security Symposium, {NDSS} 2023, San Diego, California, USA, February 27 - March 3, 2023}}. \bibinfo{publisher}{The Internet Society}.
\newblock
\urldef\tempurl%
\url{https://www.ndss-symposium.org/ndss-paper/no-grammar-no-problem-towards-fuzzing-the-linux-kernel-without-system-call-descriptions/}
\showURL{%
\tempurl}


\bibitem[Busch et~al\mbox{.}(2023)]%
        {marcel2023teezz}
\bibfield{author}{\bibinfo{person}{Marcel Busch}, \bibinfo{person}{Aravind Machiry}, \bibinfo{person}{Chad Spensky}, \bibinfo{person}{Giovanni Vigna}, \bibinfo{person}{Christopher Kruegel}, {and} \bibinfo{person}{Mathias Payer}.} \bibinfo{year}{2023}\natexlab{}.
\newblock \showarticletitle{TEEzz: Fuzzing Trusted Applications on COTS Android Devices}. In \bibinfo{booktitle}{\emph{2023 IEEE Symposium on Security and Privacy (SP)}}. \bibinfo{pages}{1204--1219}.
\newblock
\href{https://doi.org/10.1109/SP46215.2023.10179302}{doi:\nolinkurl{10.1109/SP46215.2023.10179302}}


\bibitem[Cao et~al\mbox{.}(2020)]%
        {cao2020device}
\bibfield{author}{\bibinfo{person}{Chen Cao}, \bibinfo{person}{Le Guan}, \bibinfo{person}{Jiang Ming}, {and} \bibinfo{person}{Peng Liu}.} \bibinfo{year}{2020}\natexlab{}.
\newblock \showarticletitle{Device-agnostic firmware execution is possible: A concolic execution approach for peripheral emulation}. In \bibinfo{booktitle}{\emph{Annual Computer Security Applications Conference}}. \bibinfo{pages}{746--759}.
\newblock


\bibitem[Cao et~al\mbox{.}(2008)]%
        {cao2008liteos}
\bibfield{author}{\bibinfo{person}{Qing Cao}, \bibinfo{person}{Tarek Abdelzaher}, \bibinfo{person}{John Stankovic}, {and} \bibinfo{person}{Tian He}.} \bibinfo{year}{2008}\natexlab{}.
\newblock \showarticletitle{The LiteOS Operating System: Towards Unix-Like Abstractions for Wireless Sensor Networks}. In \bibinfo{booktitle}{\emph{2008 International Conference on Information Processing in Sensor Networks (ipsn 2008)}}. \bibinfo{pages}{233--244}.
\newblock
\href{https://doi.org/10.1109/IPSN.2008.54}{doi:\nolinkurl{10.1109/IPSN.2008.54}}


\bibitem[Chen et~al\mbox{.}(2020)]%
        {chen2020test}
\bibfield{author}{\bibinfo{person}{Dongjie Chen}, \bibinfo{person}{Yanyan Jiang}, \bibinfo{person}{Chang Xu}, \bibinfo{person}{Xiaoxing Ma}, {and} \bibinfo{person}{Jian Lu}.} \bibinfo{year}{2020}\natexlab{}.
\newblock \showarticletitle{Testing file system implementations on layered models}. In \bibinfo{booktitle}{\emph{Proceedings of the ACM/IEEE 42nd International Conference on Software Engineering}} (Seoul, South Korea) \emph{(\bibinfo{series}{ICSE '20})}. \bibinfo{publisher}{Association for Computing Machinery}, \bibinfo{address}{New York, NY, USA}, \bibinfo{pages}{1483–1495}.
\newblock
\showISBNx{9781450371216}
\href{https://doi.org/10.1145/3377811.3380350}{doi:\nolinkurl{10.1145/3377811.3380350}}


\bibitem[Chen et~al\mbox{.}(2018)]%
        {chen2018iotfuzzer}
\bibfield{author}{\bibinfo{person}{Jiongyi Chen}, \bibinfo{person}{Wenrui Diao}, \bibinfo{person}{Qingchuan Zhao}, \bibinfo{person}{Chaoshun Zuo}, \bibinfo{person}{Zhiqiang Lin}, \bibinfo{person}{XiaoFeng Wang}, \bibinfo{person}{Wing~Cheong Lau}, \bibinfo{person}{Menghan Sun}, \bibinfo{person}{Ronghai Yang}, {and} \bibinfo{person}{Kehuan Zhang}.} \bibinfo{year}{2018}\natexlab{}.
\newblock \showarticletitle{IoTFuzzer: Discovering Memory Corruptions in IoT Through App-based Fuzzing}. In \bibinfo{booktitle}{\emph{25th Annual Network and Distributed System Security Symposium, {NDSS} 2018, San Diego, California, USA, February 18-21, 2018}}. \bibinfo{publisher}{The Internet Society}.
\newblock
\urldef\tempurl%
\url{https://www.ndss-symposium.org/wp-content/uploads/2018/02/ndss2018\_01A-1\_Chen\_paper.pdf}
\showURL{%
\tempurl}


\bibitem[Chen et~al\mbox{.}(2022)]%
        {chen2022sfuzz}
\bibfield{author}{\bibinfo{person}{Libo Chen}, \bibinfo{person}{Quanpu Cai}, \bibinfo{person}{Zhenbang Ma}, \bibinfo{person}{Yanhao Wang}, \bibinfo{person}{Hong Hu}, \bibinfo{person}{Minghang Shen}, \bibinfo{person}{Yue Liu}, \bibinfo{person}{Shanqing Guo}, \bibinfo{person}{Haixin Duan}, \bibinfo{person}{Kaida Jiang}, {and} \bibinfo{person}{Zhi Xue}.} \bibinfo{year}{2022}\natexlab{}.
\newblock \showarticletitle{SFuzz: Slice-Based Fuzzing for Real-Time Operating Systems}. In \bibinfo{booktitle}{\emph{Proceedings of the 2022 ACM SIGSAC Conference on Computer and Communications Security}} (Los Angeles, CA, USA) \emph{(\bibinfo{series}{CCS '22})}. \bibinfo{publisher}{Association for Computing Machinery}, \bibinfo{address}{New York, NY, USA}, \bibinfo{pages}{485–498}.
\newblock
\showISBNx{9781450394505}
\href{https://doi.org/10.1145/3548606.3559367}{doi:\nolinkurl{10.1145/3548606.3559367}}


\bibitem[Chen et~al\mbox{.}(2021)]%
        {chen2021syzgen}
\bibfield{author}{\bibinfo{person}{Weiteng Chen}, \bibinfo{person}{Yu Wang}, \bibinfo{person}{Zheng Zhang}, {and} \bibinfo{person}{Zhiyun Qian}.} \bibinfo{year}{2021}\natexlab{}.
\newblock \showarticletitle{SyzGen: Automated Generation of Syscall Specification of Closed-Source macOS Drivers}. In \bibinfo{booktitle}{\emph{Proceedings of the 2021 ACM SIGSAC Conference on Computer and Communications Security}} (Virtual Event, Republic of Korea) \emph{(\bibinfo{series}{CCS '21})}. \bibinfo{publisher}{Association for Computing Machinery}, \bibinfo{address}{New York, NY, USA}, \bibinfo{pages}{749–763}.
\newblock
\showISBNx{9781450384544}
\href{https://doi.org/10.1145/3460120.3484564}{doi:\nolinkurl{10.1145/3460120.3484564}}


\bibitem[Cho et~al\mbox{.}(2023)]%
        {cho2023bokasan}
\bibfield{author}{\bibinfo{person}{Mingi Cho}, \bibinfo{person}{Dohyeon An}, \bibinfo{person}{Hoyong Jin}, {and} \bibinfo{person}{Taekyoung Kwon}.} \bibinfo{year}{2023}\natexlab{}.
\newblock \showarticletitle{{BoKASAN}: Binary-only Kernel Address Sanitizer for Effective Kernel Fuzzing}. In \bibinfo{booktitle}{\emph{32nd USENIX Security Symposium (USENIX Security 23)}}. \bibinfo{publisher}{USENIX Association}, \bibinfo{address}{Anaheim, CA}, \bibinfo{pages}{4985--5002}.
\newblock
\showISBNx{978-1-939133-37-3}
\urldef\tempurl%
\url{https://www.usenix.org/conference/usenixsecurity23/presentation/cho}
\showURL{%
\tempurl}


\bibitem[Choi et~al\mbox{.}(2021)]%
        {choi2021ntfuzz}
\bibfield{author}{\bibinfo{person}{Jaeseung Choi}, \bibinfo{person}{Kangsu Kim}, \bibinfo{person}{Daejin Lee}, {and} \bibinfo{person}{Sang~Kil Cha}.} \bibinfo{year}{2021}\natexlab{}.
\newblock \showarticletitle{NtFuzz: Enabling Type-Aware Kernel Fuzzing on Windows with Static Binary Analysis}. In \bibinfo{booktitle}{\emph{2021 IEEE Symposium on Security and Privacy (SP)}}. \bibinfo{pages}{677--693}.
\newblock
\href{https://doi.org/10.1109/SP40001.2021.00114}{doi:\nolinkurl{10.1109/SP40001.2021.00114}}


\bibitem[Clements et~al\mbox{.}(2020)]%
        {clements2020halucinator}
\bibfield{author}{\bibinfo{person}{Abraham~A Clements}, \bibinfo{person}{Eric Gustafson}, \bibinfo{person}{Tobias Scharnowski}, \bibinfo{person}{Paul Grosen}, \bibinfo{person}{David Fritz}, \bibinfo{person}{Christopher Kruegel}, \bibinfo{person}{Giovanni Vigna}, \bibinfo{person}{Saurabh Bagchi}, {and} \bibinfo{person}{Mathias Payer}.} \bibinfo{year}{2020}\natexlab{}.
\newblock \showarticletitle{{HALucinator}: Firmware Re-hosting Through Abstraction Layer Emulation}. In \bibinfo{booktitle}{\emph{29th USENIX Security Symposium (USENIX Security 20)}}. \bibinfo{publisher}{USENIX Association}, \bibinfo{pages}{1201--1218}.
\newblock
\showISBNx{978-1-939133-17-5}
\urldef\tempurl%
\url{https://www.usenix.org/conference/usenixsecurity20/presentation/clements}
\showURL{%
\tempurl}


\bibitem[Corina et~al\mbox{.}(2017)]%
        {jake2017difuze}
\bibfield{author}{\bibinfo{person}{Jake Corina}, \bibinfo{person}{Aravind Machiry}, \bibinfo{person}{Christopher Salls}, \bibinfo{person}{Yan Shoshitaishvili}, \bibinfo{person}{Shuang Hao}, \bibinfo{person}{Christopher Kruegel}, {and} \bibinfo{person}{Giovanni Vigna}.} \bibinfo{year}{2017}\natexlab{}.
\newblock \showarticletitle{DIFUZE: Interface Aware Fuzzing for Kernel Drivers}. In \bibinfo{booktitle}{\emph{Proceedings of the 2017 ACM SIGSAC Conference on Computer and Communications Security}} (Dallas, Texas, USA) \emph{(\bibinfo{series}{CCS '17})}. \bibinfo{publisher}{Association for Computing Machinery}, \bibinfo{address}{New York, NY, USA}, \bibinfo{pages}{2123–2138}.
\newblock
\showISBNx{9781450349468}
\href{https://doi.org/10.1145/3133956.3134069}{doi:\nolinkurl{10.1145/3133956.3134069}}


\bibitem[Corporation(1993)]%
        {windows}
\bibfield{author}{\bibinfo{person}{Microsoft Corporation}.} \bibinfo{year}{1993}\natexlab{}.
\newblock \bibinfo{title}{Windows Kernel}.
\newblock \bibinfo{howpublished}{\url{https://learn.microsoft.com/en-us/windows-hardware/drivers/kernel/}}.
\newblock


\bibitem[Dawoud and Bugiel(2021)]%
        {Dawoud2021bring}
\bibfield{author}{\bibinfo{person}{Abdallah Dawoud} {and} \bibinfo{person}{Sven Bugiel}.} \bibinfo{year}{2021}\natexlab{}.
\newblock \showarticletitle{Bringing Balance to the Force: Dynamic Analysis of the Android Application Framework}. In \bibinfo{booktitle}{\emph{28th Annual Network and Distributed System Security Symposium, {NDSS} 2021, virtually, February 21-25, 2021}}. \bibinfo{publisher}{The Internet Society}.
\newblock
\urldef\tempurl%
\url{https://www.ndss-symposium.org/ndss-paper/bringing-balance-to-the-force-dynamic-analysis-of-the-android-application-framework/}
\showURL{%
\tempurl}


\bibitem[Dinesh et~al\mbox{.}(2020)]%
        {dinesh2020retro}
\bibfield{author}{\bibinfo{person}{Sushant Dinesh}, \bibinfo{person}{Nathan Burow}, \bibinfo{person}{Dongyan Xu}, {and} \bibinfo{person}{Mathias Payer}.} \bibinfo{year}{2020}\natexlab{}.
\newblock \showarticletitle{RetroWrite: Statically Instrumenting COTS Binaries for Fuzzing and Sanitization}. In \bibinfo{booktitle}{\emph{2020 IEEE Symposium on Security and Privacy (SP)}}. \bibinfo{pages}{1497--1511}.
\newblock
\href{https://doi.org/10.1109/SP40000.2020.00009}{doi:\nolinkurl{10.1109/SP40000.2020.00009}}


\bibitem[Drozdovskyi and Moliavko(2019)]%
        {taras2019mtower}
\bibfield{author}{\bibinfo{person}{Taras~A. Drozdovskyi} {and} \bibinfo{person}{Oleksandr~S. Moliavko}.} \bibinfo{year}{2019}\natexlab{}.
\newblock \showarticletitle{mTower: Trusted Execution Environment for MCU-based devices}.
\newblock \bibinfo{journal}{\emph{Journal of Open Source Software}} \bibinfo{volume}{4}, \bibinfo{number}{40} (\bibinfo{year}{2019}), \bibinfo{pages}{1494}.
\newblock
\href{https://doi.org/10.21105/joss.01494}{doi:\nolinkurl{10.21105/joss.01494}}


\bibitem[Eisele et~al\mbox{.}(2023)]%
        {eisele2023fuzzing}
\bibfield{author}{\bibinfo{person}{Max Eisele}, \bibinfo{person}{Daniel Ebert}, \bibinfo{person}{Christopher Huth}, {and} \bibinfo{person}{Andreas Zeller}.} \bibinfo{year}{2023}\natexlab{}.
\newblock \showarticletitle{Fuzzing Embedded Systems using Debug Interfaces}. In \bibinfo{booktitle}{\emph{Proceedings of the 32nd ACM SIGSOFT International Symposium on Software Testing and Analysis}} (Seattle, WA, USA) \emph{(\bibinfo{series}{ISSTA 2023})}. \bibinfo{publisher}{Association for Computing Machinery}, \bibinfo{address}{New York, NY, USA}, \bibinfo{pages}{1031–1042}.
\newblock
\showISBNx{9798400702211}
\href{https://doi.org/10.1145/3597926.3598115}{doi:\nolinkurl{10.1145/3597926.3598115}}


\bibitem[Eom et~al\mbox{.}(2024)]%
        {jueon2024fuzzing}
\bibfield{author}{\bibinfo{person}{Jueon Eom}, \bibinfo{person}{Seyeon Jeong}, {and} \bibinfo{person}{Taekyoung Kwon}.} \bibinfo{year}{2024}\natexlab{}.
\newblock \showarticletitle{Fuzzing JavaScript Interpreters with Coverage-Guided Reinforcement Learning for LLM-Based Mutation}. In \bibinfo{booktitle}{\emph{Proceedings of the 33rd ACM SIGSOFT International Symposium on Software Testing and Analysis}} (Vienna, Austria) \emph{(\bibinfo{series}{ISSTA 2024})}. \bibinfo{publisher}{Association for Computing Machinery}, \bibinfo{address}{New York, NY, USA}, \bibinfo{pages}{1656–1668}.
\newblock
\showISBNx{9798400706127}
\href{https://doi.org/10.1145/3650212.3680389}{doi:\nolinkurl{10.1145/3650212.3680389}}


\bibitem[Fasano et~al\mbox{.}(2021)]%
        {fasano2021sok}
\bibfield{author}{\bibinfo{person}{Andrew Fasano}, \bibinfo{person}{Tiemoko Ballo}, \bibinfo{person}{Marius Muench}, \bibinfo{person}{Tim Leek}, \bibinfo{person}{Alexander Bulekov}, \bibinfo{person}{Brendan Dolan-Gavitt}, \bibinfo{person}{Manuel Egele}, \bibinfo{person}{Aur{\'e}lien Francillon}, \bibinfo{person}{Long Lu}, \bibinfo{person}{Nick Gregory}, {et~al\mbox{.}}} \bibinfo{year}{2021}\natexlab{}.
\newblock \showarticletitle{Sok: Enabling security analyses of embedded systems via rehosting}. In \bibinfo{booktitle}{\emph{Proceedings of the 2021 ACM Asia conference on computer and communications security}}. \bibinfo{pages}{687--701}.
\newblock


\bibitem[Feng et~al\mbox{.}(2020)]%
        {feng2020p2im}
\bibfield{author}{\bibinfo{person}{Bo Feng}, \bibinfo{person}{Alejandro Mera}, {and} \bibinfo{person}{Long Lu}.} \bibinfo{year}{2020}\natexlab{}.
\newblock \showarticletitle{{P2IM}: Scalable and Hardware-independent Firmware Testing via Automatic Peripheral Interface Modeling}. In \bibinfo{booktitle}{\emph{29th USENIX Security Symposium (USENIX Security 20)}}. \bibinfo{publisher}{USENIX Association}, \bibinfo{pages}{1237--1254}.
\newblock
\showISBNx{978-1-939133-17-5}
\urldef\tempurl%
\url{https://www.usenix.org/conference/usenixsecurity20/presentation/feng}
\showURL{%
\tempurl}


\bibitem[Fleischer et~al\mbox{.}(2023)]%
        {marius2023actor}
\bibfield{author}{\bibinfo{person}{Marius Fleischer}, \bibinfo{person}{Dipanjan Das}, \bibinfo{person}{Priyanka Bose}, \bibinfo{person}{Weiheng Bai}, \bibinfo{person}{Kangjie Lu}, \bibinfo{person}{Mathias Payer}, \bibinfo{person}{Christopher Kruegel}, {and} \bibinfo{person}{Giovanni Vigna}.} \bibinfo{year}{2023}\natexlab{}.
\newblock \showarticletitle{{ACTOR}: {Action-Guided} Kernel Fuzzing}. In \bibinfo{booktitle}{\emph{32nd USENIX Security Symposium (USENIX Security 23)}}. \bibinfo{publisher}{USENIX Association}, \bibinfo{address}{Anaheim, CA}, \bibinfo{pages}{5003--5020}.
\newblock
\showISBNx{978-1-939133-37-3}
\urldef\tempurl%
\url{https://www.usenix.org/conference/usenixsecurity23/presentation/fleischer}
\showURL{%
\tempurl}


\bibitem[Fonseca et~al\mbox{.}(2014)]%
        {pedro2014ski}
\bibfield{author}{\bibinfo{person}{Pedro Fonseca}, \bibinfo{person}{Rodrigo Rodrigues}, {and} \bibinfo{person}{Bj{\"o}rn~B. Brandenburg}.} \bibinfo{year}{2014}\natexlab{}.
\newblock \showarticletitle{{SKI}: Exposing Kernel Concurrency Bugs through Systematic Schedule Exploration}. In \bibinfo{booktitle}{\emph{11th USENIX Symposium on Operating Systems Design and Implementation (OSDI 14)}}. \bibinfo{publisher}{USENIX Association}, \bibinfo{address}{Broomfield, CO}, \bibinfo{pages}{415--431}.
\newblock
\showISBNx{978-1-931971-16-4}
\urldef\tempurl%
\url{https://www.usenix.org/conference/osdi14/technical-sessions/presentation/fonseca}
\showURL{%
\tempurl}


\bibitem[Gong et~al\mbox{.}(2021)]%
        {gong2021snowboard}
\bibfield{author}{\bibinfo{person}{Sishuai Gong}, \bibinfo{person}{Deniz Altinb\"{u}ken}, \bibinfo{person}{Pedro Fonseca}, {and} \bibinfo{person}{Petros Maniatis}.} \bibinfo{year}{2021}\natexlab{}.
\newblock \showarticletitle{Snowboard: Finding Kernel Concurrency Bugs through Systematic Inter-Thread Communication Analysis}. In \bibinfo{booktitle}{\emph{Proceedings of the ACM SIGOPS 28th Symposium on Operating Systems Principles}} (Virtual Event, Germany) \emph{(\bibinfo{series}{SOSP '21})}. \bibinfo{publisher}{Association for Computing Machinery}, \bibinfo{address}{New York, NY, USA}, \bibinfo{pages}{66–83}.
\newblock
\showISBNx{9781450387095}
\href{https://doi.org/10.1145/3477132.3483549}{doi:\nolinkurl{10.1145/3477132.3483549}}


\bibitem[Gong et~al\mbox{.}(2023)]%
        {gong2023snowcat}
\bibfield{author}{\bibinfo{person}{Sishuai Gong}, \bibinfo{person}{Dinglan Peng}, \bibinfo{person}{Deniz Alt\i{}nb\"{u}ken}, \bibinfo{person}{Pedro Fonseca}, {and} \bibinfo{person}{Petros Maniatis}.} \bibinfo{year}{2023}\natexlab{}.
\newblock \showarticletitle{Snowcat: Efficient Kernel Concurrency Testing using a Learned Coverage Predictor}. In \bibinfo{booktitle}{\emph{Proceedings of the 29th Symposium on Operating Systems Principles}} (Koblenz, Germany) \emph{(\bibinfo{series}{SOSP '23})}. \bibinfo{publisher}{Association for Computing Machinery}, \bibinfo{address}{New York, NY, USA}, \bibinfo{pages}{35–51}.
\newblock
\showISBNx{9798400702297}
\href{https://doi.org/10.1145/3600006.3613148}{doi:\nolinkurl{10.1145/3600006.3613148}}


\bibitem[Gong et~al\mbox{.}(2025)]%
        {gong2025snowflow}
\bibfield{author}{\bibinfo{person}{Sishuai Gong}, \bibinfo{person}{Wang Rui}, \bibinfo{person}{Deniz Altinb\"{u}ken}, \bibinfo{person}{Pedro Fonseca}, {and} \bibinfo{person}{Petros Maniatis}.} \bibinfo{year}{2025}\natexlab{}.
\newblock \showarticletitle{Snowplow: Effective Kernel Fuzzing with a Learned White-box Test Mutator}. In \bibinfo{booktitle}{\emph{Proceedings of the 30th ACM International Conference on Architectural Support for Programming Languages and Operating Systems, Volume 2}} (Rotterdam, Netherlands) \emph{(\bibinfo{series}{ASPLOS '25})}. \bibinfo{publisher}{Association for Computing Machinery}, \bibinfo{address}{New York, NY, USA}, \bibinfo{pages}{1124–1138}.
\newblock
\showISBNx{9798400710797}
\href{https://doi.org/10.1145/3676641.3716019}{doi:\nolinkurl{10.1145/3676641.3716019}}


\bibitem[Google(2008)]%
        {android}
\bibfield{author}{\bibinfo{person}{Google}.} \bibinfo{year}{2008}\natexlab{}.
\newblock \bibinfo{title}{Android}.
\newblock \bibinfo{howpublished}{\url{https://www.android.com/}}.
\newblock


\bibitem[{Google}(2015)]%
        {Syzkaller}
\bibfield{author}{\bibinfo{person}{{Google}}.} \bibinfo{year}{2015}\natexlab{}.
\newblock \bibinfo{title}{syzkaller: an Unsupervised Coverage-Guided Kernel Fuzzer}.
\newblock
\newblock
\shownote{\url{https://github.com/google/syzkaller}}.


\bibitem[Google(2022)]%
        {gvisor}
\bibfield{author}{\bibinfo{person}{Google}.} \bibinfo{year}{2022}\natexlab{}.
\newblock \bibinfo{title}{gVisor}.
\newblock \bibinfo{howpublished}{\url{https://github.com/google/ gvisor}}.
\newblock


\bibitem[Gustafson et~al\mbox{.}(2019)]%
        {gustafson2019toward}
\bibfield{author}{\bibinfo{person}{Eric Gustafson}, \bibinfo{person}{Marius Muench}, \bibinfo{person}{Chad Spensky}, \bibinfo{person}{Nilo Redini}, \bibinfo{person}{Aravind Machiry}, \bibinfo{person}{Yanick Fratantonio}, \bibinfo{person}{Davide Balzarotti}, \bibinfo{person}{Aur{\'e}lien Francillon}, \bibinfo{person}{Yung~Ryn Choe}, \bibinfo{person}{Christophe Kruegel}, {et~al\mbox{.}}} \bibinfo{year}{2019}\natexlab{}.
\newblock \showarticletitle{Toward the analysis of embedded firmware through automated re-hosting}. In \bibinfo{booktitle}{\emph{22nd International Symposium on Research in Attacks, Intrusions and Defenses (RAID 2019)}}. \bibinfo{pages}{135--150}.
\newblock


\bibitem[Hahm et~al\mbox{.}(2016)]%
        {hahm2016os}
\bibfield{author}{\bibinfo{person}{Oliver Hahm}, \bibinfo{person}{Emmanuel Baccelli}, \bibinfo{person}{Hauke Petersen}, {and} \bibinfo{person}{Nicolas Tsiftes}.} \bibinfo{year}{2016}\natexlab{}.
\newblock \showarticletitle{Operating Systems for Low-End Devices in the Internet of Things: A Survey}.
\newblock \bibinfo{journal}{\emph{IEEE Internet of Things Journal}} \bibinfo{volume}{3}, \bibinfo{number}{5} (\bibinfo{year}{2016}), \bibinfo{pages}{720--734}.
\newblock
\href{https://doi.org/10.1109/JIOT.2015.2505901}{doi:\nolinkurl{10.1109/JIOT.2015.2505901}}


\bibitem[Han and Cha(2017)]%
        {han2017imf}
\bibfield{author}{\bibinfo{person}{HyungSeok Han} {and} \bibinfo{person}{Sang~Kil Cha}.} \bibinfo{year}{2017}\natexlab{}.
\newblock \showarticletitle{IMF: Inferred Model-Based Fuzzer}. In \bibinfo{booktitle}{\emph{Proceedings of the 2017 ACM SIGSAC Conference on Computer and Communications Security}} (Dallas, Texas, USA) \emph{(\bibinfo{series}{CCS '17})}. \bibinfo{publisher}{Association for Computing Machinery}, \bibinfo{address}{New York, NY, USA}, \bibinfo{pages}{2345–2358}.
\newblock
\showISBNx{9781450349468}
\href{https://doi.org/10.1145/3133956.3134103}{doi:\nolinkurl{10.1145/3133956.3134103}}


\bibitem[Hao et~al\mbox{.}(2023)]%
        {hao2023syzdiscribe}
\bibfield{author}{\bibinfo{person}{Yu Hao}, \bibinfo{person}{Guoren Li}, \bibinfo{person}{Xiaochen Zou}, \bibinfo{person}{Weiteng Chen}, \bibinfo{person}{Shitong Zhu}, \bibinfo{person}{Zhiyun Qian}, {and} \bibinfo{person}{Ardalan~Amiri Sani}.} \bibinfo{year}{2023}\natexlab{}.
\newblock \showarticletitle{SyzDescribe: Principled, Automated, Static Generation of Syscall Descriptions for Kernel Drivers}. In \bibinfo{booktitle}{\emph{2023 IEEE Symposium on Security and Privacy (SP)}}. \bibinfo{pages}{3262--3278}.
\newblock
\href{https://doi.org/10.1109/SP46215.2023.10179298}{doi:\nolinkurl{10.1109/SP46215.2023.10179298}}


\bibitem[Hao et~al\mbox{.}(2022)]%
        {hao2022demy}
\bibfield{author}{\bibinfo{person}{Yu Hao}, \bibinfo{person}{Hang Zhang}, \bibinfo{person}{Guoren Li}, \bibinfo{person}{Xingyun Du}, \bibinfo{person}{Zhiyun Qian}, {and} \bibinfo{person}{Ardalan~Amiri Sani}.} \bibinfo{year}{2022}\natexlab{}.
\newblock \showarticletitle{Demystifying the Dependency Challenge in Kernel Fuzzing}. In \bibinfo{booktitle}{\emph{Proceedings of the 44th International Conference on Software Engineering}} (Pittsburgh, Pennsylvania) \emph{(\bibinfo{series}{ICSE '22})}. \bibinfo{publisher}{Association for Computing Machinery}, \bibinfo{address}{New York, NY, USA}, \bibinfo{pages}{659–671}.
\newblock
\showISBNx{9781450392211}
\href{https://doi.org/10.1145/3510003.3510126}{doi:\nolinkurl{10.1145/3510003.3510126}}


\bibitem[Harrison et~al\mbox{.}(2020a)]%
        {harrison2020partemu}
\bibfield{author}{\bibinfo{person}{Lee Harrison}, \bibinfo{person}{Hayawardh Vijayakumar}, \bibinfo{person}{Rohan Padhye}, \bibinfo{person}{Koushik Sen}, {and} \bibinfo{person}{Michael Grace}.} \bibinfo{year}{2020}\natexlab{a}.
\newblock \showarticletitle{{PARTEMU}: Enabling Dynamic Analysis of {Real-World} {TrustZone} Software Using Emulation}. In \bibinfo{booktitle}{\emph{29th USENIX Security Symposium (USENIX Security 20)}}. \bibinfo{publisher}{USENIX Association}, \bibinfo{pages}{789--806}.
\newblock
\showISBNx{978-1-939133-17-5}
\urldef\tempurl%
\url{https://www.usenix.org/conference/usenixsecurity20/presentation/harrison}
\showURL{%
\tempurl}


\bibitem[Harrison et~al\mbox{.}(2020b)]%
        {lee2020partemu}
\bibfield{author}{\bibinfo{person}{Lee Harrison}, \bibinfo{person}{Hayawardh Vijayakumar}, \bibinfo{person}{Rohan Padhye}, \bibinfo{person}{Koushik Sen}, {and} \bibinfo{person}{Michael Grace}.} \bibinfo{year}{2020}\natexlab{b}.
\newblock \showarticletitle{{PARTEMU}: Enabling Dynamic Analysis of {Real-World} {TrustZone} Software Using Emulation}. In \bibinfo{booktitle}{\emph{29th USENIX Security Symposium (USENIX Security 20)}}. \bibinfo{publisher}{USENIX Association}, \bibinfo{pages}{789--806}.
\newblock
\showISBNx{978-1-939133-17-5}
\urldef\tempurl%
\url{https://www.usenix.org/conference/usenixsecurity20/presentation/harrison}
\showURL{%
\tempurl}


\bibitem[Hasanov et~al\mbox{.}(2025)]%
        {hasanov2025alittle}
\bibfield{author}{\bibinfo{person}{Sanan Hasanov}, \bibinfo{person}{Stefan Nagy}, {and} \bibinfo{person}{Paul Gazzillo}.} \bibinfo{year}{2025}\natexlab{}.
\newblock \showarticletitle{{ A Little Goes a Long Way: Tuning Configuration Selection for Continuous Kernel Fuzzing }}. In \bibinfo{booktitle}{\emph{2025 IEEE/ACM 47th International Conference on Software Engineering (ICSE)}}. \bibinfo{publisher}{IEEE Computer Society}, \bibinfo{address}{Los Alamitos, CA, USA}, \bibinfo{pages}{521--533}.
\newblock
\showISSN{1558-1225}
\href{https://doi.org/10.1109/ICSE55347.2025.00042}{doi:\nolinkurl{10.1109/ICSE55347.2025.00042}}


\bibitem[Hazimeh et~al\mbox{.}(2022)]%
        {hazimeh2021magma}
\bibfield{author}{\bibinfo{person}{Ahmad Hazimeh}, \bibinfo{person}{Adrian Herrera}, {and} \bibinfo{person}{Mathias Payer}.} \bibinfo{year}{2022}\natexlab{}.
\newblock \showarticletitle{Magma: A Ground-Truth Fuzzing Benchmark}.
\newblock \bibinfo{journal}{\emph{SIGMETRICS Perform. Eval. Rev.}} \bibinfo{volume}{49}, \bibinfo{number}{1} (\bibinfo{date}{jun} \bibinfo{year}{2022}), \bibinfo{pages}{81–82}.
\newblock
\showISSN{0163-5999}
\href{https://doi.org/10.1145/3543516.3456276}{doi:\nolinkurl{10.1145/3543516.3456276}}


\bibitem[Henkel(2006)]%
        {henkel2006embeded}
\bibfield{author}{\bibinfo{person}{Joachim Henkel}.} \bibinfo{year}{2006}\natexlab{}.
\newblock \showarticletitle{Selective revealing in open innovation processes: The case of embedded Linux}.
\newblock \bibinfo{journal}{\emph{Research Policy}} \bibinfo{volume}{35}, \bibinfo{number}{7} (\bibinfo{year}{2006}), \bibinfo{pages}{953--969}.
\newblock
\showISSN{0048-7333}
\href{https://doi.org/10.1016/j.respol.2006.04.010}{doi:\nolinkurl{10.1016/j.respol.2006.04.010}}


\bibitem[Hu et~al\mbox{.}(2021)]%
        {hu2021achyb}
\bibfield{author}{\bibinfo{person}{Yang Hu}, \bibinfo{person}{Wenxi Wang}, \bibinfo{person}{Casen Hunger}, \bibinfo{person}{Riley Wood}, \bibinfo{person}{Sarfraz Khurshid}, {and} \bibinfo{person}{Mohit Tiwari}.} \bibinfo{year}{2021}\natexlab{}.
\newblock \showarticletitle{ACHyb: a hybrid analysis approach to detect kernel access control vulnerabilities} \emph{(\bibinfo{series}{ESEC/FSE 2021})}. \bibinfo{publisher}{Association for Computing Machinery}, \bibinfo{address}{New York, NY, USA}, \bibinfo{pages}{316–327}.
\newblock
\showISBNx{9781450385626}
\href{https://doi.org/10.1145/3468264.3468627}{doi:\nolinkurl{10.1145/3468264.3468627}}


\bibitem[Hung and Amiri~Sani(2024)]%
        {hung2024brf}
\bibfield{author}{\bibinfo{person}{Hsin-Wei Hung} {and} \bibinfo{person}{Ardalan Amiri~Sani}.} \bibinfo{year}{2024}\natexlab{}.
\newblock \showarticletitle{BRF: Fuzzing the eBPF Runtime}.
\newblock \bibinfo{journal}{\emph{Proc. ACM Softw. Eng.}} \bibinfo{volume}{1}, \bibinfo{number}{FSE}, Article \bibinfo{articleno}{52} (\bibinfo{date}{jul} \bibinfo{year}{2024}), \bibinfo{numpages}{20}~pages.
\newblock
\href{https://doi.org/10.1145/3643778}{doi:\nolinkurl{10.1145/3643778}}


\bibitem[ICORE({[n.\,d.]})]%
        {core2023}
\bibfield{author}{\bibinfo{person}{ICORE}.} \bibinfo{year}{[n.\,d.]}\natexlab{}.
\newblock \bibinfo{title}{CORE2023}.
\newblock \bibinfo{howpublished}{\url{https://www.core.edu.au/conference-portal}}.
\newblock


\bibitem[Inc.(2001)]%
        {macos}
\bibfield{author}{\bibinfo{person}{Apple Inc.}} \bibinfo{year}{2001}\natexlab{}.
\newblock \bibinfo{title}{XNU Kernel (macOS)}.
\newblock \bibinfo{howpublished}{\url{https://opensource.apple.com/source/xnu/}}.
\newblock


\bibitem[Jang et~al\mbox{.}(2023)]%
        {jang2023reusb}
\bibfield{author}{\bibinfo{person}{Jisoo Jang}, \bibinfo{person}{Minsuk Kang}, {and} \bibinfo{person}{Dokyung Song}.} \bibinfo{year}{2023}\natexlab{}.
\newblock \showarticletitle{ReUSB: Replay-Guided USB Driver Fuzzing}. In \bibinfo{booktitle}{\emph{32nd USENIX Security Symposium (USENIX Security 23)}}. \bibinfo{publisher}{USENIX Association}, \bibinfo{pages}{2921--2938}.
\newblock
\showISBNx{978-1-939133-37-3}
\urldef\tempurl%
\url{https://www.usenix.org/conference/usenixsecurity23/presentation/jang}
\showURL{%
\tempurl}


\bibitem[Jeon et~al\mbox{.}(2020)]%
        {jeon2020fuzzan}
\bibfield{author}{\bibinfo{person}{Yuseok Jeon}, \bibinfo{person}{WookHyun Han}, \bibinfo{person}{Nathan Burow}, {and} \bibinfo{person}{Mathias Payer}.} \bibinfo{year}{2020}\natexlab{}.
\newblock \showarticletitle{{FuZZan}: Efficient Sanitizer Metadata Design for Fuzzing}. In \bibinfo{booktitle}{\emph{2020 USENIX Annual Technical Conference (USENIX ATC 20)}}. \bibinfo{publisher}{USENIX Association}, \bibinfo{pages}{249--263}.
\newblock
\showISBNx{978-1-939133-14-4}
\urldef\tempurl%
\url{https://www.usenix.org/conference/atc20/presentation/jeon}
\showURL{%
\tempurl}


\bibitem[Jeong et~al\mbox{.}(2024)]%
        {dae2024ozz}
\bibfield{author}{\bibinfo{person}{Dae~R. Jeong}, \bibinfo{person}{Yewon Choi}, \bibinfo{person}{Byoungyoung Lee}, \bibinfo{person}{Insik Shin}, {and} \bibinfo{person}{Youngjin Kwon}.} \bibinfo{year}{2024}\natexlab{}.
\newblock \showarticletitle{OZZ: Identifying Kernel Out-of-Order Concurrency Bugs with In-Vivo Memory Access Reordering}. In \bibinfo{booktitle}{\emph{Proceedings of the ACM SIGOPS 30th Symposium on Operating Systems Principles}} (Austin, TX, USA) \emph{(\bibinfo{series}{SOSP '24})}. \bibinfo{publisher}{Association for Computing Machinery}, \bibinfo{address}{New York, NY, USA}, \bibinfo{pages}{229–248}.
\newblock
\showISBNx{9798400712517}
\href{https://doi.org/10.1145/3694715.3695944}{doi:\nolinkurl{10.1145/3694715.3695944}}


\bibitem[Jeong et~al\mbox{.}(2019)]%
        {jeong2019razzer}
\bibfield{author}{\bibinfo{person}{Dae~R. Jeong}, \bibinfo{person}{Kyungtae Kim}, \bibinfo{person}{Basavesh Shivakumar}, \bibinfo{person}{Byoungyoung Lee}, {and} \bibinfo{person}{Insik Shin}.} \bibinfo{year}{2019}\natexlab{}.
\newblock \showarticletitle{Razzer: Finding Kernel Race Bugs through Fuzzing}. In \bibinfo{booktitle}{\emph{2019 IEEE Symposium on Security and Privacy (SP)}}. \bibinfo{pages}{754--768}.
\newblock
\href{https://doi.org/10.1109/SP.2019.00017}{doi:\nolinkurl{10.1109/SP.2019.00017}}


\bibitem[Jeong et~al\mbox{.}(2023)]%
        {jeong2023segfuzz}
\bibfield{author}{\bibinfo{person}{Dae~R. Jeong}, \bibinfo{person}{Byoungyoung Lee}, \bibinfo{person}{Insik Shin}, {and} \bibinfo{person}{Youngjin Kwon}.} \bibinfo{year}{2023}\natexlab{}.
\newblock \showarticletitle{SegFuzz: Segmentizing Thread Interleaving to Discover Kernel Concurrency Bugs through Fuzzing}. In \bibinfo{booktitle}{\emph{2023 IEEE Symposium on Security and Privacy (SP)}}. \bibinfo{pages}{2104--2121}.
\newblock
\href{https://doi.org/10.1109/SP46215.2023.10179398}{doi:\nolinkurl{10.1109/SP46215.2023.10179398}}


\bibitem[Jia et~al\mbox{.}(2023)]%
        {jia2023programmable}
\bibfield{author}{\bibinfo{person}{Jinghao Jia}, \bibinfo{person}{YiFei Zhu}, \bibinfo{person}{Dan Williams}, \bibinfo{person}{Andrea Arcangeli}, \bibinfo{person}{Claudio Canella}, \bibinfo{person}{Hubertus Franke}, \bibinfo{person}{Tobin Feldman-Fitzthum}, \bibinfo{person}{Dimitrios Skarlatos}, \bibinfo{person}{Daniel Gruss}, {and} \bibinfo{person}{Tianyin Xu}.} \bibinfo{year}{2023}\natexlab{}.
\newblock \bibinfo{title}{Programmable System Call Security with eBPF}.
\newblock
\showeprint[arxiv]{2302.10366}~[cs.OS]
\urldef\tempurl%
\url{https://arxiv.org/abs/2302.10366}
\showURL{%
\tempurl}


\bibitem[Jiang et~al\mbox{.}(2021)]%
        {jiang2021ecmo}
\bibfield{author}{\bibinfo{person}{Muhui Jiang}, \bibinfo{person}{Lin Ma}, \bibinfo{person}{Yajin Zhou}, \bibinfo{person}{Qiang Liu}, \bibinfo{person}{Cen Zhang}, \bibinfo{person}{Zhi Wang}, \bibinfo{person}{Xiapu Luo}, \bibinfo{person}{Lei Wu}, {and} \bibinfo{person}{Kui Ren}.} \bibinfo{year}{2021}\natexlab{}.
\newblock \showarticletitle{ECMO: Peripheral Transplantation to Rehost Embedded Linux Kernels}. In \bibinfo{booktitle}{\emph{Proceedings of the 2021 ACM SIGSAC Conference on Computer and Communications Security}} (Virtual Event, Republic of Korea) \emph{(\bibinfo{series}{CCS '21})}. \bibinfo{publisher}{Association for Computing Machinery}, \bibinfo{address}{New York, NY, USA}, \bibinfo{pages}{734–748}.
\newblock
\showISBNx{9781450384544}
\href{https://doi.org/10.1145/3460120.3484753}{doi:\nolinkurl{10.1145/3460120.3484753}}


\bibitem[Jiang et~al\mbox{.}(2022)]%
        {jiang2022context}
\bibfield{author}{\bibinfo{person}{Zu{-}Ming Jiang}, \bibinfo{person}{Jia{-}Ju Bai}, \bibinfo{person}{Kangjie Lu}, {and} \bibinfo{person}{Shi{-}Min Hu}.} \bibinfo{year}{2022}\natexlab{}.
\newblock \showarticletitle{Context-Sensitive and Directional Concurrency Fuzzing for Data-Race Detection}. In \bibinfo{booktitle}{\emph{29th Annual Network and Distributed System Security Symposium, {NDSS} 2022, San Diego, California, USA, April 24-28, 2022}}. \bibinfo{publisher}{The Internet Society}.
\newblock
\urldef\tempurl%
\url{https://www.ndss-symposium.org/ndss-paper/auto-draft-198/}
\showURL{%
\tempurl}


\bibitem[Johnson et~al\mbox{.}(2021)]%
        {johnson2021jetset}
\bibfield{author}{\bibinfo{person}{Evan Johnson}, \bibinfo{person}{Maxwell Bland}, \bibinfo{person}{YiFei Zhu}, \bibinfo{person}{Joshua Mason}, \bibinfo{person}{Stephen Checkoway}, \bibinfo{person}{Stefan Savage}, {and} \bibinfo{person}{Kirill Levchenko}.} \bibinfo{year}{2021}\natexlab{}.
\newblock \showarticletitle{Jetset: Targeted Firmware Rehosting for Embedded Systems}. In \bibinfo{booktitle}{\emph{30th USENIX Security Symposium (USENIX Security 21)}}. \bibinfo{publisher}{USENIX Association}, \bibinfo{pages}{321--338}.
\newblock
\showISBNx{978-1-939133-24-3}
\urldef\tempurl%
\url{https://www.usenix.org/conference/usenixsecurity21/presentation/johnson}
\showURL{%
\tempurl}


\bibitem[Jung et~al\mbox{.}(2025)]%
        {jung2025moneta}
\bibfield{author}{\bibinfo{person}{Joonkyo Jung}, \bibinfo{person}{Jisoo Jang}, \bibinfo{person}{Yongwan Jo}, \bibinfo{person}{Jonas Vinck}, \bibinfo{person}{Alexios Voulimeneas}, \bibinfo{person}{Stijn Volckaert}, {and} \bibinfo{person}{Dokyung Song}.} \bibinfo{year}{2025}\natexlab{}.
\newblock \showarticletitle{Moneta: Ex-Vivo {GPU} Driver Fuzzing by Recalling In-Vivo Execution States}. In \bibinfo{booktitle}{\emph{32nd Annual Network and Distributed System Security Symposium, {NDSS} 2025, San Diego, California, USA, February 24-28, 2025}}. \bibinfo{publisher}{The Internet Society}.
\newblock
\urldef\tempurl%
\url{https://www.ndss-symposium.org/ndss-paper/moneta-ex-vivo-gpu-driver-fuzzing-by-recalling-in-vivo-execution-states/}
\showURL{%
\tempurl}


\bibitem[Keil and Kolbitsch(2007)]%
        {keil2007stateful}
\bibfield{author}{\bibinfo{person}{Sylvester Keil} {and} \bibinfo{person}{Clemens Kolbitsch}.} \bibinfo{year}{2007}\natexlab{}.
\newblock \showarticletitle{Stateful fuzzing of wireless device drivers in an emulated environment}.
\newblock \bibinfo{journal}{\emph{Black Hat Japan}} (\bibinfo{year}{2007}).
\newblock


\bibitem[Khan et~al\mbox{.}(2022)]%
        {khan2022virtualization}
\bibfield{author}{\bibinfo{person}{Rida Khan}, \bibinfo{person}{Nouf AlHarbi}, \bibinfo{person}{Ghadi AlGhamdi}, {and} \bibinfo{person}{Lamia Berriche}.} \bibinfo{year}{2022}\natexlab{}.
\newblock \showarticletitle{Virtualization software security: oracle VM VirtualBox}. In \bibinfo{booktitle}{\emph{2022 Fifth International Conference of Women in Data Science at Prince Sultan University (WiDS PSU)}}. IEEE, \bibinfo{pages}{58--60}.
\newblock


\bibitem[Kim et~al\mbox{.}(2020)]%
        {kim2020hfl}
\bibfield{author}{\bibinfo{person}{Kyungtae Kim}, \bibinfo{person}{Dae~R. Jeong}, \bibinfo{person}{Chung~Hwan Kim}, \bibinfo{person}{Yeongjin Jang}, \bibinfo{person}{Insik Shin}, {and} \bibinfo{person}{Byoungyoung Lee}.} \bibinfo{year}{2020}\natexlab{}.
\newblock \showarticletitle{{HFL:} Hybrid Fuzzing on the Linux Kernel}. In \bibinfo{booktitle}{\emph{27th Annual Network and Distributed System Security Symposium, {NDSS} 2020, San Diego, California, USA, February 23-26, 2020}}. \bibinfo{publisher}{The Internet Society}.
\newblock
\urldef\tempurl%
\url{https://www.ndss-symposium.org/ndss-paper/hfl-hybrid-fuzzing-on-the-linux-kernel/}
\showURL{%
\tempurl}


\bibitem[Kim et~al\mbox{.}(2022)]%
        {kim2022fuzzusb}
\bibfield{author}{\bibinfo{person}{Kyungtae Kim}, \bibinfo{person}{Taegyu Kim}, \bibinfo{person}{Ertza Warraich}, \bibinfo{person}{Byoungyoung Lee}, \bibinfo{person}{Kevin R.~B. Butler}, \bibinfo{person}{Antonio Bianchi}, {and} \bibinfo{person}{Dave Jing~Tian}.} \bibinfo{year}{2022}\natexlab{}.
\newblock \showarticletitle{FuzzUSB: Hybrid Stateful Fuzzing of USB Gadget Stacks}. In \bibinfo{booktitle}{\emph{2022 IEEE Symposium on Security and Privacy (SP)}}. \bibinfo{pages}{2212--2229}.
\newblock
\href{https://doi.org/10.1109/SP46214.2022.9833593}{doi:\nolinkurl{10.1109/SP46214.2022.9833593}}


\bibitem[Kim and Kim(2022)]%
        {kim2022robofuzz}
\bibfield{author}{\bibinfo{person}{Seulbae Kim} {and} \bibinfo{person}{Taesoo Kim}.} \bibinfo{year}{2022}\natexlab{}.
\newblock \showarticletitle{RoboFuzz: fuzzing robotic systems over robot operating system (ROS) for finding correctness bugs}. In \bibinfo{booktitle}{\emph{Proceedings of the 30th ACM Joint European Software Engineering Conference and Symposium on the Foundations of Software Engineering}} (Singapore, Singapore) \emph{(\bibinfo{series}{ESEC/FSE 2022})}. \bibinfo{publisher}{Association for Computing Machinery}, \bibinfo{address}{New York, NY, USA}, \bibinfo{pages}{447–458}.
\newblock
\showISBNx{9781450394130}
\href{https://doi.org/10.1145/3540250.3549164}{doi:\nolinkurl{10.1145/3540250.3549164}}


\bibitem[Kim et~al\mbox{.}(2019)]%
        {kim2019finding}
\bibfield{author}{\bibinfo{person}{Seulbae Kim}, \bibinfo{person}{Meng Xu}, \bibinfo{person}{Sanidhya Kashyap}, \bibinfo{person}{Jungyeon Yoon}, \bibinfo{person}{Wen Xu}, {and} \bibinfo{person}{Taesoo Kim}.} \bibinfo{year}{2019}\natexlab{}.
\newblock \showarticletitle{Finding semantic bugs in file systems with an extensible fuzzing framework}. In \bibinfo{booktitle}{\emph{Proceedings of the 27th ACM Symposium on Operating Systems Principles}} (Huntsville, Ontario, Canada) \emph{(\bibinfo{series}{SOSP '19})}. \bibinfo{publisher}{Association for Computing Machinery}, \bibinfo{address}{New York, NY, USA}, \bibinfo{pages}{147–161}.
\newblock
\showISBNx{9781450368735}
\href{https://doi.org/10.1145/3341301.3359662}{doi:\nolinkurl{10.1145/3341301.3359662}}


\bibitem[Kim et~al\mbox{.}(2017)]%
        {atc17cabfuzz}
\bibfield{author}{\bibinfo{person}{Su~Yong Kim}, \bibinfo{person}{Sangho Lee}, \bibinfo{person}{Insu Yun}, \bibinfo{person}{Wen Xu}, \bibinfo{person}{Byoungyoung Lee}, \bibinfo{person}{Youngtae Yun}, {and} \bibinfo{person}{Taesoo Kim}.} \bibinfo{year}{2017}\natexlab{}.
\newblock \showarticletitle{{CAB-Fuzz}: Practical Concolic Testing Techniques for {COTS} Operating Systems}. In \bibinfo{booktitle}{\emph{2017 USENIX Annual Technical Conference (USENIX ATC 17)}}. \bibinfo{publisher}{USENIX Association}, \bibinfo{address}{Santa Clara, CA}, \bibinfo{pages}{689--701}.
\newblock
\showISBNx{978-1-931971-38-6}
\urldef\tempurl%
\url{https://www.usenix.org/conference/atc17/technical-sessions/presentation/kim}
\showURL{%
\tempurl}


\bibitem[Klees et~al\mbox{.}(2018)]%
        {klees2018evaluating}
\bibfield{author}{\bibinfo{person}{George Klees}, \bibinfo{person}{Andrew Ruef}, \bibinfo{person}{Benji Cooper}, \bibinfo{person}{Shiyi Wei}, {and} \bibinfo{person}{Michael Hicks}.} \bibinfo{year}{2018}\natexlab{}.
\newblock \showarticletitle{Evaluating Fuzz Testing}. In \bibinfo{booktitle}{\emph{Proceedings of the 2018 ACM SIGSAC Conference on Computer and Communications Security}} (Toronto, Canada) \emph{(\bibinfo{series}{CCS '18})}. \bibinfo{publisher}{Association for Computing Machinery}, \bibinfo{address}{New York, NY, USA}, \bibinfo{pages}{2123–2138}.
\newblock
\showISBNx{9781450356930}
\href{https://doi.org/10.1145/3243734.3243804}{doi:\nolinkurl{10.1145/3243734.3243804}}


\bibitem[Koschel(2025)]%
        {jakob2025uncovering}
\bibfield{author}{\bibinfo{person}{Jakob Koschel}.} \bibinfo{year}{2025}\natexlab{}.
\newblock \emph{\bibinfo{title}{Uncovering New Classes of Kernel Vulnerabilities}}.
\newblock PhD-Thesis - Research and graduation internal. \bibinfo{school}{Vrije Universiteit Amsterdam}.
\newblock
\href{https://doi.org/10.5463/thesis.833}{doi:\nolinkurl{10.5463/thesis.833}}


\bibitem[Koscher et~al\mbox{.}(2015)]%
        {koscher2015surrogates}
\bibfield{author}{\bibinfo{person}{Karl Koscher}, \bibinfo{person}{Tadayoshi Kohno}, {and} \bibinfo{person}{David Molnar}.} \bibinfo{year}{2015}\natexlab{}.
\newblock \showarticletitle{$\{$SURROGATES$\}$: Enabling $\{$Near-Real-Time$\}$ dynamic analyses of embedded systems}. In \bibinfo{booktitle}{\emph{9th USENIX Workshop on Offensive Technologies (WOOT 15)}}.
\newblock


\bibitem[Lan et~al\mbox{.}(2023)]%
        {yang2024thunder}
\bibfield{author}{\bibinfo{person}{Yang Lan}, \bibinfo{person}{Di Jin}, \bibinfo{person}{Zhun Wang}, \bibinfo{person}{Wende Tan}, \bibinfo{person}{Zheyu Ma}, {and} \bibinfo{person}{Chao Zhang}.} \bibinfo{year}{2023}\natexlab{}.
\newblock \showarticletitle{Thunderkaller: Profiling and Improving the Performance of Syzkaller}. In \bibinfo{booktitle}{\emph{2023 38th IEEE/ACM International Conference on Automated Software Engineering (ASE)}}. \bibinfo{pages}{1567--1578}.
\newblock
\href{https://doi.org/10.1109/ASE56229.2023.00124}{doi:\nolinkurl{10.1109/ASE56229.2023.00124}}


\bibitem[Lee et~al\mbox{.}(2021)]%
        {lee2021constraint}
\bibfield{author}{\bibinfo{person}{Gwangmu Lee}, \bibinfo{person}{Woochul Shim}, {and} \bibinfo{person}{Byoungyoung Lee}.} \bibinfo{year}{2021}\natexlab{}.
\newblock \showarticletitle{Constraint-guided Directed Greybox Fuzzing}. In \bibinfo{booktitle}{\emph{30th USENIX Security Symposium (USENIX Security 21)}}. \bibinfo{publisher}{USENIX Association}, \bibinfo{pages}{3559--3576}.
\newblock
\showISBNx{978-1-939133-24-3}
\urldef\tempurl%
\url{https://www.usenix.org/conference/usenixsecurity21/presentation/lee-gwangmu}
\showURL{%
\tempurl}


\bibitem[Lee et~al\mbox{.}(2024)]%
        {lee2024syzrisk}
\bibfield{author}{\bibinfo{person}{Gwangmu Lee}, \bibinfo{person}{Duo Xu}, \bibinfo{person}{Solmaz Salimi}, \bibinfo{person}{Byoungyoung Lee}, {and} \bibinfo{person}{Mathias Payer}.} \bibinfo{year}{2024}\natexlab{}.
\newblock \showarticletitle{SyzRisk: A Change-Pattern-Based Continuous Kernel Regression Fuzzer}. In \bibinfo{booktitle}{\emph{Proceedings of the 19th ACM Asia Conference on Computer and Communications Security}} (Singapore, Singapore) \emph{(\bibinfo{series}{ASIA CCS '24})}. \bibinfo{publisher}{Association for Computing Machinery}, \bibinfo{address}{New York, NY, USA}, \bibinfo{pages}{1480–1494}.
\newblock
\showISBNx{9798400704826}
\href{https://doi.org/10.1145/3634737.3637642}{doi:\nolinkurl{10.1145/3634737.3637642}}


\bibitem[Levis(2012)]%
        {tinyos}
\bibfield{author}{\bibinfo{person}{Philip Levis}.} \bibinfo{year}{2012}\natexlab{}.
\newblock \showarticletitle{Experiences from a Decade of {TinyOS} Development}. In \bibinfo{booktitle}{\emph{10th USENIX Symposium on Operating Systems Design and Implementation (OSDI 12)}}. \bibinfo{publisher}{USENIX Association}, \bibinfo{address}{Hollywood, CA}, \bibinfo{pages}{207--220}.
\newblock
\showISBNx{978-1-931971-96-6}
\urldef\tempurl%
\url{https://www.usenix.org/conference/osdi12/technical-sessions/presentation/levis}
\showURL{%
\tempurl}


\bibitem[Li et~al\mbox{.}(2024b)]%
        {li2024empirical}
\bibfield{author}{\bibinfo{person}{Hongyu Li}, \bibinfo{person}{Liwei Guo}, \bibinfo{person}{Yexuan Yang}, \bibinfo{person}{Shangguang Wang}, {and} \bibinfo{person}{Mengwei Xu}.} \bibinfo{year}{2024}\natexlab{b}.
\newblock \showarticletitle{An Empirical Study of {Rust-for-Linux}: The Success, Dissatisfaction, and Compromise}. In \bibinfo{booktitle}{\emph{2024 USENIX Annual Technical Conference (USENIX ATC 24)}}. \bibinfo{publisher}{USENIX Association}, \bibinfo{address}{Santa Clara, CA}, \bibinfo{pages}{425--443}.
\newblock
\showISBNx{978-1-939133-41-0}
\urldef\tempurl%
\url{https://www.usenix.org/conference/atc24/presentation/li-hongyu}
\showURL{%
\tempurl}


\bibitem[Li et~al\mbox{.}(2024c)]%
        {li2024enhance}
\bibfield{author}{\bibinfo{person}{Haonan Li}, \bibinfo{person}{Yu Hao}, \bibinfo{person}{Yizhuo Zhai}, {and} \bibinfo{person}{Zhiyun Qian}.} \bibinfo{year}{2024}\natexlab{c}.
\newblock \showarticletitle{Enhancing Static Analysis for Practical Bug Detection: An LLM-Integrated Approach}.
\newblock \bibinfo{journal}{\emph{Proc. ACM Program. Lang.}} \bibinfo{volume}{8}, \bibinfo{number}{OOPSLA1}, Article \bibinfo{articleno}{111} (\bibinfo{date}{apr} \bibinfo{year}{2024}), \bibinfo{numpages}{26}~pages.
\newblock
\href{https://doi.org/10.1145/3649828}{doi:\nolinkurl{10.1145/3649828}}


\bibitem[Li et~al\mbox{.}(2021)]%
        {li2021library}
\bibfield{author}{\bibinfo{person}{Wenqiang Li}, \bibinfo{person}{Le Guan}, \bibinfo{person}{Jingqiang Lin}, \bibinfo{person}{Jiameng Shi}, {and} \bibinfo{person}{Fengjun Li}.} \bibinfo{year}{2021}\natexlab{}.
\newblock \showarticletitle{From library portability to para-rehosting: Natively executing microcontroller software on commodity hardware}.
\newblock \bibinfo{journal}{\emph{arXiv preprint arXiv:2107.12867}} (\bibinfo{year}{2021}).
\newblock


\bibitem[Li et~al\mbox{.}(2022)]%
        {li2022μAFL}
\bibfield{author}{\bibinfo{person}{Wenqiang Li}, \bibinfo{person}{Jiameng Shi}, \bibinfo{person}{Fengjun Li}, \bibinfo{person}{Jingqiang Lin}, \bibinfo{person}{Wei Wang}, {and} \bibinfo{person}{Le Guan}.} \bibinfo{year}{2022}\natexlab{}.
\newblock \showarticletitle{$\mu$AFL: Non-Intrusive Feedback-Driven Fuzzing for Microcontroller Firmware}. In \bibinfo{booktitle}{\emph{Proceedings of the 44th International Conference on Software Engineering}} (Pittsburgh, Pennsylvania) \emph{(\bibinfo{series}{ICSE '22})}. \bibinfo{publisher}{Association for Computing Machinery}, \bibinfo{address}{New York, NY, USA}, \bibinfo{pages}{1–12}.
\newblock
\showISBNx{9781450392211}
\href{https://doi.org/10.1145/3510003.3510208}{doi:\nolinkurl{10.1145/3510003.3510208}}


\bibitem[Li et~al\mbox{.}(2025)]%
        {li2025yesterday}
\bibfield{author}{\bibinfo{person}{Xingwei Li}, \bibinfo{person}{Yan Kang}, \bibinfo{person}{Chenggang Wu}, \bibinfo{person}{Danjun Liu}, \bibinfo{person}{Jiming Wang}, \bibinfo{person}{Yue Sun}, \bibinfo{person}{Zehui Wu}, \bibinfo{person}{Yunchao Wang}, \bibinfo{person}{Rongkuan Ma}, {and} \bibinfo{person}{Qiang Wei}.} \bibinfo{year}{2025}\natexlab{}.
\newblock \showarticletitle{Yesterday Once MorE: Facilitating Linux Kernel Bug Reproduction via Reverse Fuzzing}.
\newblock \bibinfo{journal}{\emph{IEEE Transactions on Information Forensics and Security}} (\bibinfo{year}{2025}), \bibinfo{pages}{1--1}.
\newblock
\href{https://doi.org/10.1109/TIFS.2025.3562704}{doi:\nolinkurl{10.1109/TIFS.2025.3562704}}


\bibitem[Li et~al\mbox{.}(2024a)]%
        {li2023gfuzz}
\bibfield{author}{\bibinfo{person}{Yuwei Li}, \bibinfo{person}{Yuan Chen}, \bibinfo{person}{Shouling Ji}, \bibinfo{person}{Xuhong Zhang}, \bibinfo{person}{Guanglu Yan}, \bibinfo{person}{Alex~X. Liu}, \bibinfo{person}{Chunming Wu}, \bibinfo{person}{Zulie Pan}, {and} \bibinfo{person}{Peng Lin}.} \bibinfo{year}{2024}\natexlab{a}.
\newblock \showarticletitle{G-Fuzz: A Directed Fuzzing Framework for gVisor}.
\newblock \bibinfo{journal}{\emph{IEEE Transactions on Dependable and Secure Computing}} \bibinfo{volume}{21}, \bibinfo{number}{1} (\bibinfo{year}{2024}), \bibinfo{pages}{168--185}.
\newblock
\href{https://doi.org/10.1109/TDSC.2023.3244825}{doi:\nolinkurl{10.1109/TDSC.2023.3244825}}


\bibitem[Li et~al\mbox{.}(2024d)]%
        {li2024rust}
\bibfield{author}{\bibinfo{person}{Zhaofeng Li}, \bibinfo{person}{Vikram Narayanan}, \bibinfo{person}{Xiangdong Chen}, \bibinfo{person}{Jerry Zhang}, {and} \bibinfo{person}{Anton Burtsev}.} \bibinfo{year}{2024}\natexlab{d}.
\newblock \showarticletitle{Rust for Linux: Understanding the Security Impact of Rust in the Linux Kernel}. In \bibinfo{booktitle}{\emph{2024 Annual Computer Security Applications Conference (ACSAC)}}. \bibinfo{pages}{548--562}.
\newblock
\href{https://doi.org/10.1109/ACSAC63791.2024.00054}{doi:\nolinkurl{10.1109/ACSAC63791.2024.00054}}


\bibitem[Lin et~al\mbox{.}(2022)]%
        {lin2022grebe}
\bibfield{author}{\bibinfo{person}{Zhenpeng Lin}, \bibinfo{person}{Yueqi Chen}, \bibinfo{person}{Yuhang Wu}, \bibinfo{person}{Dongliang Mu}, \bibinfo{person}{Chensheng Yu}, \bibinfo{person}{Xinyu Xing}, {and} \bibinfo{person}{Kang Li}.} \bibinfo{year}{2022}\natexlab{}.
\newblock \showarticletitle{GREBE: Unveiling Exploitation Potential for Linux Kernel Bugs}. In \bibinfo{booktitle}{\emph{2022 IEEE Symposium on Security and Privacy (SP)}}. \bibinfo{pages}{2078--2095}.
\newblock
\href{https://doi.org/10.1109/SP46214.2022.9833683}{doi:\nolinkurl{10.1109/SP46214.2022.9833683}}


\bibitem[Linux(2024)]%
        {sched_ext}
\bibfield{author}{\bibinfo{person}{Linux}.} \bibinfo{year}{2024}\natexlab{}.
\newblock \bibinfo{title}{Extensible Scheduler Class}.
\newblock
\urldef\tempurl%
\url{https://www.kernel.org/doc/html/next/scheduler/sched-ext.html}
\showURL{%
\tempurl}


\bibitem[Liu et~al\mbox{.}(2020)]%
        {liu2020fans}
\bibfield{author}{\bibinfo{person}{Baozheng Liu}, \bibinfo{person}{Chao Zhang}, \bibinfo{person}{Guang Gong}, \bibinfo{person}{Yishun Zeng}, \bibinfo{person}{Haifeng Ruan}, {and} \bibinfo{person}{Jianwei Zhuge}.} \bibinfo{year}{2020}\natexlab{}.
\newblock \showarticletitle{{FANS}: Fuzzing Android Native System Services via Automated Interface Analysis}. In \bibinfo{booktitle}{\emph{29th USENIX Security Symposium (USENIX Security 20)}}. \bibinfo{publisher}{USENIX Association}, \bibinfo{pages}{307--323}.
\newblock
\showISBNx{978-1-939133-17-5}
\urldef\tempurl%
\url{https://www.usenix.org/conference/usenixsecurity20/presentation/liu}
\showURL{%
\tempurl}


\bibitem[Liu et~al\mbox{.}(2024a)]%
        {liu2024improve}
\bibfield{author}{\bibinfo{person}{Dinghao Liu}, \bibinfo{person}{Shouling Ji}, \bibinfo{person}{Kangjie Lu}, {and} \bibinfo{person}{Qinming He}.} \bibinfo{year}{2024}\natexlab{a}.
\newblock \showarticletitle{Improving {Indirect-Call} Analysis in {LLVM} with Type and {Data-Flow} {Co-Analysis}}. In \bibinfo{booktitle}{\emph{33rd USENIX Security Symposium (USENIX Security 24)}}. \bibinfo{publisher}{USENIX Association}, \bibinfo{address}{Philadelphia, PA}, \bibinfo{pages}{5895--5912}.
\newblock
\showISBNx{978-1-939133-44-1}
\urldef\tempurl%
\url{https://www.usenix.org/conference/usenixsecurity24/presentation/liu-dinghao-improving}
\showURL{%
\tempurl}


\bibitem[Liu et~al\mbox{.}(2024b)]%
        {liu2024leverage}
\bibfield{author}{\bibinfo{person}{Jianzhong Liu}, \bibinfo{person}{Yuheng Shen}, \bibinfo{person}{Yiru Xu}, {and} \bibinfo{person}{Yu Jiang}.} \bibinfo{year}{2024}\natexlab{b}.
\newblock \showarticletitle{Leveraging Binary Coverage for Effective Generation Guidance in Kernel Fuzzing}. In \bibinfo{booktitle}{\emph{Proceedings of the 2024 on ACM SIGSAC Conference on Computer and Communications Security}} (Salt Lake City, UT, USA) \emph{(\bibinfo{series}{CCS '24})}. \bibinfo{publisher}{Association for Computing Machinery}, \bibinfo{address}{New York, NY, USA}, \bibinfo{pages}{3763–3777}.
\newblock
\showISBNx{9798400706363}
\href{https://doi.org/10.1145/3658644.3690232}{doi:\nolinkurl{10.1145/3658644.3690232}}


\bibitem[Liu et~al\mbox{.}(2023)]%
        {liu2023horus}
\bibfield{author}{\bibinfo{person}{Jianzhong Liu}, \bibinfo{person}{Yuheng Shen}, \bibinfo{person}{Yiru Xu}, \bibinfo{person}{Hao Sun}, {and} \bibinfo{person}{Yu Jiang}.} \bibinfo{year}{2023}\natexlab{}.
\newblock \showarticletitle{Horus: Accelerating Kernel Fuzzing through Efficient Host-VM Memory Access Procedures}.
\newblock \bibinfo{journal}{\emph{ACM Trans. Softw. Eng. Methodol.}} \bibinfo{volume}{33}, \bibinfo{number}{1}, Article \bibinfo{articleno}{11} (\bibinfo{date}{Nov.} \bibinfo{year}{2023}), \bibinfo{numpages}{25}~pages.
\newblock
\showISSN{1049-331X}
\href{https://doi.org/10.1145/3611665}{doi:\nolinkurl{10.1145/3611665}}


\bibitem[Liu et~al\mbox{.}(2021a)]%
        {liu2021ifizz}
\bibfield{author}{\bibinfo{person}{Peiyu Liu}, \bibinfo{person}{Shouling Ji}, \bibinfo{person}{Xuhong Zhang}, \bibinfo{person}{Qinming Dai}, \bibinfo{person}{Kangjie Lu}, \bibinfo{person}{Lirong Fu}, \bibinfo{person}{Wenzhi Chen}, \bibinfo{person}{Peng Cheng}, \bibinfo{person}{Wenhai Wang}, {and} \bibinfo{person}{Raheem Beyah}.} \bibinfo{year}{2021}\natexlab{a}.
\newblock \showarticletitle{IFIZZ: Deep-State and Efficient Fault-Scenario Generation to Test IoT Firmware}. In \bibinfo{booktitle}{\emph{2021 36th IEEE/ACM International Conference on Automated Software Engineering (ASE)}}. \bibinfo{pages}{805--816}.
\newblock
\href{https://doi.org/10.1109/ASE51524.2021.9678785}{doi:\nolinkurl{10.1109/ASE51524.2021.9678785}}


\bibitem[Liu et~al\mbox{.}(2021b)]%
        {liu2021firmguide}
\bibfield{author}{\bibinfo{person}{Qiang Liu}, \bibinfo{person}{Cen Zhang}, \bibinfo{person}{Lin Ma}, \bibinfo{person}{Muhui Jiang}, \bibinfo{person}{Yajin Zhou}, \bibinfo{person}{Lei Wu}, \bibinfo{person}{Wenbo Shen}, \bibinfo{person}{Xiapu Luo}, \bibinfo{person}{Yang Liu}, {and} \bibinfo{person}{Kui Ren}.} \bibinfo{year}{2021}\natexlab{b}.
\newblock \showarticletitle{FirmGuide: Boosting the Capability of Rehosting Embedded Linux Kernels through Model-Guided Kernel Execution}. In \bibinfo{booktitle}{\emph{2021 36th IEEE/ACM International Conference on Automated Software Engineering (ASE)}}. \bibinfo{pages}{792--804}.
\newblock
\href{https://doi.org/10.1109/ASE51524.2021.9678653}{doi:\nolinkurl{10.1109/ASE51524.2021.9678653}}


\bibitem[Lyu et~al\mbox{.}(2019)]%
        {lyu2019mopt}
\bibfield{author}{\bibinfo{person}{Chenyang Lyu}, \bibinfo{person}{Shouling Ji}, \bibinfo{person}{Chao Zhang}, \bibinfo{person}{Yuwei Li}, \bibinfo{person}{Wei-Han Lee}, \bibinfo{person}{Yu Song}, {and} \bibinfo{person}{Raheem Beyah}.} \bibinfo{year}{2019}\natexlab{}.
\newblock \showarticletitle{{MOPT}: Optimized Mutation Scheduling for Fuzzers}. In \bibinfo{booktitle}{\emph{28th USENIX Security Symposium (USENIX Security 19)}}. \bibinfo{publisher}{USENIX Association}, \bibinfo{address}{Santa Clara, CA}, \bibinfo{pages}{1949--1966}.
\newblock
\showISBNx{978-1-939133-06-9}
\urldef\tempurl%
\url{https://www.usenix.org/conference/usenixsecurity19/presentation/lyu}
\showURL{%
\tempurl}


\bibitem[Lyu et~al\mbox{.}(2024)]%
        {lyu2024monarch}
\bibfield{author}{\bibinfo{person}{Tao Lyu}, \bibinfo{person}{Liyi Zhang}, \bibinfo{person}{Zhiyao Feng}, \bibinfo{person}{Yueyang Pan}, \bibinfo{person}{Yujie Ren}, \bibinfo{person}{Meng Xu}, \bibinfo{person}{Mathias Payer}, {and} \bibinfo{person}{Sanidhya Kashyap}.} \bibinfo{year}{2024}\natexlab{}.
\newblock \showarticletitle{Monarch: A Fuzzing Framework for Distributed File Systems}. In \bibinfo{booktitle}{\emph{2024 USENIX Annual Technical Conference (USENIX ATC 24)}}. \bibinfo{publisher}{USENIX Association}, \bibinfo{address}{Santa Clara, CA}, \bibinfo{pages}{529--543}.
\newblock
\showISBNx{978-1-939133-41-0}
\urldef\tempurl%
\url{https://www.usenix.org/conference/atc24/presentation/lyu}
\showURL{%
\tempurl}


\bibitem[Ma et~al\mbox{.}(2022)]%
        {ma2022print}
\bibfield{author}{\bibinfo{person}{Zheyu Ma}, \bibinfo{person}{Bodong Zhao}, \bibinfo{person}{Letu Ren}, \bibinfo{person}{Zheming Li}, \bibinfo{person}{Siqi Ma}, \bibinfo{person}{Xiapu Luo}, {and} \bibinfo{person}{Chao Zhang}.} \bibinfo{year}{2022}\natexlab{}.
\newblock \showarticletitle{PrIntFuzz: Fuzzing Linux Drivers via Automated Virtual Device Simulation}. In \bibinfo{booktitle}{\emph{Proceedings of the 31st ACM SIGSOFT International Symposium on Software Testing and Analysis}} (Virtual, South Korea) \emph{(\bibinfo{series}{ISSTA 2022})}. \bibinfo{publisher}{Association for Computing Machinery}, \bibinfo{address}{New York, NY, USA}, \bibinfo{pages}{404–416}.
\newblock
\showISBNx{9781450393799}
\href{https://doi.org/10.1145/3533767.3534226}{doi:\nolinkurl{10.1145/3533767.3534226}}


\bibitem[Maier et~al\mbox{.}(2019)]%
        {maier2019unicorefuzz}
\bibfield{author}{\bibinfo{person}{Dominik Maier}, \bibinfo{person}{Benedikt Radtke}, {and} \bibinfo{person}{Bastian Harren}.} \bibinfo{year}{2019}\natexlab{}.
\newblock \showarticletitle{Unicorefuzz: On the Viability of Emulation for Kernelspace Fuzzing}. In \bibinfo{booktitle}{\emph{13th USENIX Workshop on Offensive Technologies (WOOT 19)}}. \bibinfo{publisher}{USENIX Association}, \bibinfo{address}{Santa Clara, CA}.
\newblock
\urldef\tempurl%
\url{https://www.usenix.org/conference/woot19/presentation/maier}
\showURL{%
\tempurl}


\bibitem[Maier and Toepfer(2021)]%
        {maier2021bsod}
\bibfield{author}{\bibinfo{person}{Dominik Maier} {and} \bibinfo{person}{Fabian Toepfer}.} \bibinfo{year}{2021}\natexlab{}.
\newblock \showarticletitle{BSOD: Binary-Only Scalable Fuzzing Of Device Drivers}. In \bibinfo{booktitle}{\emph{Proceedings of the 24th International Symposium on Research in Attacks, Intrusions and Defenses}} (San Sebastian, Spain) \emph{(\bibinfo{series}{RAID '21})}. \bibinfo{publisher}{Association for Computing Machinery}, \bibinfo{address}{New York, NY, USA}, \bibinfo{pages}{48–61}.
\newblock
\showISBNx{9781450390583}
\href{https://doi.org/10.1145/3471621.3471863}{doi:\nolinkurl{10.1145/3471621.3471863}}


\bibitem[Manès et~al\mbox{.}(2021)]%
        {manes2021art}
\bibfield{author}{\bibinfo{person}{Valentin~J.M. Manès}, \bibinfo{person}{HyungSeok Han}, \bibinfo{person}{Choongwoo Han}, \bibinfo{person}{Sang~Kil Cha}, \bibinfo{person}{Manuel Egele}, \bibinfo{person}{Edward~J. Schwartz}, {and} \bibinfo{person}{Maverick Woo}.} \bibinfo{year}{2021}\natexlab{}.
\newblock \showarticletitle{The Art, Science, and Engineering of Fuzzing: A Survey}.
\newblock \bibinfo{journal}{\emph{IEEE Transactions on Software Engineering}} \bibinfo{volume}{47}, \bibinfo{number}{11} (\bibinfo{year}{2021}), \bibinfo{pages}{2312--2331}.
\newblock
\href{https://doi.org/10.1109/TSE.2019.2946563}{doi:\nolinkurl{10.1109/TSE.2019.2946563}}


\bibitem[Mera et~al\mbox{.}(2021)]%
        {mera2021dice}
\bibfield{author}{\bibinfo{person}{Alejandro Mera}, \bibinfo{person}{Bo Feng}, \bibinfo{person}{Long Lu}, {and} \bibinfo{person}{Engin Kirda}.} \bibinfo{year}{2021}\natexlab{}.
\newblock \showarticletitle{DICE: Automatic emulation of dma input channels for dynamic firmware analysis}. In \bibinfo{booktitle}{\emph{2021 IEEE Symposium on Security and Privacy (SP)}}. \bibinfo{pages}{1938--1954}.
\newblock


\bibitem[Mera et~al\mbox{.}(2024)]%
        {mera2024shift}
\bibfield{author}{\bibinfo{person}{Alejandro Mera}, \bibinfo{person}{Changming Liu}, \bibinfo{person}{Ruimin Sun}, \bibinfo{person}{Engin Kirda}, {and} \bibinfo{person}{Long Lu}.} \bibinfo{year}{2024}\natexlab{}.
\newblock \showarticletitle{SHiFT: Semi-hosted Fuzz Testing for Embedded Applications}. In \bibinfo{booktitle}{\emph{33rd USENIX Security Symposium (USENIX Security 24)}}. \bibinfo{publisher}{USENIX Association}, \bibinfo{address}{Philadelphia, PA}.
\newblock
\urldef\tempurl%
\url{https://www.usenix.org/conference/usenixsecurity24/presentation/wu-jianliang}
\showURL{%
\tempurl}


\bibitem[Metzman et~al\mbox{.}(2021)]%
        {metzman2021fuzzbench}
\bibfield{author}{\bibinfo{person}{Jonathan Metzman}, \bibinfo{person}{László Szekeres}, \bibinfo{person}{Laurent Maurice~Romain Simon}, \bibinfo{person}{Read~Trevelin Sprabery}, {and} \bibinfo{person}{Abhishek Arya}.} \bibinfo{year}{2021}\natexlab{}.
\newblock \showarticletitle{FuzzBench: An Open Fuzzer Benchmarking Platform and Service}. In \bibinfo{booktitle}{\emph{Proceedings of the 29th ACM Joint Meeting on European Software Engineering Conference and Symposium on the Foundations of Software Engineering}}. \bibinfo{address}{New York, NY, USA}.
\newblock


\bibitem[Mu et~al\mbox{.}(2022)]%
        {mu2022indepth}
\bibfield{author}{\bibinfo{person}{Dongliang Mu}, \bibinfo{person}{Yuhang Wu}, \bibinfo{person}{Yueqi Chen}, \bibinfo{person}{Zhenpeng Lin}, \bibinfo{person}{Chensheng Yu}, \bibinfo{person}{Xinyu Xing}, {and} \bibinfo{person}{Gang Wang}.} \bibinfo{year}{2022}\natexlab{}.
\newblock \showarticletitle{An In-depth Analysis of Duplicated Linux Kernel Bug Reports}. In \bibinfo{booktitle}{\emph{29th Annual Network and Distributed System Security Symposium, {NDSS} 2022, San Diego, California, USA, April 24-28, 2022}}. \bibinfo{publisher}{The Internet Society}.
\newblock
\urldef\tempurl%
\url{https://www.ndss-symposium.org/ndss-paper/auto-draft-246/}
\showURL{%
\tempurl}


\bibitem[Muench et~al\mbox{.}(2024)]%
        {marius2018what}
\bibfield{author}{\bibinfo{person}{Marius Muench}, \bibinfo{person}{Jan Stijohann}, \bibinfo{person}{Frank Kargl}, \bibinfo{person}{Aur{\'e}lien Francillon}, {and} \bibinfo{person}{Davide Balzarotti}.} \bibinfo{year}{2024}\natexlab{}.
\newblock \showarticletitle{What You Corrupt Is Not What You Crash: Challenges in Fuzzing Embedded Devices}. In \bibinfo{booktitle}{\emph{Network and Distributed System Security Symposium, {NDSS} 2018, San Diego, California, USA, February 18 - 26, 2024}}. \bibinfo{publisher}{The Internet Society}.
\newblock


\bibitem[Nagy et~al\mbox{.}(2021)]%
        {nagy2021zafl}
\bibfield{author}{\bibinfo{person}{Stefan Nagy}, \bibinfo{person}{Anh Nguyen-Tuong}, \bibinfo{person}{Jason~D. Hiser}, \bibinfo{person}{Jack~W. Davidson}, {and} \bibinfo{person}{Matthew Hicks}.} \bibinfo{year}{2021}\natexlab{}.
\newblock \showarticletitle{Breaking Through Binaries: Compiler-quality Instrumentation for Better Binary-only Fuzzing}. In \bibinfo{booktitle}{\emph{30th USENIX Security Symposium (USENIX Security 21)}}. \bibinfo{publisher}{USENIX Association}, \bibinfo{pages}{1683--1700}.
\newblock
\showISBNx{978-1-939133-24-3}
\urldef\tempurl%
\url{https://www.usenix.org/conference/usenixsecurity21/presentation/nagy}
\showURL{%
\tempurl}


\bibitem[Natella and Pham(2021)]%
        {natella2021profuzzbench}
\bibfield{author}{\bibinfo{person}{Roberto Natella} {and} \bibinfo{person}{Van-Thuan Pham}.} \bibinfo{year}{2021}\natexlab{}.
\newblock \showarticletitle{ProFuzzBench: a benchmark for stateful protocol fuzzing}. In \bibinfo{booktitle}{\emph{Proceedings of the 30th ACM SIGSOFT International Symposium on Software Testing and Analysis}} (Virtual, Denmark) \emph{(\bibinfo{series}{ISSTA 2021})}. \bibinfo{publisher}{Association for Computing Machinery}, \bibinfo{address}{New York, NY, USA}, \bibinfo{pages}{662–665}.
\newblock
\showISBNx{9781450384599}
\href{https://doi.org/10.1145/3460319.3469077}{doi:\nolinkurl{10.1145/3460319.3469077}}


\bibitem[Nogikh(2023)]%
        {alek2023syzbot}
\bibfield{author}{\bibinfo{person}{Aleksandr Nogikh}.} \bibinfo{year}{2023}\natexlab{}.
\newblock \showarticletitle{Syzbot: 7 years of continuous kernel fuzzing}. In \bibinfo{booktitle}{\emph{Linux Plumbers Conference 2023}}.
\newblock
\urldef\tempurl%
\url{https://lpc.events/event/17/contributions/1521/}
\showURL{%
\tempurl}


\bibitem[Pabla(2009)]%
        {pabla2009completely}
\bibfield{author}{\bibinfo{person}{Chandandeep~Singh Pabla}.} \bibinfo{year}{2009}\natexlab{}.
\newblock \showarticletitle{Completely fair scheduler}.
\newblock \bibinfo{journal}{\emph{Linux Journal}} \bibinfo{volume}{2009}, \bibinfo{number}{184} (\bibinfo{year}{2009}), \bibinfo{pages}{4}.
\newblock


\bibitem[Pailoor et~al\mbox{.}(2018)]%
        {shankara2018moonshine}
\bibfield{author}{\bibinfo{person}{Shankara Pailoor}, \bibinfo{person}{Andrew Aday}, {and} \bibinfo{person}{Suman Jana}.} \bibinfo{year}{2018}\natexlab{}.
\newblock \showarticletitle{{MoonShine}: Optimizing {OS} Fuzzer Seed Selection with Trace Distillation}. In \bibinfo{booktitle}{\emph{27th USENIX Security Symposium (USENIX Security 18)}}. \bibinfo{publisher}{USENIX Association}, \bibinfo{address}{Baltimore, MD}, \bibinfo{pages}{729--743}.
\newblock
\showISBNx{978-1-939133-04-5}
\urldef\tempurl%
\url{https://www.usenix.org/conference/usenixsecurity18/presentation/pailoor}
\showURL{%
\tempurl}


\bibitem[Pan et~al\mbox{.}(2017)]%
        {pan2017digtool}
\bibfield{author}{\bibinfo{person}{Jianfeng Pan}, \bibinfo{person}{Guanglu Yan}, {and} \bibinfo{person}{Xiaocao Fan}.} \bibinfo{year}{2017}\natexlab{}.
\newblock \showarticletitle{Digtool: A $\{$virtualization-based$\}$ framework for detecting kernel vulnerabilities}. In \bibinfo{booktitle}{\emph{26th USENIX Security Symposium (USENIX Security 17)}}. \bibinfo{pages}{149--165}.
\newblock


\bibitem[Pandey et~al\mbox{.}(2019)]%
        {pandey2019triforce}
\bibfield{author}{\bibinfo{person}{Pallavi Pandey}, \bibinfo{person}{Anupam Sarkar}, {and} \bibinfo{person}{Ansuman Banerjee}.} \bibinfo{year}{2019}\natexlab{}.
\newblock \showarticletitle{Triforce QNX Syscall Fuzzer}. In \bibinfo{booktitle}{\emph{2019 IEEE International Symposium on Software Reliability Engineering Workshops (ISSREW)}}. \bibinfo{pages}{59--60}.
\newblock
\href{https://doi.org/10.1109/ISSREW.2019.00043}{doi:\nolinkurl{10.1109/ISSREW.2019.00043}}


\bibitem[Peng and Payer(2020)]%
        {peng2020usb}
\bibfield{author}{\bibinfo{person}{Hui Peng} {and} \bibinfo{person}{Mathias Payer}.} \bibinfo{year}{2020}\natexlab{}.
\newblock \showarticletitle{{USBFuzz}: A Framework for Fuzzing {USB} Drivers by Device Emulation}. In \bibinfo{booktitle}{\emph{29th USENIX Security Symposium (USENIX Security 20)}}. \bibinfo{publisher}{USENIX Association}, \bibinfo{pages}{2559--2575}.
\newblock
\showISBNx{978-1-939133-17-5}
\urldef\tempurl%
\url{https://www.usenix.org/conference/usenixsecurity20/presentation/peng}
\showURL{%
\tempurl}


\bibitem[Pustogarov et~al\mbox{.}(2020)]%
        {pu2020exvivo}
\bibfield{author}{\bibinfo{person}{Ivan Pustogarov}, \bibinfo{person}{Qian Wu}, {and} \bibinfo{person}{David Lie}.} \bibinfo{year}{2020}\natexlab{}.
\newblock \showarticletitle{Ex-vivo dynamic analysis framework for Android device drivers}. In \bibinfo{booktitle}{\emph{2020 IEEE Symposium on Security and Privacy (SP)}}. \bibinfo{pages}{1088--1105}.
\newblock
\href{https://doi.org/10.1109/SP40000.2020.00094}{doi:\nolinkurl{10.1109/SP40000.2020.00094}}


\bibitem[Qinying et~al\mbox{.}(2024)]%
        {wang2024syztrust}
\bibfield{author}{\bibinfo{person}{Wang Qinying}, \bibinfo{person}{Chang Boyu}, \bibinfo{person}{Ji Shouling}, \bibinfo{person}{Tian Yuan}, \bibinfo{person}{Zhang Xuhong}, \bibinfo{person}{Zhao Binbin}, \bibinfo{person}{Pan Gaoning}, \bibinfo{person}{Lyu Chenyang}, \bibinfo{person}{Payer Mathias}, \bibinfo{person}{Wang Wenhai}, {and} \bibinfo{person}{Beyah Reheem}.} \bibinfo{year}{2024}\natexlab{}.
\newblock \showarticletitle{SyzTrust: State-aware Fuzzing on Trusted OS Designed for IoT Devices}. In \bibinfo{booktitle}{\emph{2024 IEEE Symposium on Security and Privacy (SP)}}.
\newblock


\bibitem[Renzelmann et~al\mbox{.}(2012)]%
        {renzelmann2012symdrive}
\bibfield{author}{\bibinfo{person}{Matthew~J Renzelmann}, \bibinfo{person}{Asim Kadav}, {and} \bibinfo{person}{Michael~M Swift}.} \bibinfo{year}{2012}\natexlab{}.
\newblock \showarticletitle{$\{$SymDrive$\}$: Testing Drivers without Devices}. In \bibinfo{booktitle}{\emph{10th USENIX Symposium on Operating Systems Design and Implementation (OSDI 12)}}. \bibinfo{pages}{279--292}.
\newblock


\bibitem[Ryan et~al\mbox{.}(2023)]%
        {ryan2023precise}
\bibfield{author}{\bibinfo{person}{Gabriel Ryan}, \bibinfo{person}{Abhishek Shah}, \bibinfo{person}{Dongdong She}, {and} \bibinfo{person}{Suman Jana}.} \bibinfo{year}{2023}\natexlab{}.
\newblock \showarticletitle{Precise Detection of Kernel Data Races with Probabilistic Lockset Analysis}. In \bibinfo{booktitle}{\emph{2023 IEEE Symposium on Security and Privacy (SP)}}. \bibinfo{pages}{2086--2103}.
\newblock
\href{https://doi.org/10.1109/SP46215.2023.10179366}{doi:\nolinkurl{10.1109/SP46215.2023.10179366}}


\bibitem[Scharnowski et~al\mbox{.}(2023)]%
        {Scharnowski2023hoedur}
\bibfield{author}{\bibinfo{person}{Tobias Scharnowski}, \bibinfo{person}{Simon Wörner}, \bibinfo{person}{Felix Buchmann}, \bibinfo{person}{Moritz~Schloegel Nils~Bars}, {and} \bibinfo{person}{Thorsten}.} \bibinfo{year}{2023}\natexlab{}.
\newblock \showarticletitle{Hoedur: Embedded Firmware Fuzzing using Multi-Stream Inputs}. In \bibinfo{booktitle}{\emph{32nd USENIX Security Symposium (USENIX Security 23)}}. \bibinfo{publisher}{USENIX Association}, \bibinfo{address}{Anaheim, CA}, \bibinfo{pages}{2885--2902}.
\newblock
\showISBNx{978-1-939133-37-}
\urldef\tempurl%
\url{https://www.usenix.org/conference/usenixsecurity23/presentation/tay}
\showURL{%
\tempurl}


\bibitem[Schiller et~al\mbox{.}(2023)]%
        {nico2023drone}
\bibfield{author}{\bibinfo{person}{Nico Schiller}, \bibinfo{person}{Merlin Chlosta}, \bibinfo{person}{Moritz Schloegel}, \bibinfo{person}{Nils Bars}, \bibinfo{person}{Thorsten Eisenhofer}, \bibinfo{person}{Tobias Scharnowski}, \bibinfo{person}{Felix Domke}, \bibinfo{person}{Lea Sch{\"{o}}nherr}, {and} \bibinfo{person}{Thorsten Holz}.} \bibinfo{year}{2023}\natexlab{}.
\newblock \showarticletitle{Drone Security and the Mysterious Case of DJI's DroneID}. In \bibinfo{booktitle}{\emph{30th Annual Network and Distributed System Security Symposium, {NDSS} 2023, San Diego, California, USA, February 27 - March 3, 2023}}. \bibinfo{publisher}{The Internet Society}.
\newblock
\urldef\tempurl%
\url{https://www.ndss-symposium.org/ndss-paper/drone-security-and-the-mysterious-case-of-djis-droneid/}
\showURL{%
\tempurl}


\bibitem[Schloegel et~al\mbox{.}(2024)]%
        {schl2024prudent}
\bibfield{author}{\bibinfo{person}{M. Schloegel}, \bibinfo{person}{N. Bars}, \bibinfo{person}{N. Schiller}, \bibinfo{person}{L. Bernhard}, \bibinfo{person}{T. Scharnowski}, \bibinfo{person}{A. Crump}, \bibinfo{person}{A. Ale-Ebrahim}, \bibinfo{person}{N. Bissantz}, \bibinfo{person}{M. Muench}, {and} \bibinfo{person}{T. Holz}.} \bibinfo{year}{2024}\natexlab{}.
\newblock \showarticletitle{SoK: Prudent Evaluation Practices for Fuzzing}. In \bibinfo{booktitle}{\emph{2024 IEEE Symposium on Security and Privacy (SP)}}. \bibinfo{publisher}{IEEE Computer Society}, \bibinfo{address}{Los Alamitos, CA, USA}, \bibinfo{pages}{140--140}.
\newblock
\showISSN{2375-1207}
\href{https://doi.org/10.1109/SP54263.2024.00137}{doi:\nolinkurl{10.1109/SP54263.2024.00137}}


\bibitem[Schumilo et~al\mbox{.}(2021)]%
        {schumilo2021nyx}
\bibfield{author}{\bibinfo{person}{Sergej Schumilo}, \bibinfo{person}{Cornelius Aschermann}, \bibinfo{person}{Ali Abbasi}, \bibinfo{person}{Simon W{\"o}r-ner}, {and} \bibinfo{person}{Thorsten Holz}.} \bibinfo{year}{2021}\natexlab{}.
\newblock \showarticletitle{Nyx: Greybox Hypervisor Fuzzing using Fast Snapshots and Affine Types}. In \bibinfo{booktitle}{\emph{30th USENIX Security Symposium (USENIX Security 21)}}. \bibinfo{publisher}{USENIX Association}, \bibinfo{pages}{2597--2614}.
\newblock
\showISBNx{978-1-939133-24-3}
\urldef\tempurl%
\url{https://www.usenix.org/conference/usenixsecurity21/presentation/schumilo}
\showURL{%
\tempurl}


\bibitem[Schumilo et~al\mbox{.}(2017)]%
        {schumilo2017kafl}
\bibfield{author}{\bibinfo{person}{Sergej Schumilo}, \bibinfo{person}{Cornelius Aschermann}, \bibinfo{person}{Robert Gawlik}, \bibinfo{person}{Sebastian Schinzel}, {and} \bibinfo{person}{Thorsten Holz}.} \bibinfo{year}{2017}\natexlab{}.
\newblock \showarticletitle{{kAFL}: {Hardware-Assisted} Feedback Fuzzing for {OS} Kernels}. In \bibinfo{booktitle}{\emph{26th USENIX Security Symposium (USENIX Security 17)}}. \bibinfo{publisher}{USENIX Association}, \bibinfo{address}{Vancouver, BC}, \bibinfo{pages}{167--182}.
\newblock
\showISBNx{978-1-931971-40-9}
\urldef\tempurl%
\url{https://www.usenix.org/conference/usenixsecurity17/technical-sessions/presentation/schumilo}
\showURL{%
\tempurl}


\bibitem[Schumilo et~al\mbox{.}(2014)]%
        {schumilo2014don}
\bibfield{author}{\bibinfo{person}{Sergej Schumilo}, \bibinfo{person}{Ralf Spenneberg}, {and} \bibinfo{person}{Hendrik Schwartke}.} \bibinfo{year}{2014}\natexlab{}.
\newblock \showarticletitle{Don’t trust your USB! How to find bugs in USB device drivers}.
\newblock \bibinfo{journal}{\emph{Blackhat Europe}} (\bibinfo{year}{2014}).
\newblock


\bibitem[Shen et~al\mbox{.}(2024)]%
        {shen2024enhancing}
\bibfield{author}{\bibinfo{person}{Yuheng Shen}, \bibinfo{person}{Jianzhong Liu}, \bibinfo{person}{Yiru Xu}, \bibinfo{person}{Hao Sun}, \bibinfo{person}{Mingzhe Wang}, \bibinfo{person}{Nan Guan}, \bibinfo{person}{Heyuan Shi}, {and} \bibinfo{person}{Yu Jiang}.} \bibinfo{year}{2024}\natexlab{}.
\newblock \showarticletitle{Enhancing ROS System Fuzzing through Callback Tracing}. In \bibinfo{booktitle}{\emph{Proceedings of the 33rd ACM SIGSOFT International Symposium on Software Testing and Analysis}} (Vienna, Austria) \emph{(\bibinfo{series}{ISSTA 2024})}. \bibinfo{publisher}{Association for Computing Machinery}, \bibinfo{address}{New York, NY, USA}.
\newblock


\bibitem[Shen et~al\mbox{.}(2021)]%
        {shen2021rtkaller}
\bibfield{author}{\bibinfo{person}{Yuheng Shen}, \bibinfo{person}{Hao Sun}, \bibinfo{person}{Yu Jiang}, \bibinfo{person}{Heyuan Shi}, \bibinfo{person}{Yixiao Yang}, {and} \bibinfo{person}{Wanli Chang}.} \bibinfo{year}{2021}\natexlab{}.
\newblock \showarticletitle{Rtkaller: State-aware Task Generation for RTOS Fuzzing}.
\newblock \bibinfo{journal}{\emph{ACM Trans. Embed. Comput. Syst.}} \bibinfo{volume}{20}, \bibinfo{number}{5s}, Article \bibinfo{articleno}{83} (\bibinfo{date}{Sept.} \bibinfo{year}{2021}), \bibinfo{numpages}{22}~pages.
\newblock
\showISSN{1539-9087}
\href{https://doi.org/10.1145/3477014}{doi:\nolinkurl{10.1145/3477014}}


\bibitem[Shen et~al\mbox{.}(2022b)]%
        {shen2022tardis}
\bibfield{author}{\bibinfo{person}{Yuheng Shen}, \bibinfo{person}{Yiru Xu}, \bibinfo{person}{Hao Sun}, \bibinfo{person}{Jianzhong Liu}, \bibinfo{person}{Zichen Xu}, \bibinfo{person}{Aiguo Cui}, \bibinfo{person}{Heyuan Shi}, {and} \bibinfo{person}{Yu Jiang}.} \bibinfo{year}{2022}\natexlab{b}.
\newblock \showarticletitle{Tardis: Coverage-Guided Embedded Operating System Fuzzing}.
\newblock \bibinfo{journal}{\emph{IEEE Transactions on Computer-Aided Design of Integrated Circuits and Systems}} \bibinfo{volume}{41}, \bibinfo{number}{11} (\bibinfo{year}{2022}), \bibinfo{pages}{4563--4574}.
\newblock
\href{https://doi.org/10.1109/TCAD.2022.3198910}{doi:\nolinkurl{10.1109/TCAD.2022.3198910}}


\bibitem[Shen et~al\mbox{.}(2022a)]%
        {shen2022drifuzz}
\bibfield{author}{\bibinfo{person}{Zekun Shen}, \bibinfo{person}{Ritik Roongta}, {and} \bibinfo{person}{Brendan Dolan-Gavitt}.} \bibinfo{year}{2022}\natexlab{a}.
\newblock \showarticletitle{Drifuzz: Harvesting Bugs in Device Drivers from Golden Seeds}. In \bibinfo{booktitle}{\emph{31st USENIX Security Symposium (USENIX Security 22)}}. \bibinfo{publisher}{USENIX Association}, \bibinfo{address}{Boston, MA}, \bibinfo{pages}{1275--1290}.
\newblock
\showISBNx{978-1-939133-31-1}
\urldef\tempurl%
\url{https://www.usenix.org/conference/usenixsecurity22/presentation/shen-zekun}
\showURL{%
\tempurl}


\bibitem[Shi et~al\mbox{.}(2024a)]%
        {shi2024industry}
\bibfield{author}{\bibinfo{person}{Heyuan Shi}, \bibinfo{person}{Shijun Chen}, \bibinfo{person}{Runzhe Wang}, \bibinfo{person}{Yuhan Chen}, \bibinfo{person}{Weibo Zhang}, \bibinfo{person}{Qiang Zhang}, \bibinfo{person}{Yuheng Shen}, \bibinfo{person}{Xiaohai Shi}, \bibinfo{person}{Chao Hu}, {and} \bibinfo{person}{Yu Jiang}.} \bibinfo{year}{2024}\natexlab{a}.
\newblock \showarticletitle{Industry Practice of Directed Kernel Fuzzing for Open-source Linux Distribution}. In \bibinfo{booktitle}{\emph{Proceedings of the 39th IEEE/ACM International Conference on Automated Software Engineering}} (Sacramento, CA, USA) \emph{(\bibinfo{series}{ASE '24})}. \bibinfo{publisher}{Association for Computing Machinery}, \bibinfo{address}{New York, NY, USA}, \bibinfo{pages}{2159–2169}.
\newblock
\showISBNx{9798400712487}
\urldef\tempurl%
\url{https://doi.org/10.1145/3691620.3695278}
\showURL{%
\tempurl}


\bibitem[Shi et~al\mbox{.}(2019)]%
        {shi2019industry}
\bibfield{author}{\bibinfo{person}{Heyuan Shi}, \bibinfo{person}{Runzhe Wang}, \bibinfo{person}{Ying Fu}, \bibinfo{person}{Mingzhe Wang}, \bibinfo{person}{Xiaohai Shi}, \bibinfo{person}{Xun Jiao}, \bibinfo{person}{Houbing Song}, \bibinfo{person}{Yu Jiang}, {and} \bibinfo{person}{Jiaguang Sun}.} \bibinfo{year}{2019}\natexlab{}.
\newblock \showarticletitle{Industry practice of coverage-guided enterprise Linux kernel fuzzing}. In \bibinfo{booktitle}{\emph{Proceedings of the 2019 27th ACM Joint Meeting on European Software Engineering Conference and Symposium on the Foundations of Software Engineering}} (Tallinn, Estonia) \emph{(\bibinfo{series}{ESEC/FSE 2019})}. \bibinfo{publisher}{Association for Computing Machinery}, \bibinfo{address}{New York, NY, USA}, \bibinfo{pages}{986–995}.
\newblock
\showISBNx{9781450355728}
\href{https://doi.org/10.1145/3338906.3340460}{doi:\nolinkurl{10.1145/3338906.3340460}}


\bibitem[Shi et~al\mbox{.}(2024b)]%
        {shi2024facilitating}
\bibfield{author}{\bibinfo{person}{Jiameng Shi}, \bibinfo{person}{Wenqiang Li}, \bibinfo{person}{Wenwen Wang}, {and} \bibinfo{person}{Le Guan}.} \bibinfo{year}{2024}\natexlab{b}.
\newblock \showarticletitle{Facilitating Non-Intrusive In-Vivo Firmware Testing with Stateless Instrumentation}. In \bibinfo{booktitle}{\emph{31st Annual Network and Distributed System Security Symposium, {NDSS} 2024, San Diego, California, USA, February 26 - March 1, 2024}}. \bibinfo{publisher}{The Internet Society}.
\newblock


\bibitem[SimonKagstrom(2010)]%
        {kcov}
\bibfield{author}{\bibinfo{person}{SimonKagstrom}.} \bibinfo{year}{2010}\natexlab{}.
\newblock \bibinfo{title}{KCOV: code coverage for fuzzing}.
\newblock \bibinfo{howpublished}{\url{https://docs.kernel.org/dev-tools/kcov.html}}.
\newblock


\bibitem[Song et~al\mbox{.}(2019)]%
        {song2019periscope}
\bibfield{author}{\bibinfo{person}{Dokyung Song}, \bibinfo{person}{Felicitas Hetzelt}, \bibinfo{person}{Dipanjan Das}, \bibinfo{person}{Chad Spensky}, \bibinfo{person}{Yeoul Na}, \bibinfo{person}{Stijn Volckaert}, \bibinfo{person}{Giovanni Vigna}, \bibinfo{person}{Christopher Kruegel}, \bibinfo{person}{Jean{-}Pierre Seifert}, {and} \bibinfo{person}{Michael Franz}.} \bibinfo{year}{2019}\natexlab{}.
\newblock \showarticletitle{PeriScope: An Effective Probing and Fuzzing Framework for the Hardware-OS Boundary}. In \bibinfo{booktitle}{\emph{26th Annual Network and Distributed System Security Symposium, {NDSS} 2019, San Diego, California, USA, February 24-27, 2019}}. \bibinfo{publisher}{The Internet Society}.
\newblock
\urldef\tempurl%
\url{https://www.ndss-symposium.org/ndss-paper/periscope-an-effective-probing-and-fuzzing-framework-for-the-hardware-os-boundary/}
\showURL{%
\tempurl}


\bibitem[Song et~al\mbox{.}(2020)]%
        {song2020agamotto}
\bibfield{author}{\bibinfo{person}{Dokyung Song}, \bibinfo{person}{Felicitas Hetzelt}, \bibinfo{person}{Jonghwan Kim}, \bibinfo{person}{Brent~ByungHoon Kang}, \bibinfo{person}{Jean-Pierre Seifert}, {and} \bibinfo{person}{Michael Franz}.} \bibinfo{year}{2020}\natexlab{}.
\newblock \showarticletitle{Agamotto: Accelerating Kernel Driver Fuzzing with Lightweight Virtual Machine Checkpoints}. In \bibinfo{booktitle}{\emph{29th USENIX Security Symposium (USENIX Security 20)}}. \bibinfo{publisher}{USENIX Association}, \bibinfo{pages}{2541--2557}.
\newblock
\showISBNx{978-1-939133-17-5}
\urldef\tempurl%
\url{https://www.usenix.org/conference/usenixsecurity20/presentation/song}
\showURL{%
\tempurl}


\bibitem[Sun et~al\mbox{.}(2022)]%
        {sun2022ksg}
\bibfield{author}{\bibinfo{person}{Hao Sun}, \bibinfo{person}{Yuheng Shen}, \bibinfo{person}{Jianzhong Liu}, \bibinfo{person}{Yiru Xu}, {and} \bibinfo{person}{Yu Jiang}.} \bibinfo{year}{2022}\natexlab{}.
\newblock \showarticletitle{{KSG}: Augmenting Kernel Fuzzing with System Call Specification Generation}. In \bibinfo{booktitle}{\emph{2022 USENIX Annual Technical Conference (USENIX ATC 22)}}. \bibinfo{publisher}{USENIX Association}, \bibinfo{address}{Carlsbad, CA}, \bibinfo{pages}{351--366}.
\newblock
\showISBNx{978-1-939133-29-20}
\urldef\tempurl%
\url{https://www.usenix.org/conference/atc22/presentation/sun}
\showURL{%
\tempurl}


\bibitem[Sun et~al\mbox{.}(2021)]%
        {sun2021healer}
\bibfield{author}{\bibinfo{person}{Hao Sun}, \bibinfo{person}{Yuheng Shen}, \bibinfo{person}{Cong Wang}, \bibinfo{person}{Jianzhong Liu}, \bibinfo{person}{Yu Jiang}, \bibinfo{person}{Ting Chen}, {and} \bibinfo{person}{Aiguo Cui}.} \bibinfo{year}{2021}\natexlab{}.
\newblock \showarticletitle{HEALER: Relation Learning Guided Kernel Fuzzing}. In \bibinfo{booktitle}{\emph{Proceedings of the ACM SIGOPS 28th Symposium on Operating Systems Principles}} (Virtual Event, Germany) \emph{(\bibinfo{series}{SOSP '21})}. \bibinfo{publisher}{Association for Computing Machinery}, \bibinfo{address}{New York, NY, USA}, \bibinfo{pages}{344--358}.
\newblock
\showISBNx{9781450387095}
\href{https://doi.org/10.1145/3477132.3483547}{doi:\nolinkurl{10.1145/3477132.3483547}}


\bibitem[Sun and Su(2024)]%
        {sun2024validating}
\bibfield{author}{\bibinfo{person}{Hao Sun} {and} \bibinfo{person}{Zhendong Su}.} \bibinfo{year}{2024}\natexlab{}.
\newblock \showarticletitle{Validating the {eBPF} Verifier via State Embedding}. In \bibinfo{booktitle}{\emph{18th USENIX Symposium on Operating Systems Design and Implementation (OSDI 24)}}. \bibinfo{publisher}{USENIX Association}, \bibinfo{address}{Santa Clara, CA}, \bibinfo{pages}{615--628}.
\newblock
\showISBNx{978-1-939133-40-3}
\urldef\tempurl%
\url{https://www.usenix.org/conference/osdi24/presentation/sun-hao}
\showURL{%
\tempurl}


\bibitem[Sun et~al\mbox{.}(2024)]%
        {sun2024finding}
\bibfield{author}{\bibinfo{person}{Hao Sun}, \bibinfo{person}{Yiru Xu}, \bibinfo{person}{Jianzhong Liu}, \bibinfo{person}{Yuheng Shen}, \bibinfo{person}{Nan Guan}, {and} \bibinfo{person}{Yu Jiang}.} \bibinfo{year}{2024}\natexlab{}.
\newblock \showarticletitle{Finding Correctness Bugs in eBPF Verifier with Structured and Sanitized Program}. In \bibinfo{booktitle}{\emph{Proceedings of the Nineteenth European Conference on Computer Systems}} (Athens, Greece) \emph{(\bibinfo{series}{EuroSys '24})}. \bibinfo{publisher}{Association for Computing Machinery}, \bibinfo{address}{New York, NY, USA}, \bibinfo{pages}{689–703}.
\newblock
\showISBNx{9798400704376}
\href{https://doi.org/10.1145/3627703.3629562}{doi:\nolinkurl{10.1145/3627703.3629562}}


\bibitem[Sun et~al\mbox{.}(2025)]%
        {sun2025syzparam}
\bibfield{author}{\bibinfo{person}{Yue Sun}, \bibinfo{person}{Yan Kang}, \bibinfo{person}{Chenggang Wu}, \bibinfo{person}{Kangjie Lu}, \bibinfo{person}{Jiming Wang}, \bibinfo{person}{Xingwei Li}, \bibinfo{person}{Yuhao Hu}, \bibinfo{person}{Jikai Ren}, \bibinfo{person}{Yuanming Lai}, \bibinfo{person}{Mengyao Xie}, {and} \bibinfo{person}{Zhe Wang}.} \bibinfo{year}{2025}\natexlab{}.
\newblock \bibinfo{title}{SyzParam: Introducing Runtime Parameters into Kernel Driver Fuzzing}.
\newblock
\showeprint[arxiv]{2501.10002}~[cs.CR]
\urldef\tempurl%
\url{https://arxiv.org/abs/2501.10002}
\showURL{%
\tempurl}


\bibitem[Sönke~Huster and Classen(2024)]%
        {huster2024boldly}
\bibfield{author}{\bibinfo{person}{Matthias~Hollick Sönke~Huster} {and} \bibinfo{person}{Jiska Classen}.} \bibinfo{year}{2024}\natexlab{}.
\newblock \showarticletitle{To Boldly Go Where No Fuzzer Has Gone Before: Finding Bugs in Linux' Wireless Stacks through VirtIO Devices}. In \bibinfo{booktitle}{\emph{2024 IEEE Symposium on Security and Privacy (SP)}}.
\newblock


\bibitem[Talebi et~al\mbox{.}(2018)]%
        {talebi2018charm}
\bibfield{author}{\bibinfo{person}{Seyed Mohammadjavad~Seyed Talebi}, \bibinfo{person}{Hamid Tavakoli}, \bibinfo{person}{Hang Zhang}, \bibinfo{person}{Zheng Zhang}, \bibinfo{person}{Ardalan~Amiri Sani}, {and} \bibinfo{person}{Zhiyun Qian}.} \bibinfo{year}{2018}\natexlab{}.
\newblock \showarticletitle{Charm: Facilitating dynamic analysis of device drivers of mobile systems}. In \bibinfo{booktitle}{\emph{27th USENIX Security Symposium (USENIX Security 18)}}. \bibinfo{pages}{291--307}.
\newblock


\bibitem[Tan et~al\mbox{.}(2023)]%
        {tan2023syzdirect}
\bibfield{author}{\bibinfo{person}{Xin Tan}, \bibinfo{person}{Yuan Zhang}, \bibinfo{person}{Jiadong Lu}, \bibinfo{person}{Xin Xiong}, \bibinfo{person}{Zhuang Liu}, {and} \bibinfo{person}{Min Yang}.} \bibinfo{year}{2023}\natexlab{}.
\newblock \showarticletitle{SyzDirect: Directed Greybox Fuzzing for Linux Kernel}. In \bibinfo{booktitle}{\emph{Proceedings of the 2023 ACM SIGSAC Conference on Computer and Communications Security}} (Copenhagen, Denmark) \emph{(\bibinfo{series}{CCS '23})}. \bibinfo{publisher}{Association for Computing Machinery}, \bibinfo{address}{New York, NY, USA}, \bibinfo{pages}{1630–1644}.
\newblock
\showISBNx{9798400700507}
\href{https://doi.org/10.1145/3576915.3623146}{doi:\nolinkurl{10.1145/3576915.3623146}}


\bibitem[Tay et~al\mbox{.}(2023)]%
        {hui2023greehouse}
\bibfield{author}{\bibinfo{person}{Hui~Jun Tay}, \bibinfo{person}{Kyle Zeng}, \bibinfo{person}{Jayakrishna~Menon Vadayath}, \bibinfo{person}{Arvind~S Raj}, \bibinfo{person}{Audrey Dutcher}, \bibinfo{person}{Tejesh Reddy}, \bibinfo{person}{Wil Gibbs}, \bibinfo{person}{Zion~Leonahenahe Basque}, \bibinfo{person}{Fangzhou Dong}, \bibinfo{person}{Zack Smith}, \bibinfo{person}{Adam Doup{\'e}}, \bibinfo{person}{Tiffany Bao}, \bibinfo{person}{Yan Shoshitaishvili}, {and} \bibinfo{person}{Ruoyu Wang}.} \bibinfo{year}{2023}\natexlab{}.
\newblock \showarticletitle{Greenhouse: {Single-Service} Rehosting of {Linux-Based} Firmware Binaries in {User-Space} Emulation}. In \bibinfo{booktitle}{\emph{32nd USENIX Security Symposium (USENIX Security 23)}}. \bibinfo{publisher}{USENIX Association}, \bibinfo{address}{Anaheim, CA}, \bibinfo{pages}{5791--5808}.
\newblock
\showISBNx{978-1-939133-37-3}
\urldef\tempurl%
\url{https://www.usenix.org/conference/usenixsecurity23/presentation/tay}
\showURL{%
\tempurl}


\bibitem[Tian et~al\mbox{.}(2025)]%
        {zhao2025fix}
\bibfield{author}{\bibinfo{person}{Zhao Tian}, \bibinfo{person}{Junjie Chen}, {and} \bibinfo{person}{Xiangyu Zhang}.} \bibinfo{year}{2025}\natexlab{}.
\newblock \showarticletitle{Fixing Large Language Models' Specification Misunderstanding for Better Code Generation}. In \bibinfo{booktitle}{\emph{2025 IEEE/ACM 47th International Conference on Software Engineering (ICSE)}}. \bibinfo{pages}{1514--1526}.
\newblock
\href{https://doi.org/10.1109/ICSE55347.2025.00108}{doi:\nolinkurl{10.1109/ICSE55347.2025.00108}}


\bibitem[Torvalds and the Linux Kernel~Community(1991)]%
        {linus1991linux}
\bibfield{author}{\bibinfo{person}{Linus Torvalds} {and} \bibinfo{person}{the Linux Kernel~Community}.} \bibinfo{year}{1991}\natexlab{}.
\newblock \bibinfo{title}{Linux Kernel}.
\newblock \bibinfo{howpublished}{\url{https://www.kernel.org}}.
\newblock


\bibitem[Vyukov({[n.\,d.]})]%
        {syzbot}
\bibfield{author}{\bibinfo{person}{Dmitry Vyukov}.} \bibinfo{year}{[n.\,d.]}\natexlab{}.
\newblock \bibinfo{title}{syzbot}.
\newblock \bibinfo{howpublished}{\url{https://syzkaller.appspot.com/upstream}}.
\newblock


\bibitem[Walters(1999)]%
        {walters1999vmware}
\bibfield{author}{\bibinfo{person}{Brian Walters}.} \bibinfo{year}{1999}\natexlab{}.
\newblock \showarticletitle{VMware virtual platform}.
\newblock \bibinfo{journal}{\emph{Linux journal}} \bibinfo{volume}{1999}, \bibinfo{number}{63es} (\bibinfo{year}{1999}), \bibinfo{pages}{6--es}.
\newblock


\bibitem[Wang et~al\mbox{.}(2021b)]%
        {wang2021syzvegas}
\bibfield{author}{\bibinfo{person}{Daimeng Wang}, \bibinfo{person}{Zheng Zhang}, \bibinfo{person}{Hang Zhang}, \bibinfo{person}{Zhiyun Qian}, \bibinfo{person}{Srikanth~V. Krishnamurthy}, {and} \bibinfo{person}{Nael Abu-Ghazaleh}.} \bibinfo{year}{2021}\natexlab{b}.
\newblock \showarticletitle{{SyzVegas}: Beating Kernel Fuzzing Odds with Reinforcement Learning}. In \bibinfo{booktitle}{\emph{30th USENIX Security Symposium (USENIX Security 21)}}. \bibinfo{publisher}{USENIX Association}, \bibinfo{pages}{2741--2758}.
\newblock
\showISBNx{978-1-939133-24-3}
\urldef\tempurl%
\url{https://www.usenix.org/conference/usenixsecurity21/presentation/wang-daimeng}
\showURL{%
\tempurl}


\bibitem[Wang et~al\mbox{.}(2019)]%
        {wang2019sensitive}
\bibfield{author}{\bibinfo{person}{Jinghan Wang}, \bibinfo{person}{Yue Duan}, \bibinfo{person}{Wei Song}, \bibinfo{person}{Heng Yin}, {and} \bibinfo{person}{Chengyu Song}.} \bibinfo{year}{2019}\natexlab{}.
\newblock \showarticletitle{Be Sensitive and Collaborative: Analyzing Impact of Coverage Metrics in Greybox Fuzzing}. In \bibinfo{booktitle}{\emph{22nd International Symposium on Research in Attacks, Intrusions and Defenses (RAID 2019)}}. \bibinfo{publisher}{USENIX Association}, \bibinfo{address}{Chaoyang District, Beijing}, \bibinfo{pages}{1--15}.
\newblock
\showISBNx{978-1-939133-07-6}
\urldef\tempurl%
\url{https://www.usenix.org/conference/raid2019/presentation/wang}
\showURL{%
\tempurl}


\bibitem[Wang et~al\mbox{.}(2021a)]%
        {wang2021rein}
\bibfield{author}{\bibinfo{person}{Jinghan Wang}, \bibinfo{person}{Chengyu Song}, {and} \bibinfo{person}{Heng Yin}.} \bibinfo{year}{2021}\natexlab{a}.
\newblock \showarticletitle{Reinforcement Learning-based Hierarchical Seed Scheduling for Greybox Fuzzing}. In \bibinfo{booktitle}{\emph{28th Annual Network and Distributed System Security Symposium, {NDSS} 2021, virtually, February 21-25, 2021}}. \bibinfo{publisher}{The Internet Society}.
\newblock
\urldef\tempurl%
\url{https://www.ndss-symposium.org/ndss-paper/reinforcement-learning-based-hierarchical-seed-scheduling-for-greybox-fuzzing/}
\showURL{%
\tempurl}


\bibitem[Wang et~al\mbox{.}(2020)]%
        {wang2020sok}
\bibfield{author}{\bibinfo{person}{Pengfei Wang}, \bibinfo{person}{Xu Zhou}, \bibinfo{person}{Kai Lu}, \bibinfo{person}{Tai Yue}, {and} \bibinfo{person}{Yingying Liu}.} \bibinfo{year}{2020}\natexlab{}.
\newblock \showarticletitle{Sok: The progress, challenges, and perspectives of directed greybox fuzzing}.
\newblock \bibinfo{journal}{\emph{Challenges, and Perspectives of Directed Greybox Fuzzing}} (\bibinfo{year}{2020}).
\newblock


\bibitem[Weiteng et~al\mbox{.}(2024)]%
        {chen2024syzgenplus}
\bibfield{author}{\bibinfo{person}{Chen Weiteng}, \bibinfo{person}{Hao Yu}, \bibinfo{person}{Zhang Zheng}, \bibinfo{person}{Zou Xiaochen}, \bibinfo{person}{Kirat Dhilung}, \bibinfo{person}{Mishra Shachee}, \bibinfo{person}{Schales Douglas}, \bibinfo{person}{Jang Jiyong}, {and} \bibinfo{person}{Qian Zhiyun}.} \bibinfo{year}{2024}\natexlab{}.
\newblock \showarticletitle{SyzGen++: Dependency Inference for Augmenting Kernel Driver Fuzzing}. In \bibinfo{booktitle}{\emph{2024 IEEE Symposium on Security and Privacy (SP)}}.
\newblock


\bibitem[Wu et~al\mbox{.}(2023a)]%
        {wu2023mitigating}
\bibfield{author}{\bibinfo{person}{Yuhang Wu}, \bibinfo{person}{Zhenpeng Lin}, \bibinfo{person}{Yueqi Chen}, \bibinfo{person}{Dang~K Le}, \bibinfo{person}{Dongliang Mu}, {and} \bibinfo{person}{Xinyu Xing}.} \bibinfo{year}{2023}\natexlab{a}.
\newblock \showarticletitle{Mitigating Security Risks in Linux with {KLAUS}: A Method for Evaluating Patch Correctness}. In \bibinfo{booktitle}{\emph{32nd USENIX Security Symposium (USENIX Security 23)}}. \bibinfo{publisher}{USENIX Association}, \bibinfo{address}{Anaheim, CA}, \bibinfo{pages}{4247--4264}.
\newblock
\showISBNx{978-1-939133-37-3}
\urldef\tempurl%
\url{https://www.usenix.org/conference/usenixsecurity23/presentation/wu-yuhang}
\showURL{%
\tempurl}


\bibitem[Wu et~al\mbox{.}(2023b)]%
        {wu23devfuzz}
\bibfield{author}{\bibinfo{person}{Yilun Wu}, \bibinfo{person}{Tong Zhang}, \bibinfo{person}{Changhee Jung}, {and} \bibinfo{person}{Dongyoon Lee}.} \bibinfo{year}{2023}\natexlab{b}.
\newblock \showarticletitle{DevFuzz: Automatic Device Model-Guided Device Driver Fuzzing}. In \bibinfo{booktitle}{\emph{2023 IEEE Symposium on Security and Privacy (SP)}}. \bibinfo{pages}{3246--3261}.
\newblock
\href{https://doi.org/10.1109/SP46215.2023.10179293}{doi:\nolinkurl{10.1109/SP46215.2023.10179293}}


\bibitem[Xia et~al\mbox{.}(2024)]%
        {xia2024fuzz4all}
\bibfield{author}{\bibinfo{person}{Chunqiu~Steven Xia}, \bibinfo{person}{Matteo Paltenghi}, \bibinfo{person}{Jia Le~Tian}, \bibinfo{person}{Michael Pradel}, {and} \bibinfo{person}{Lingming Zhang}.} \bibinfo{year}{2024}\natexlab{}.
\newblock \showarticletitle{Fuzz4All: Universal Fuzzing with Large Language Models}. In \bibinfo{booktitle}{\emph{Proceedings of the IEEE/ACM 46th International Conference on Software Engineering}} (Lisbon, Portugal) \emph{(\bibinfo{series}{ICSE '24})}. \bibinfo{publisher}{Association for Computing Machinery}, \bibinfo{address}{New York, NY, USA}, Article \bibinfo{articleno}{126}, \bibinfo{numpages}{13}~pages.
\newblock
\showISBNx{9798400702174}
\href{https://doi.org/10.1145/3597503.3639121}{doi:\nolinkurl{10.1145/3597503.3639121}}


\bibitem[Xu et~al\mbox{.}(2025a)]%
        {xu2025ckgfuzzer}
\bibfield{author}{\bibinfo{person}{Hanxiang Xu}, \bibinfo{person}{Wei Ma}, \bibinfo{person}{Ting Zhou}, \bibinfo{person}{Yanjie Zhao}, \bibinfo{person}{Kai Chen}, \bibinfo{person}{Qiang Hu}, \bibinfo{person}{Yang Liu}, {and} \bibinfo{person}{Haoyu Wang}.} \bibinfo{year}{2025}\natexlab{a}.
\newblock \showarticletitle{CKGFuzzer: LLM-Based Fuzz Driver Generation Enhanced By Code Knowledge Graph}. In \bibinfo{booktitle}{\emph{2025 IEEE/ACM 47th International Conference on Software Engineering: Companion Proceedings (ICSE-Companion)}}. \bibinfo{pages}{243--254}.
\newblock
\href{https://doi.org/10.1109/ICSE-Companion66252.2025.00079}{doi:\nolinkurl{10.1109/ICSE-Companion66252.2025.00079}}


\bibitem[Xu et~al\mbox{.}(2025b)]%
        {xu2025concur}
\bibfield{author}{\bibinfo{person}{Jiacheng Xu}, \bibinfo{person}{Dylan Wolff}, \bibinfo{person}{Xing~Yi Han}, \bibinfo{person}{Jialin Li}, {and} \bibinfo{person}{Abhik Roychoudhury}.} \bibinfo{year}{2025}\natexlab{b}.
\newblock \bibinfo{title}{Concurrency Testing in the Linux Kernel via eBPF}.
\newblock
\showeprint[arxiv]{2504.21394}~[cs.OS]
\urldef\tempurl%
\url{https://arxiv.org/abs/2504.21394}
\showURL{%
\tempurl}


\bibitem[Xu et~al\mbox{.}(2024)]%
        {xu2024mock}
\bibfield{author}{\bibinfo{person}{Jiacheng Xu}, \bibinfo{person}{Xuhong Zhang}, \bibinfo{person}{Shouling Ji}, \bibinfo{person}{Yuan Tian}, \bibinfo{person}{Binbin Zhao}, \bibinfo{person}{Qinying Wang}, \bibinfo{person}{Peng Cheng}, {and} \bibinfo{person}{Jiming Chen}.} \bibinfo{year}{2024}\natexlab{}.
\newblock \showarticletitle{MOCK: Optimizing Kernel Fuzzing Mutation with Context-aware Dependency}. In \bibinfo{booktitle}{\emph{31st Annual Network and Distributed System Security Symposium, {NDSS} 2024, San Diego, California, USA, February 26 - March 1, 2024}}. \bibinfo{publisher}{The Internet Society}.
\newblock


\bibitem[Xu et~al\mbox{.}(2020)]%
        {meng2020krace}
\bibfield{author}{\bibinfo{person}{Meng Xu}, \bibinfo{person}{Sanidhya Kashyap}, \bibinfo{person}{Hanqing Zhao}, {and} \bibinfo{person}{Taesoo Kim}.} \bibinfo{year}{2020}\natexlab{}.
\newblock \showarticletitle{Krace: Data Race Fuzzing for Kernel File Systems}. In \bibinfo{booktitle}{\emph{2020 IEEE Symposium on Security and Privacy (SP)}}. \bibinfo{pages}{1643--1660}.
\newblock
\href{https://doi.org/10.1109/SP40000.2020.00078}{doi:\nolinkurl{10.1109/SP40000.2020.00078}}


\bibitem[Xu and Huang(2025)]%
        {xu2025opt}
\bibfield{author}{\bibinfo{person}{Qingxiao Xu} {and} \bibinfo{person}{Jeff Huang}.} \bibinfo{year}{2025}\natexlab{}.
\newblock \showarticletitle{Optimizing Type Migration for LLM-Based C-to-Rust Translation: A Data Flow Graph Approach}. In \bibinfo{booktitle}{\emph{Proceedings of the 14th ACM SIGPLAN International Workshop on the State Of the Art in Program Analysis}} (Seoul, Republic of Korea) \emph{(\bibinfo{series}{SOAP '25})}. \bibinfo{publisher}{Association for Computing Machinery}, \bibinfo{address}{New York, NY, USA}, \bibinfo{pages}{8–14}.
\newblock
\showISBNx{9798400719226}
\href{https://doi.org/10.1145/3735544.3735582}{doi:\nolinkurl{10.1145/3735544.3735582}}


\bibitem[Xu et~al\mbox{.}(2019)]%
        {kim2019janus}
\bibfield{author}{\bibinfo{person}{Wen Xu}, \bibinfo{person}{Hyungon Moon}, \bibinfo{person}{Sanidhya Kashyap}, \bibinfo{person}{Po-Ning Tseng}, {and} \bibinfo{person}{Taesoo Kim}.} \bibinfo{year}{2019}\natexlab{}.
\newblock \showarticletitle{Fuzzing File Systems via Two-Dimensional Input Space Exploration}. In \bibinfo{booktitle}{\emph{2019 IEEE Symposium on Security and Privacy (SP)}}. \bibinfo{pages}{818--834}.
\newblock
\href{https://doi.org/10.1109/SP.2019.00035}{doi:\nolinkurl{10.1109/SP.2019.00035}}


\bibitem[Yang et~al\mbox{.}(2025)]%
        {yang2025kernelgpt}
\bibfield{author}{\bibinfo{person}{Chenyuan Yang}, \bibinfo{person}{Zijie Zhao}, {and} \bibinfo{person}{Lingming Zhang}.} \bibinfo{year}{2025}\natexlab{}.
\newblock \showarticletitle{KernelGPT: Enhanced Kernel Fuzzing via Large Language Models}. In \bibinfo{booktitle}{\emph{Proceedings of the 30th ACM International Conference on Architectural Support for Programming Languages and Operating Systems, Volume 2}} (Rotterdam, Netherlands) \emph{(\bibinfo{series}{ASPLOS '25})}. \bibinfo{publisher}{Association for Computing Machinery}, \bibinfo{address}{New York, NY, USA}, \bibinfo{pages}{560–573}.
\newblock
\showISBNx{9798400710797}
\href{https://doi.org/10.1145/3676641.3716022}{doi:\nolinkurl{10.1145/3676641.3716022}}


\bibitem[Y\i{}ld\i{}ran et~al\mbox{.}(2024)]%
        {necip2024maxmizing}
\bibfield{author}{\bibinfo{person}{Necip~Faz\i{}l Y\i{}ld\i{}ran}, \bibinfo{person}{Jeho Oh}, \bibinfo{person}{Julia Lawall}, {and} \bibinfo{person}{Paul Gazzillo}.} \bibinfo{year}{2024}\natexlab{}.
\newblock \showarticletitle{Maximizing Patch Coverage for Testing of Highly-Configurable Software without Exploding Build Times}.
\newblock \bibinfo{journal}{\emph{Proc. ACM Softw. Eng.}} \bibinfo{volume}{1}, \bibinfo{number}{FSE}, Article \bibinfo{articleno}{20} (\bibinfo{date}{July} \bibinfo{year}{2024}), \bibinfo{numpages}{23}~pages.
\newblock
\href{https://doi.org/10.1145/3643746}{doi:\nolinkurl{10.1145/3643746}}


\bibitem[Yin et~al\mbox{.}(2023)]%
        {yin2023kext}
\bibfield{author}{\bibinfo{person}{Tingting Yin}, \bibinfo{person}{Zicong Gao}, \bibinfo{person}{Zhenghang Xiao}, \bibinfo{person}{Zheyu Ma}, \bibinfo{person}{Min Zheng}, {and} \bibinfo{person}{Chao Zhang}.} \bibinfo{year}{2023}\natexlab{}.
\newblock \showarticletitle{{KextFuzz}: Fuzzing {macOS} Kernel {EXTensions} on Apple Silicon via Exploiting Mitigations}. In \bibinfo{booktitle}{\emph{32nd USENIX Security Symposium (USENIX Security 23)}}. \bibinfo{publisher}{USENIX Association}, \bibinfo{address}{Anaheim, CA}, \bibinfo{pages}{5039--5054}.
\newblock
\showISBNx{978-1-939133-37-3}
\urldef\tempurl%
\url{https://www.usenix.org/conference/usenixsecurity23/presentation/yin}
\showURL{%
\tempurl}


\bibitem[Yiru et~al\mbox{.}(2024)]%
        {xu2024saturn}
\bibfield{author}{\bibinfo{person}{Xu Yiru}, \bibinfo{person}{Sun Hao}, \bibinfo{person}{Liu Jianzhong}, \bibinfo{person}{Shen Yuheng}, {and} \bibinfo{person}{Jiang Yu}.} \bibinfo{year}{2024}\natexlab{}.
\newblock \showarticletitle{SATURN: Host-Gadget Synergistic USB Driver Fuzzing}. In \bibinfo{booktitle}{\emph{2024 IEEE Symposium on Security and Privacy (SP)}}.
\newblock


\bibitem[You et~al\mbox{.}(2017)]%
        {you2017semfuzz}
\bibfield{author}{\bibinfo{person}{Wei You}, \bibinfo{person}{Peiyuan Zong}, \bibinfo{person}{Kai Chen}, \bibinfo{person}{XiaoFeng Wang}, \bibinfo{person}{Xiaojing Liao}, \bibinfo{person}{Pan Bian}, {and} \bibinfo{person}{Bin Liang}.} \bibinfo{year}{2017}\natexlab{}.
\newblock \showarticletitle{SemFuzz: Semantics-based Automatic Generation of Proof-of-Concept Exploits}. In \bibinfo{booktitle}{\emph{Proceedings of the 2017 ACM SIGSAC Conference on Computer and Communications Security}} (Dallas, Texas, USA) \emph{(\bibinfo{series}{CCS '17})}. \bibinfo{publisher}{Association for Computing Machinery}, \bibinfo{address}{New York, NY, USA}, \bibinfo{pages}{2139–2154}.
\newblock
\showISBNx{9781450349468}
\href{https://doi.org/10.1145/3133956.3134085}{doi:\nolinkurl{10.1145/3133956.3134085}}


\bibitem[Yu et~al\mbox{.}(2020)]%
        {yu2020ava}
\bibfield{author}{\bibinfo{person}{Hangchen Yu}, \bibinfo{person}{Arthur~Michener Peters}, \bibinfo{person}{Amogh Akshintala}, {and} \bibinfo{person}{Christopher~J. Rossbach}.} \bibinfo{year}{2020}\natexlab{}.
\newblock \showarticletitle{AvA: Accelerated Virtualization of Accelerators}. In \bibinfo{booktitle}{\emph{Proceedings of the Twenty-Fifth International Conference on Architectural Support for Programming Languages and Operating Systems}} (Lausanne, Switzerland) \emph{(\bibinfo{series}{ASPLOS '20})}. \bibinfo{publisher}{Association for Computing Machinery}, \bibinfo{address}{New York, NY, USA}, \bibinfo{pages}{807–825}.
\newblock
\showISBNx{9781450371025}
\href{https://doi.org/10.1145/3373376.3378466}{doi:\nolinkurl{10.1145/3373376.3378466}}


\bibitem[Yuan et~al\mbox{.}(2023)]%
        {yuan2023ddrace}
\bibfield{author}{\bibinfo{person}{Ming Yuan}, \bibinfo{person}{Bodong Zhao}, \bibinfo{person}{Penghui Li}, \bibinfo{person}{Jiashuo Liang}, \bibinfo{person}{Xinhui Han}, \bibinfo{person}{Xiapu Luo}, {and} \bibinfo{person}{Chao Zhang}.} \bibinfo{year}{2023}\natexlab{}.
\newblock \showarticletitle{DDRace: Finding Concurrency {UAF} Vulnerabilities in Linux Drivers with Directed Fuzzing}. In \bibinfo{booktitle}{\emph{32nd {USENIX} Security Symposium, {USENIX} Security 2023, Anaheim, CA, USA, August 9-11, 2023}}, \bibfield{editor}{\bibinfo{person}{Joseph~A. Calandrino} {and} \bibinfo{person}{Carmela Troncoso}} (Eds.). \bibinfo{publisher}{{USENIX} Association}.
\newblock
\urldef\tempurl%
\url{https://www.usenix.org/conference/usenixsecurity23/presentation/yuan-ming}
\showURL{%
\tempurl}


\bibitem[Yue et~al\mbox{.}(2020)]%
        {yue2020ecofuzz}
\bibfield{author}{\bibinfo{person}{Tai Yue}, \bibinfo{person}{Pengfei Wang}, \bibinfo{person}{Yong Tang}, \bibinfo{person}{Enze Wang}, \bibinfo{person}{Bo Yu}, \bibinfo{person}{Kai Lu}, {and} \bibinfo{person}{Xu Zhou}.} \bibinfo{year}{2020}\natexlab{}.
\newblock \showarticletitle{{EcoFuzz}: Adaptive {Energy-Saving} Greybox Fuzzing as a Variant of the Adversarial {Multi-Armed} Bandit}. In \bibinfo{booktitle}{\emph{29th USENIX Security Symposium (USENIX Security 20)}}. \bibinfo{publisher}{USENIX Association}, \bibinfo{pages}{2307--2324}.
\newblock
\showISBNx{978-1-939133-17-5}
\urldef\tempurl%
\url{https://www.usenix.org/conference/usenixsecurity20/presentation/yue}
\showURL{%
\tempurl}


\bibitem[Yun et~al\mbox{.}(2018)]%
        {yun2018qsym}
\bibfield{author}{\bibinfo{person}{Insu Yun}, \bibinfo{person}{Sangho Lee}, \bibinfo{person}{Meng Xu}, \bibinfo{person}{Yeongjin Jang}, {and} \bibinfo{person}{Taesoo Kim}.} \bibinfo{year}{2018}\natexlab{}.
\newblock \showarticletitle{{QSYM} : A Practical Concolic Execution Engine Tailored for Hybrid Fuzzing}. In \bibinfo{booktitle}{\emph{27th USENIX Security Symposium (USENIX Security 18)}}. \bibinfo{publisher}{USENIX Association}, \bibinfo{address}{Baltimore, MD}, \bibinfo{pages}{745--761}.
\newblock
\showISBNx{978-1-939133-04-5}
\urldef\tempurl%
\url{https://www.usenix.org/conference/usenixsecurity18/presentation/yun}
\showURL{%
\tempurl}


\bibitem[Yun et~al\mbox{.}(2022)]%
        {yun2022fuzzing}
\bibfield{author}{\bibinfo{person}{Joobeom Yun}, \bibinfo{person}{Fayozbek Rustamov}, \bibinfo{person}{Juhwan Kim}, {and} \bibinfo{person}{Youngjoo Shin}.} \bibinfo{year}{2022}\natexlab{}.
\newblock \showarticletitle{Fuzzing of Embedded Systems: A Survey}.
\newblock \bibinfo{journal}{\emph{Comput. Surveys}} \bibinfo{volume}{55}, \bibinfo{number}{7} (\bibinfo{year}{2022}), \bibinfo{pages}{1--33}.
\newblock


\bibitem[Zalewski(2013)]%
        {afl}
\bibfield{author}{\bibinfo{person}{Michal Zalewski}.} \bibinfo{year}{2013}\natexlab{}.
\newblock \bibinfo{title}{American Fuzzy Lop: a security-oriented fuzzer}.
\newblock \bibinfo{howpublished}{\url{https://lcamtuf.coredump.cx/afl/}}.
\newblock


\bibitem[Zhang et~al\mbox{.}(2022b)]%
        {zhang2022mobfuzz}
\bibfield{author}{\bibinfo{person}{Gen Zhang}, \bibinfo{person}{Pengfei Wang}, \bibinfo{person}{Tai Yue}, \bibinfo{person}{Xiangdong Kong}, \bibinfo{person}{Shan Huang}, \bibinfo{person}{Xu Zhou}, {and} \bibinfo{person}{Kai Lu}.} \bibinfo{year}{2022}\natexlab{b}.
\newblock \showarticletitle{MobFuzz: Adaptive Multi-objective Optimization in Gray-box Fuzzing}. In \bibinfo{booktitle}{\emph{29th Annual Network and Distributed System Security Symposium, {NDSS} 2022, San Diego, California, USA, April 24-28, 2022}}. \bibinfo{publisher}{The Internet Society}.
\newblock
\urldef\tempurl%
\url{https://www.ndss-symposium.org/ndss-paper/auto-draft-199/}
\showURL{%
\tempurl}


\bibitem[Zhang et~al\mbox{.}(2022a)]%
        {zhang2022exploit}
\bibfield{author}{\bibinfo{person}{Lei Zhang}, \bibinfo{person}{Keke Lian}, \bibinfo{person}{Haoyu Xiao}, \bibinfo{person}{Zhibo Zhang}, \bibinfo{person}{Peng Liu}, \bibinfo{person}{Yuan Zhang}, \bibinfo{person}{Min Yang}, {and} \bibinfo{person}{Haixin Duan}.} \bibinfo{year}{2022}\natexlab{a}.
\newblock \showarticletitle{Exploit the Last Straw That Breaks Android Systems}. In \bibinfo{booktitle}{\emph{2022 IEEE Symposium on Security and Privacy (SP)}}. \bibinfo{pages}{2230--2247}.
\newblock
\href{https://doi.org/10.1109/SP46214.2022.9833563}{doi:\nolinkurl{10.1109/SP46214.2022.9833563}}


\bibitem[Zhang et~al\mbox{.}(2025)]%
        {zhang2025unlocking}
\bibfield{author}{\bibinfo{person}{Zhiyu Zhang}, \bibinfo{person}{Longxing Li}, \bibinfo{person}{Ruigang Liang}, {and} \bibinfo{person}{Kai Chen}.} \bibinfo{year}{2025}\natexlab{}.
\newblock \showarticletitle{Unlocking Low Frequency Syscalls in Kernel Fuzzing with Dependency-Based RAG}.
\newblock \bibinfo{journal}{\emph{Proc. ACM Softw. Eng.}} \bibinfo{volume}{2}, \bibinfo{number}{ISSTA}, Article \bibinfo{articleno}{ISSTA038} (\bibinfo{date}{June} \bibinfo{year}{2025}), \bibinfo{numpages}{23}~pages.
\newblock
\href{https://doi.org/10.1145/3728913}{doi:\nolinkurl{10.1145/3728913}}


\bibitem[Zhang et~al\mbox{.}(2021)]%
        {zhang2021stochfuzz}
\bibfield{author}{\bibinfo{person}{Zhuo Zhang}, \bibinfo{person}{Wei You}, \bibinfo{person}{Guanhong Tao}, \bibinfo{person}{Yousra Aafer}, \bibinfo{person}{Xuwei Liu}, {and} \bibinfo{person}{Xiangyu Zhang}.} \bibinfo{year}{2021}\natexlab{}.
\newblock \showarticletitle{StochFuzz: Sound and Cost-effective Fuzzing of Stripped Binaries by Incremental and Stochastic Rewriting}. In \bibinfo{booktitle}{\emph{2021 IEEE Symposium on Security and Privacy (SP)}}. \bibinfo{pages}{659--676}.
\newblock
\href{https://doi.org/10.1109/SP40001.2021.00109}{doi:\nolinkurl{10.1109/SP40001.2021.00109}}


\bibitem[Zhao et~al\mbox{.}(2022a)]%
        {zhao2023state}
\bibfield{author}{\bibinfo{person}{Bodong Zhao}, \bibinfo{person}{Zheming Li}, \bibinfo{person}{Shisong Qin}, \bibinfo{person}{Zheyu Ma}, \bibinfo{person}{Ming Yuan}, \bibinfo{person}{Wenyu Zhu}, \bibinfo{person}{Zhihong Tian}, {and} \bibinfo{person}{Chao Zhang}.} \bibinfo{year}{2022}\natexlab{a}.
\newblock \showarticletitle{{StateFuzz}: System {Call-Based} {State-Aware} Linux Driver Fuzzing}. In \bibinfo{booktitle}{\emph{31st USENIX Security Symposium (USENIX Security 22)}}. \bibinfo{publisher}{USENIX Association}, \bibinfo{address}{Boston, MA}, \bibinfo{pages}{3273--3289}.
\newblock
\showISBNx{978-1-939133-31-1}
\urldef\tempurl%
\url{https://www.usenix.org/conference/usenixsecurity22/presentation/zhao-bodong}
\showURL{%
\tempurl}


\bibitem[Zhao et~al\mbox{.}(2022b)]%
        {zhao2022semantic}
\bibfield{author}{\bibinfo{person}{Wenjia Zhao}, \bibinfo{person}{Kangjie Lu}, \bibinfo{person}{Qiushi Wu}, {and} \bibinfo{person}{Yong Qi}.} \bibinfo{year}{2022}\natexlab{b}.
\newblock \showarticletitle{Semantic-Informed Driver Fuzzing Without Both the Hardware Devices and the Emulators}. In \bibinfo{booktitle}{\emph{29th Annual Network and Distributed System Security Symposium, {NDSS} 2022, San Diego, California, USA, April 24-28, 2022}}. \bibinfo{publisher}{The Internet Society}.
\newblock
\urldef\tempurl%
\url{https://www.ndss-symposium.org/ndss-paper/auto-draft-248/}
\showURL{%
\tempurl}


\bibitem[Zheng et~al\mbox{.}(2019)]%
        {zheng2019firm}
\bibfield{author}{\bibinfo{person}{Yaowen Zheng}, \bibinfo{person}{Ali Davanian}, \bibinfo{person}{Heng Yin}, \bibinfo{person}{Chengyu Song}, \bibinfo{person}{Hongsong Zhu}, {and} \bibinfo{person}{Limin Sun}.} \bibinfo{year}{2019}\natexlab{}.
\newblock \showarticletitle{{FIRM-AFL}: {High-Throughput} Greybox Fuzzing of {IoT} Firmware via Augmented Process Emulation}. In \bibinfo{booktitle}{\emph{28th USENIX Security Symposium (USENIX Security 19)}}. \bibinfo{publisher}{USENIX Association}, \bibinfo{address}{Santa Clara, CA}, \bibinfo{pages}{1099--1114}.
\newblock
\showISBNx{978-1-939133-06-9}
\urldef\tempurl%
\url{https://www.usenix.org/conference/usenixsecurity19/presentation/zheng}
\showURL{%
\tempurl}


\bibitem[Zheng et~al\mbox{.}(2022)]%
        {zhang2022efficent}
\bibfield{author}{\bibinfo{person}{Yaowen Zheng}, \bibinfo{person}{Yuekang Li}, \bibinfo{person}{Cen Zhang}, \bibinfo{person}{Hongsong Zhu}, \bibinfo{person}{Yang Liu}, {and} \bibinfo{person}{Limin Sun}.} \bibinfo{year}{2022}\natexlab{}.
\newblock \showarticletitle{Efficient greybox fuzzing of applications in Linux-based IoT devices via enhanced user-mode emulation}. In \bibinfo{booktitle}{\emph{Proceedings of the 31st ACM SIGSOFT International Symposium on Software Testing and Analysis}} (Virtual, South Korea) \emph{(\bibinfo{series}{ISSTA 2022})}. \bibinfo{publisher}{Association for Computing Machinery}, \bibinfo{address}{New York, NY, USA}, \bibinfo{pages}{417–428}.
\newblock
\showISBNx{9781450393799}
\href{https://doi.org/10.1145/3533767.3534414}{doi:\nolinkurl{10.1145/3533767.3534414}}


\bibitem[Zhou et~al\mbox{.}(2021)]%
        {zhou2021automatic}
\bibfield{author}{\bibinfo{person}{Wei Zhou}, \bibinfo{person}{Le Guan}, \bibinfo{person}{Peng Liu}, {and} \bibinfo{person}{Yuqing Zhang}.} \bibinfo{year}{2021}\natexlab{}.
\newblock \showarticletitle{Automatic Firmware Emulation through Invalidity-guided Knowledge Inference}. In \bibinfo{booktitle}{\emph{30th USENIX Security Symposium (USENIX Security 21)}}. \bibinfo{publisher}{USENIX Association}, \bibinfo{pages}{2007--2024}.
\newblock
\showISBNx{978-1-939133-24-3}
\urldef\tempurl%
\url{https://www.usenix.org/conference/usenixsecurity21/presentation/zhou}
\showURL{%
\tempurl}


\bibitem[Zhou et~al\mbox{.}(2025)]%
        {zhou2025benchmark}
\bibfield{author}{\bibinfo{person}{Zhenhao Zhou}, \bibinfo{person}{Zhuochen Huang}, \bibinfo{person}{Yike He}, \bibinfo{person}{Chong Wang}, \bibinfo{person}{Jiajun Wang}, \bibinfo{person}{Yijian Wu}, \bibinfo{person}{Xin Peng}, {and} \bibinfo{person}{Yiling Lou}.} \bibinfo{year}{2025}\natexlab{}.
\newblock \bibinfo{title}{Benchmarking and Enhancing LLM Agents in Localizing Linux Kernel Bugs}.
\newblock
\showeprint[arxiv]{2505.19489}~[cs.AI]
\urldef\tempurl%
\url{https://arxiv.org/abs/2505.19489}
\showURL{%
\tempurl}


\bibitem[Zhu et~al\mbox{.}(2024)]%
        {zhu2024cross}
\bibfield{author}{\bibinfo{person}{Jiaxun Zhu}, \bibinfo{person}{Minghao Lin}, \bibinfo{person}{Tingting Yin}, \bibinfo{person}{Zechao Cai}, \bibinfo{person}{Yu Wang}, \bibinfo{person}{Rui Chang}, {and} \bibinfo{person}{Wenbo Shen}.} \bibinfo{year}{2024}\natexlab{}.
\newblock \showarticletitle{CrossFire: Fuzzing macOS Cross-XPU Memory on Apple Silicon}. In \bibinfo{booktitle}{\emph{Proceedings of the 2024 on ACM SIGSAC Conference on Computer and Communications Security}} (Salt Lake City, UT, USA) \emph{(\bibinfo{series}{CCS '24})}. \bibinfo{publisher}{Association for Computing Machinery}, \bibinfo{address}{New York, NY, USA}, \bibinfo{pages}{3749–3762}.
\newblock
\showISBNx{9798400706363}
\href{https://doi.org/10.1145/3658644.3690376}{doi:\nolinkurl{10.1145/3658644.3690376}}


\bibitem[Zhu et~al\mbox{.}(2022)]%
        {zhu2022fuzzing}
\bibfield{author}{\bibinfo{person}{Xiaogang Zhu}, \bibinfo{person}{Sheng Wen}, \bibinfo{person}{Seyit Camtepe}, {and} \bibinfo{person}{Yang Xiang}.} \bibinfo{year}{2022}\natexlab{}.
\newblock \showarticletitle{Fuzzing: a survey for roadmap}.
\newblock \bibinfo{journal}{\emph{ACM Computing Surveys (CSUR)}} \bibinfo{volume}{54}, \bibinfo{number}{11s} (\bibinfo{year}{2022}), \bibinfo{pages}{1--36}.
\newblock


\bibitem[Zou et~al\mbox{.}(2024)]%
        {zou2024syzbridge}
\bibfield{author}{\bibinfo{person}{Xiaochen Zou}, \bibinfo{person}{Yu Hao}, \bibinfo{person}{Zheng Zhang}, \bibinfo{person}{Juefei Pu}, \bibinfo{person}{Weiteng Chen}, {and} \bibinfo{person}{Zhiyun Qian}.} \bibinfo{year}{2024}\natexlab{}.
\newblock \showarticletitle{SyzBridge: Bridging the Gap in Exploitability Assessment of Linux Kernel Bugs in the Linux Ecosystem}. In \bibinfo{booktitle}{\emph{31st Annual Network and Distributed System Security Symposium, {NDSS} 2024, San Diego, California, USA, February 26 - March 1, 2024}}. \bibinfo{publisher}{The Internet Society}.
\newblock


\bibitem[Zou et~al\mbox{.}(2022)]%
        {zou2022syzscope}
\bibfield{author}{\bibinfo{person}{Xiaochen Zou}, \bibinfo{person}{Guoren Li}, \bibinfo{person}{Weiteng Chen}, \bibinfo{person}{Hang Zhang}, {and} \bibinfo{person}{Zhiyun Qian}.} \bibinfo{year}{2022}\natexlab{}.
\newblock \showarticletitle{{SyzScope}: Revealing {High-Risk} Security Impacts of {Fuzzer-Exposed} Bugs in Linux kernel}. In \bibinfo{booktitle}{\emph{31st USENIX Security Symposium (USENIX Security 22)}}. \bibinfo{publisher}{USENIX Association}, \bibinfo{address}{Boston, MA}, \bibinfo{pages}{3201--3217}.
\newblock
\showISBNx{978-1-939133-31-1}
\urldef\tempurl%
\url{https://www.usenix.org/conference/usenixsecurity22/presentation/zou}
\showURL{%
\tempurl}


\end{thebibliography}

\end{document}